\documentclass[useAMS,usenatbib]{mn2e}
\usepackage{color}
\usepackage{graphicx}
\usepackage{caption}
\usepackage{amsmath}
\usepackage{multicol}
\usepackage{amssymb}
\usepackage{savesym}
\usepackage{color}
\usepackage{subfig}

\newcommand{\mnras}{MNRAS}
\newcommand{\apjl}{ApJL}
\newcommand{\apj}{ApJ}
\newcommand{\apjs}{ApJS}
\newcommand{\aj}{AJ}
\newcommand{\nat}{Nature}
\newcommand{\aap}{A\&A}

\newcommand{\pasj}{PASJ}

\newcommand{\apss}{Ap\&SS}
\newcommand{\araa}{ARA\&A}

\newcommand{\nar}{NAR}

\newcommand{\sqchi}{\hbox{$\chi^{2}$}}
\newcommand{\Msun}{$M_{\odot}$}

\newcommand{\centi}{\text{c}}
\newcommand{\kilo}{\text{k}}

\newcommand{\meter}{\text{m}}

\newcommand{\second}{\text{s}}

\newcommand{\ev}{\text{eV}}

\newcommand{\gram}{\text{g}}
\newcommand{\guass}{\text{G}}
\newcommand{\ergs}{\text{erg s$^{-1}$}}

\newcommand{\E}[1]{\times10^{#1}}
\newcommand{\Msol}{{\rm M}_{\odot}}
\newcommand{\U}[3]{$#1_{-#3}^{+#2}$}

\title[Ultraluminous supersoft sources]
  {Optically thick outflows in ultraluminous supersoft sources}
\author[R. Urquhart \& R. Soria]
  {R.~Urquhart$^1${\thanks{E-mail:
ryan.urquhart@icrar.org (RU); \newline roberto.soria@curtin.edu.au (RS)}},  R.~Soria$^1$ \\
  $^1$International Centre for Radio Astronomy Research, Curtin University, GPO Box U1987, Perth, WA 6845, Australia}
\date{Accepted 2015 October 2nd}

\pagerange{\pageref{firstpage}--\pageref{lastpage}} \pubyear{2002}

\def\LaTeX{L\kern-.36em\raise.3ex\hbox{a}\kern-.15em
    T\kern-.1667em\lower.7ex\hbox{E}\kern-.125emX}

\begin{document}

\label{firstpage}

\maketitle

\begin{abstract}
Ultraluminous supersoft sources (ULSs) are defined by a thermal spectrum with colour temperatures $\sim$0.1 keV, bolometric luminosities $\sim$ a few $10^{39}$ erg s$^{-1}$, and almost no emission above 1 keV. It has never been clear how they fit into the general scheme of accreting compact objects. To address this problem, we studied a sample of seven ULSs with extensive {\it Chandra} and {\it XMM-Newton} coverage. We find an anticorrelation between fitted temperatures and radii of the thermal emitter, and no correlation between bolometric luminosity and radius or temperature. We compare the physical parameters of ULSs with those of classical supersoft sources, thought to be surface-nuclear-burning white dwarfs, and of ultraluminous X-ray sources (ULXs), thought to be super-Eddington stellar-mass black holes. We argue that ULSs are the sub-class of ULXs seen through the densest wind, perhaps an extension of the soft-ultraluminous regime. We suggest that in ULSs, the massive disk outflow becomes effectively optically thick and forms a large photosphere, shrouding the inner regions from our view. Our model predicts that when the photosphere expands to $\ga 10^5$ km and the temperature decreases below $\approx$50 eV, ULSs become brighter in the far-UV but undetectable in X-rays. Conversely, we find that harder emission components begin to appear in ULSs when the fitted size of the thermal emitter is smallest (interpreted as a shrinking of the photosphere). The observed short-term variability and absorption edges are also consistent with clumpy outflows. We suggest that the transition between ULXs (with a harder tail) and ULSs (with only a soft thermal component) occurs at blackbody temperatures of $\approx$150 eV.
\end{abstract}

\begin{keywords}
 accretion, accretion disks -- stars: black holes -- X-rays: binaries
\end{keywords}

\section{Introduction} \label{intro}

Ultraluminous supersoft sources (ULSs) are a rare class of accreting objects in nearby galaxies, characterised by strong thermal X-ray emission below $1\,\kilo\ev$ with little, or no, flux observed at higher energies \citep{2003ApJ...590L..13K}.  This is in sharp contrast to X-ray binaries (XRBs) and classical ultraluminous X-ray sources (ULXs), which have either broadband emission over the 1--10 keV range or, for XRBs in the high/soft state, a peak disk-blackbody temperature $\sim1\,\kilo\ev$. If fitted with a blackbody model, ULSs have characteristic bolometric luminosities of $\sim$ a few $10^{39} \,\ergs$, radii $\sim10^4\,\kilo\meter$ and peak temperatures $\lesssim 100 \,\ev$. Originally discovered by {\it ROSAT} observations of nearby galaxies \citep{1996A&A...311...35V, 1997MNRAS.286..626R, 2001A&A...373..473R, 2002ApJ...574..382S}, these rare sources have been studied in more detail with {\it Chandra} and {\it XMM-Newton} over the past 15 years. These objects have all been observed at various luminosities over numerous years indicating that they cannot be explained as classical novae. 
The source of the thermal component in ULSs has been interpreted in at least three different ways: (1) nuclear burning on the surface of white dwarfs (WDs); (2) accretion disks around intermediate-mass black holes (IMBHs); or (3) outflows driven by super-Eddington accreting stellar-mass black holes (BHs).

In the first scenario, ULSs may be the extreme end of the `classical' supersoft source (SSS) class, first discovered in the Large Magellanic Cloud \citep{1981ApJ...248..925L}. Classical SSSs are likely powered by surface-nuclear-burning on WD accretors \citep{1992A&A...262...97V}. A problem with this interpretation is that ULSs exceed the Eddington limit ($\approx 2\E{38}\,\ergs$) of a $1.35\,$\Msun WD. Another possible problem is the discrepancy between typical WD radii and the larger characteristic blackbody radii fitted to ULS spectra. The larger radius may be explained if the nuclear-burning WD develops an expanding envelope or an optically thick outflow.


An alternative suggestion is that ULSs contain IMBHs in a high/soft state \citep{2004ApJ...617L..49K}. If the fitted radius comes from the innermost stable circular orbit of an accretion disc, the mass of the central object is $\sim$$10^3$--$10^4$\,\Msun. One problem with the IMBH interpretation is that such massive accretors are not expected to be in their high/soft state at the luminosities inferred for ULSs. For example, an IMBH with a luminosity $\sim 10^{39}\,\ergs$ is only accreting at $\lesssim 1\%$ of its Eddington limit, which is below what is expected for the system to be in the high/soft state. 

Finally, ULSs could be powered by stellar-mass BHs  \citep{2003ApJ...582..184M} or neutron stars (NSs) \citep{1993A&A...278L..43K, 2014Natur.514..202B} accreting at super-Eddington rates. The high-mass transfer rate results in a strong, optically thick outflow that obscures the inner, high-energy emitting region \citep{2007MNRAS.377.1187P, 2015MNRAS.447L..60S}.  The X-ray photons from the central source are reprocessed by the outflow and only the soft thermal spectrum of its photosphere is observed. This would suggest that ULSs are simply standard ULXs at higher accretion rates (and consequently, denser winds).

Here we reanalyse a large sample of \textit{Chandra} and \textit{XMM-Newton} observations of seven previously known ULSs in nearby galaxies.  We discuss some of their X-ray colour, spectral and timing properties in an attempt to determine whether they form a single physical class of sources, and identify their nature. We compare the physical properties of this population of sources with those of broad-band ULXs and classical SSSs.

\section{Target Selection}

\subsection{Definition of ULS}

 To provide a more quantitative definition of ULS for our study, we introduce a classification similar (but not identical) to the one adopted by \citet{2003ApJ...592..884D}. We define three X-ray energy bands: 0.3--1.1 keV (S), 1.1--2.5 keV (M) and 2.5--7 keV (H). We measure the net count rates in the three bands and in the total 0.3--7 keV band (T). We consider ``supersoft" all those sources that have a hardness ratio $({\mathrm M}-{\mathrm S})/{\mathrm T} \la -0.8$, and ``ultraluminous" supersoft sources those that have reached an extrapolated bolometric luminosity of the thermal component $L^{\rm bb}_{\rm bol} \gtrsim 10^{39}$ in at least one observation, during which they also had $({\mathrm M}-{\mathrm S})/{\mathrm T} \la -0.8$.

We only consider sources observed with \textit{Chandra} and \textit{XMM-Newton}, because they cover a similar energy range and provide roughly similar hardness ratios. We stress that the hardness ratio condition is a purely empirical criterion for a practical target selection. We are aware that there are small differences between the colours determined from \textit{Chandra} and \textit{XMM-Newton}, as well as between \textit{Chandra} observations taken in different years, because of the sensitivity degradation at soft energies. We are also aware that extrapolating the blackbody spectrum to the UV gives only a crude estimate of the true bolometric luminosity. At the same time, the characteristic appearance of those sources (their soft thermal spectrum) makes them sufficiently distinct from ordinary XRBs and ULXs, regardless of small changes in the choice of empirical definition, and suggests that they do belong to a physically different class or accretion state. One of the objectives of this paper is to quantify the common properties of such sources, so that we can then move from a purely empirical to a more physical identification.

\begin{figure}
    \hspace{-0.5cm}
    \centering
    \includegraphics[width=82mm]{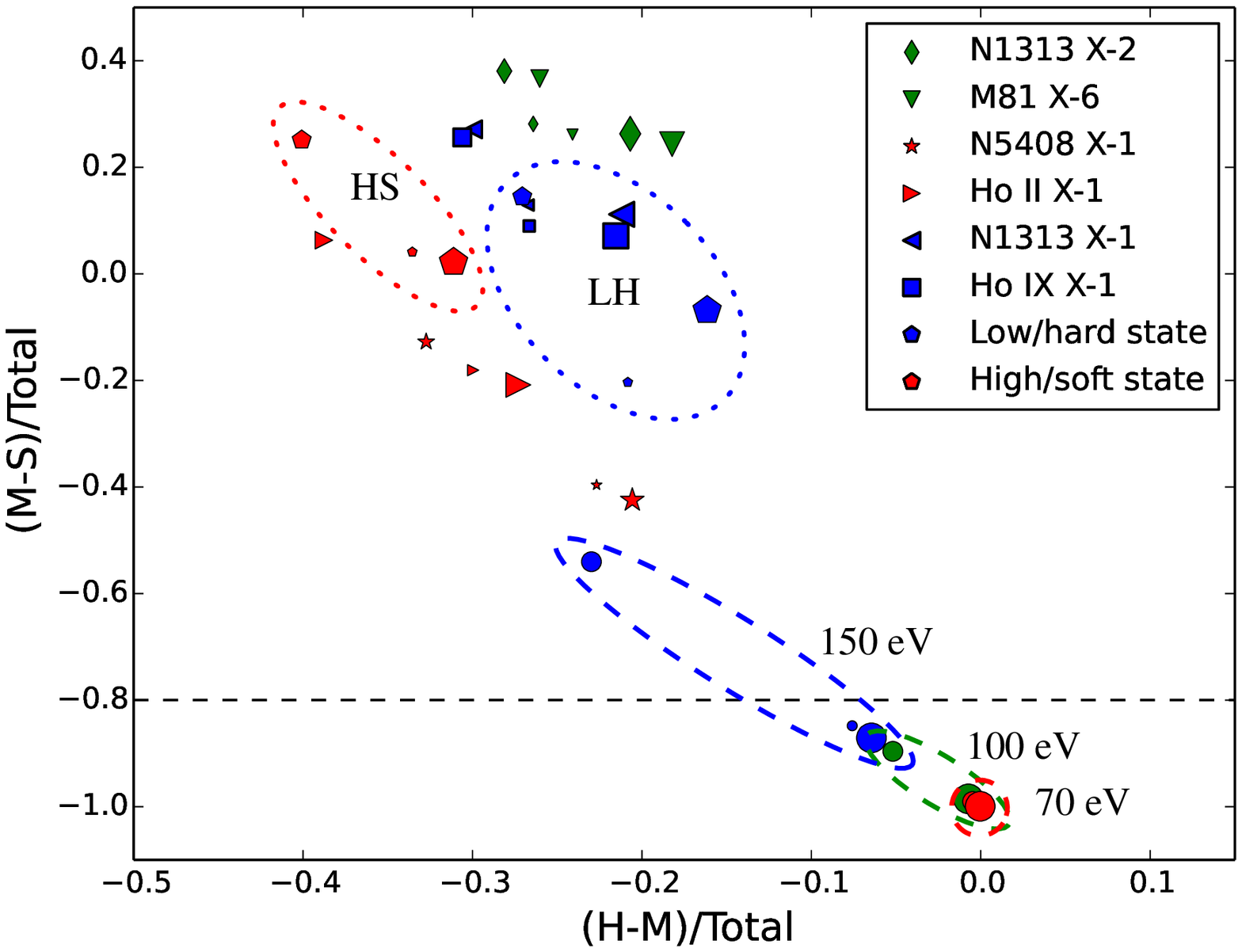}\\[-3pt]
    \hspace{-0.5cm} 
    \includegraphics[width=82mm]{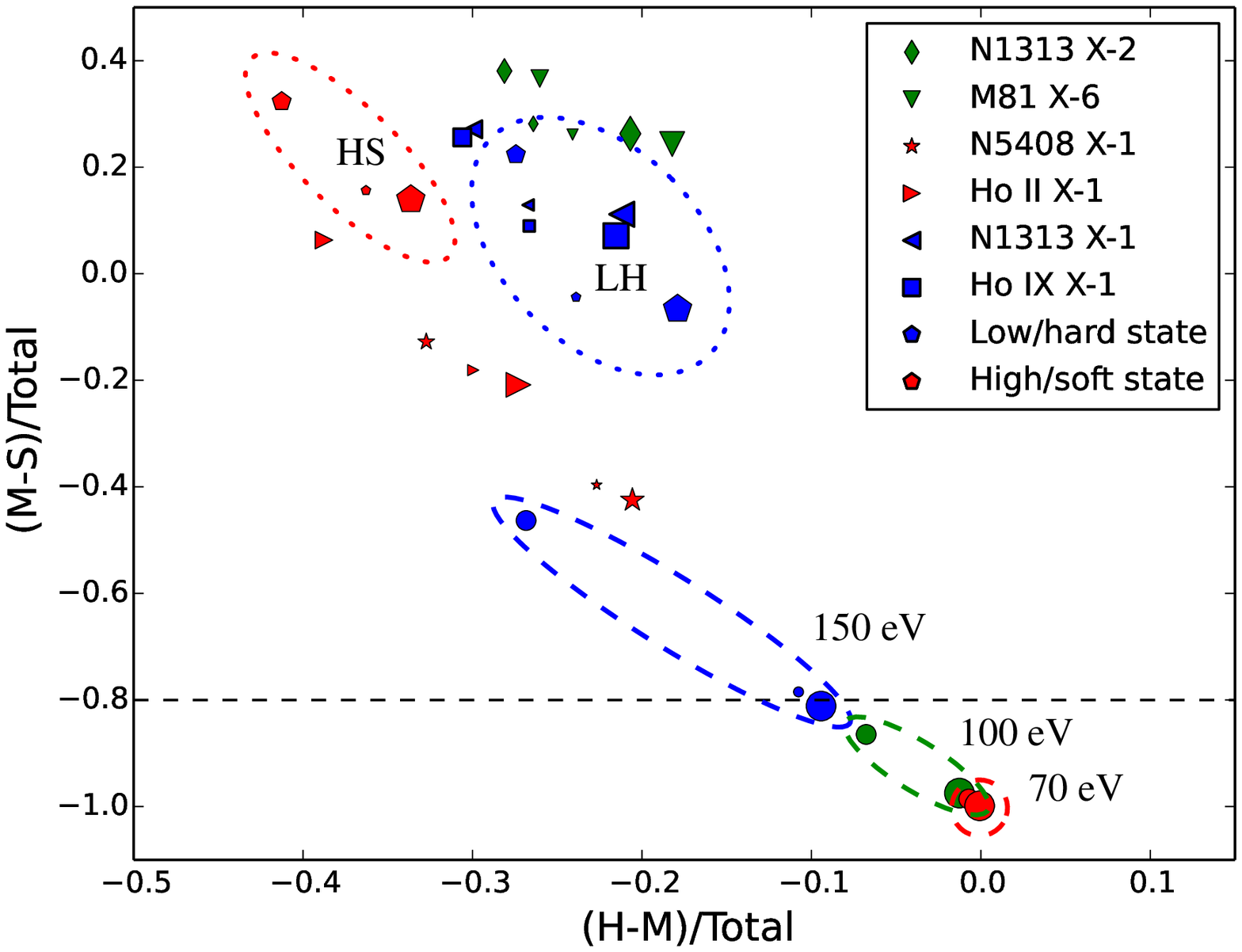}\\[-3pt]
    \hspace{-0.5cm} 
    \includegraphics[width=82mm]{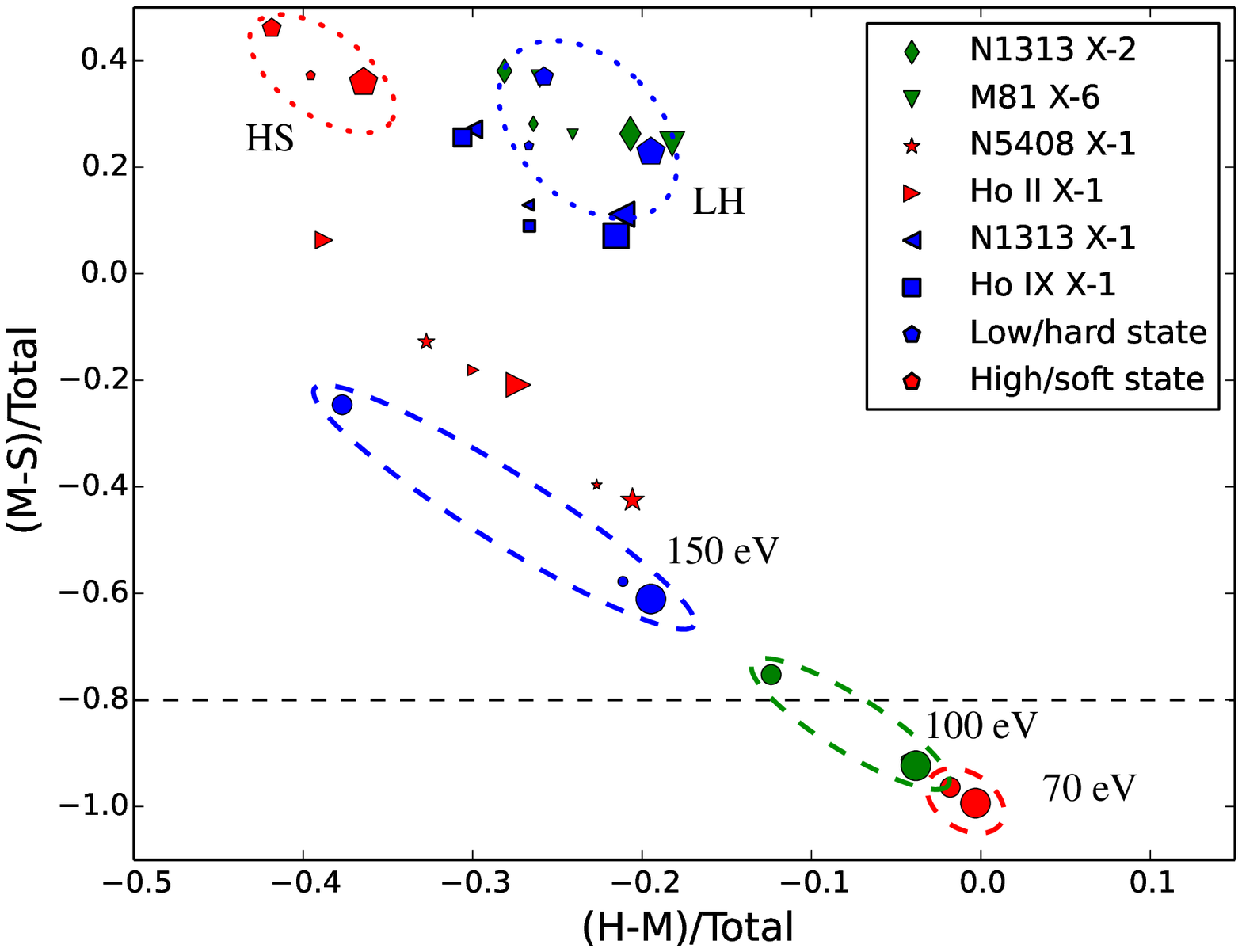}
\caption{Top panel: X-ray colours expected for five different spectral models, when observed through a column density $n_{\rm H} = 3 \times 10^{20}$ cm$^{-2}$, by three different detectors: ACIS-S in 2001 (smallest-size symbols), ACIS-S in 2014 (medium-size symbols) and EPIC-pn in 2014 (largest-size symbols). The models are: LH = low/hard state = power-law with $\Gamma=1.7$; HS=high/soft state = disk-blackbody with $kT_{\rm in} = 1$ keV; and pure blackbodies at 70 eV, 100 eV and 150 eV. Finally, we overplotted the colours expected through the same three detectors for six ULX spectral models, from \citet{2013MNRAS.435.1758S}. The dashed and or dotted regions are simply visual aids to help locating the different models. The dashed horizontal line is an empirical colour threshold for thermal sources at $kT_{\rm bb} \la 100$ eV. Middle panel: as in the top panel, but for $n_{\rm H} = 10^{21}$ cm$^{-2}$. Bottom panel: as in the top panel, but for $n_{\rm H} = 3 \times 10^{21}$ cm$^{-2}$.}
\label{colour_plot}
\end{figure}

We illustrate our selection criterion in Figure \ref{colour_plot}, for three different values of absorbing column density ($n_{\rm H} = 3\times 10^{20}$ cm$^{-2}$, $n_{\rm H} = 10^{21}$ cm$^{-2}$ and $n_{\rm H} = 3\times 10^{21}$ cm$^{-2}$ broadly representative of the range empirically found in our sample (Section 4.1). For each value of $n_{\rm H}$ we plotted the ``observable" X-ray colours of two simple spectral models representative of accreting BHs \citep{2006ARA&A..44...49R}: a power-law spectrum with photon index $\Gamma = 1.7$ (low/hard state), and a disk-blackbody with peak temperature $kT_{\rm in} = 1$ keV (high/soft state). By ``observable", we mean the colours that such model spectra would have if they had been observed by {\it Chandra}/ACIS-S in 2001, or by {\it Chandra}/ACIS-S in 2014, or by {\it XMM-Newton}/EPIC-pn (the colour evolution of this detector over the years is negligible). Then, in each of the three panels, we added the observable colours of three pure blackbody models which may be more representative of ULSs: one at $kT_{\rm bb} = 70$ eV, one at $kT_{\rm bb} = 100$ eV, and one at $kT_{\rm bb} = 150$ eV. The empirical hardness criterion $({\mathrm M}-{\mathrm S})/{\mathrm T} < -0.8$ clearly separates 70-eV and 100-eV blackbody sources from the low/hard state and high/soft state colours, for all three values of $n_{\rm H}$; for $n_{\rm H} \la 10^{21}$ cm$^{-2}$, 150-eV blackbody emitters also satisfy $({\mathrm M}-{\mathrm S})/{\mathrm T} < -0.8$ at least for early {\it Chandra} observations (no longer today). Finally, in all three plots we added the observable colours of six real ULX spectra, modelled by \citet{2013MNRAS.435.1758S}: two in the broadened-disk regime (ObsID 0405090101 of NGC\,1313 X-2 and ObsID 0200980101 of M\,81 X-6), two in the soft-ultraluminous regime (ObsID 0200470101 of Holmberg II X-1 and ObsID 0653380301 of NGC\,5408 X-1) and two in the hard-ultraluminous regime (ObsID 0200980101 of Holmberg IX X-1 and ObsID 0405090101 of NGC\,1313 X-1). Each of those six ULXs has only one value of $n_{\rm H}$ (the actual value determined from that particular observation, fitted by \citealt{2013MNRAS.435.1758S}) but is plotted three times, to represent the ``equivalent" colours such a source would have in {\it Chandra}/ACIS-S 2001, {\it Chandra}/ACIS-S 2014, and {\it XMM-Newton}/EPIC-pn. The message to take home from Figure \ref{colour_plot} is that the observed colours of standard ULXs are somewhat intermediate between those expected for BHs in the low/hard and high/soft states, but are clearly well above the threshold defined by $({\mathrm M}-{\mathrm S})/{\mathrm T} \approx -0.8$; conversely, sources observed below that threshold are clearly not in the low/hard or high/soft state, do not overlap with standard ULXs, and are likely to be blackbody emitters at $kT_{\rm bb} \la 100$ eV. It can be also noted from Figure \ref{colour_plot} that some soft thermal sources observed with {\it Chandra}/ACIS-S today (as opposed to a decade ago) would not appear supersoft, based on their X-ray colours alone. The reason is of course the loss of sensitivity of the detector at low energies, which means that the tail of the blackbody emission in the 1.1--2.5 keV band now represents a larger {\it relative} fraction of the total detected count rate (which has decreased by a factor of 3). Finally, we did not bother to plot the expected colours for $n_{\rm H} > 3 \times 10^{21}$ cm$^{-2}$ because at such column densities it becomes increasingly difficult to identify supersoft sources, so it is no longer relevant to our selection criterion.

In any classification based purely on X-ray hardness and colours (as discussed in \citealt{2003ApJ...592..884D}), the supersoft group of sources is likely to contain a mix of different physical objects, in particular nuclear-burning WDs as well as fading supernova remnants. Indeed, a deep X-ray study of the spiral galaxy M\,83 has shown \citep{2014ApJS..212...21L} two types of supersoft spectra, one dominated by optically thick thermal emission and the other by an optically thin thermal plasma. However, for this work we are considering only sources that have reached $L^{\rm bb}_{\rm bol} \gtrsim 10^{39}$, which screens out thermal supernova remnants (typically fainter than about $10^{37}$ erg s$^{-1}$). We used the extrapolated bolometric luminosity as a selection criterion, rather than the X-ray luminosity, because most of the emission falls below the sensitivity band of {\it Chandra} and {\it XMM-Newton}.

\subsection{Definition of our target sample} 

We analyzed a sample of sources that were unequivocally identified in the literature as very luminous and unusually soft, and we checked whether such sources satisfied our selection criterion defined in Section 2.1. We also examined the sample of ULXs from the catalogues of \citet{2011ApJ...741...49S} and \citet{2004ApJS..154..519S}, looking for any other unrecognized ULS candidates. We limited our search to sources that had enough counts for at least a rough spectral fitting ($\gtrsim$50 net counts, using Cash statistics) in at least one epoch, in order to have a meaningful estimate of temperature and bolometric luminosity. For this pilot study, we did not intend to define a statistically complete sample or analyse every single archival X-ray observation of each target source; however, our sample is large enough that we can draw general conclusions on the nature of ULSs as a possibly distinct population.

We selected the following seven targets, summarized below with the traditional nomenclature more often found in the literature (see also Table \ref{obs_table}):\\
{\it(i) CXOAnt J120151.6-185231.9}. Located in the Antennae Galaxies, it has been observed by {\it Chandra} multiple times. In 2002 it had a best-fit blackbody temperature of $\approx$ 90 eV; the observed 0.1--2 keV luminosity increased from $\approx$$2\E{38}\,\ergs$ in December 1999 to $\approx$$1\E{39}\,\ergs$ in May 2002 \citep{2003ApJ...591..843F}.  Corrected for absorption, the bright state equates to an intrinsic blackbody-fitted bolometric luminosity of $\approx$$2\E{40}\,\ergs$ \citep[][scaled to the different distance adopted in this paper]{2003ApJ...591..843F}.\\
{\it(ii) NGC\,4631 X1}. First detected in {\it ROSAT/PSPC} observations taken from 1991 December 15 to 1992 January 4 \citep{1996A&A...311...35V}. \citet{2007A&A...471L..55C} observed the source in 2002 June using {\it XMM-Newton} and found a best-fitted blackbody temperature $\approx$$67\,\ev$ and bolometric luminosity of $\approx$$3.2\E{40}\,\ergs$. The source also appeared to have an $\approx$4 hour modulation \citep{2007A&A...471L..55C}. \citet{2009ApJ...696..287S} reexamined the same {\it XMM-Newton} data and found that when an absorption edge ($kT_{\rm edge}\approx 1.0\,\kilo\ev$) was added to the model, the fit significantly improved, with a blackbody temperature of $\approx$$90\,\ev$ and intrinsic bolometric luminosity of $\approx$ $4\E{39}\,\ergs$.\\
{\it(iii) NGC\,247 ULX}. Observed in 2009 with {\it XMM-Newton}, NGC\,247 ULX has a dominant thermal component with temperature of $\approx$$0.1\,\kilo\ev$ along with a harder component, modelled as a power-law tail by \citet{2011ApJ...737...87J}. Based on its X-ray spectral appearance, it has been interpreted as a candidate IMBH with a suggested black hole mass of $\approx$$600\,\Msol$ \citep{2012ApJ...758...85T}. \\
{\it(iv) NGC\,300 SSS1}. Originally found in a {\it ROSAT} observation from 1992 \citep{2001A&A...373..473R}, this source exhibits bright and dim states. For example, it was not seen by {\it ROSAT} in later observations but was seen by {\it XMM-Newton} in a bright state in 2000 December and became an order of magnitude dimmer in another {\it XMM-Newton} observation only 6 days later . During its bright state, it appeared to have a temperature of $\approx$$60\,\ev$ and a bolometric luminosity of a few times $10^{39}\,\ergs$.\\
{\it(v) M\,81 N1}. One of the best known ULSs, M\,81 N1 was observed by {\it ROSAT} several times in the 1990's, by {\it Chandra} in a long observation in 2000 May \citep{2002ApJ...574..382S} and again in a series of shorter {\it Chandra} observations over 2005--2008 \citep{ 2008ApJ...674L..73L, 2008ApJS..177..181L}. 
In the 2000 May observation it was fitted with a blackbody model and had a temperature of $\approx$$70\,\ev$ and bolometric luminosity of $\approx$$2.5\E{39}\,\ergs$. The source shows significant flux variability within individual observations as well as between observations \citep{2002ApJ...574..382S}.\\
{\it(vi) M\,101 ULX-1}. Observed more than 20 times by {\it Chandra} between 2000 and 2005, it has short- and long-term flux and spectral variability, with fitted blackbody temperatures ranging between $\approx$ 50--150\,eV \citep{2004ApJ...617L..49K, 2005ApJ...634.1085M, 2005ApJ...632L.107K}.  The interpretation of this source has been actively debated in recent years. A Wolf-Rayet companion was identified from {\it Hubble Space Telescope} observations by \citet{2009ApJ...704.1628L}. A claimed period of $8.2$ days led to the suggestion that this ULS is powered by a stellar-mass black hole \citep{2013Natur.503..500L}.\\
{\it(vii) CXOM51 J132943.3+471135}. Located in M\,51, this is source 12 in the catalogue of \citet{2003ApJ...592..884D}, and source 9 in that of \citet{2004ApJ...601..735T}. 

As is the case with any purely empirical selection criterion, there may be a few other supersoft sources that just missed out on our ULS selection (reaching a maximum $L^{\rm bb}_{\rm bol}$ just below $10^{39}$ erg s$^{-1}$) but are still more luminous than expected for a nuclear-burning WD. We found two of them:\\
{\it(i) NGC\,300 SSS2}. It was not seen by {\it ROSAT} in the 1990s or {\it XMM-Newton} in 2000--2001, but was discovered in 2005 with {\it XMM-Newton} \citep{2006A&A...458..747C}. We processed and analysed both {\it XMM-Newton} observations from 2005 (ObsID 0305860401 and 0305860301), and fitted its spectrum with blackbody models. We found a blackbody temperature of $\approx$$50\,\ev$ and a bolometric luminosity of $\approx$$7\E{38}\,\ergs$ in the first observation and $\approx$$6\E{38}\,\ergs$ in the second one, taken six months later.  \\
{\it(ii) r2-12}. An SSS in M31 that has reached a bolometric luminosity up to $\approx$$8\E{38}\ergs$ \citep{2008ApJ...676.1218T,2010ApJ...717..739O,2014MNRAS.443.1821C}. It was first observed with {\it Einstein} \citep{1991ApJ...382...82T} and later with {\it Chandra} \citep{2002ApJ...577..738K} and {\it XMM-Newton} \citep{2005A&A...434..483P}.

Although we shall not present a detailed study of those two sources in this paper, we have included them in some of our plots to highlight their intermediate physical properties between ULSs and classical SSSs. We used our own spectral analysis to obtain temperatures, radii and luminosities of NGC\,300 SSS2, and we used the fit results of \citet{2008ApJ...676.1218T} for r2-12.

\begin{table*}
\caption{List of {\it Chandra} and {\it XMM-Newton} observations used in our sample study of seven ULSs. Numbered epochs (column 1) designate observations for which we fitted a spectrum; in a few cases, numbers are repeated because a single spectrum was fitted to a stack of observations in the same state. Non-numbered epochs designate observations with too few counts for spectral analysis but still used for our X-ray colour-colour plot. }
\begin{scriptsize}
\begin{center}
\begin{tabular}{clccccccc}
\hline\hline\\[-7pt]
Spectrum No. & Galaxy & Distance (Mpc) & Source RA & Source Dec & Telescope & Obs ID & Exp Time (ks) & Date\\[-1pt]
\hline\\[-9pt]
1 & NGC\,4038/39 & $22.3\pm 2.8^a$ & 12:01:51.62 & -18:52:31.87 & {\it Chandra} & 315 & 72.24 & 1999-12-01\\
2 &&&&&& 3040 & 69.04 & 2001-12-29\\
3 &&&&&& 3041 & 72.91 & 2002-11-22\\
4 &&&&&& 3042 & 67.28 & 2002-05-31\\
5 &&&&&& 3043 & 67.10 & 2002-04-18\\
6 &&&&&& 3044 & 36.50 & 2002-07-10\\
7 &&&&&& 3718 & 34.72 & 2002-07-13\\[-2pt]
\hline\\[-9pt]
8 & NGC\,4631 & $7.4\pm 0.2^b$ & 12:42:15.96 & +32:32:49.4 & {\it XMM-Newton} & 0110900201 & 53.76 & 2002-06-28\\[-2pt]
\hline\\[-9pt]
9 & NGC\,247 & $3.4\pm 0.1^c$ & 00:47:03.90 & -20:47:43.8 & {\it Chandra} & 12437 & 4.99 & 2011-02-01\\
10 &&&&&& 17547 & 5.01 & 2014-11-12\\[-2pt]
\hline\\[-9pt]
11 & NGC\,300 & $1.9\pm0.1^d$ & 00:55:10.7 & -37:38:55.0 & {\it XMM-Newton} & 0112800201 & 34.05 & 2000-12-26\\
12 &&&&&  & 0112800101 & 43.80 & 2001-01-01\\
13 &&&&& {\it Chandra} & 9883 & 10.07 & 2008-07-08\\[-2pt]
\hline\\[-9pt]
14 & M\,81 &$3.6\pm0.2^e$ & 09:55:42.15 & 69:03:36.2 & {\it Chandra} & 390 & 2.38 & 2000-03-21\\
15 &&&&&& 735 & 50.02 & 2000-05-07\\
16 &&&&&& 5938 & 11.81 & 2005-06-03\\
&&&&&& 5936 & 11.41 & 2005-05-28\\
&&&&&& 5937 & 12.01 & 2005-06-01\\
17 &&&&&& 5939 & 11.81 & 2005-06-06\\
18 &&&&&& 5940 & 11.97 & 2005-06-09\\
19 &&&&&& 5941 & 11.81 & 2005-06-11\\
20 &&&&&& 5942 & 11.95 & 2005-06-15\\
21 &&&&&& 5943 & 12.01 & 2005-06-18\\
22 &&&&&& 5944 & 11.81 & 2005-06-21\\
23 &&&&&& 5945 & 11.58 & 2005-06-24\\
24 &&&&&& 5946 & 12.02 & 2005-06-26\\
25 &&&&&& 5947 & 10.70 & 2005-06-29\\
26 &&&&&& 5948 & 12.03 & 2005-07-03\\
27 &&&&&& 5949 & 12.02 & 2005-07-06\\
28 &&&&&& 9805 & 5.11 & 2007-12-21\\
29 &&&&&& 9122 & 9.91 & 2008-02-01\\[-2pt]
\hline\\[-9pt]
30 & M\,101 & $6.4\pm 0.5^f$ &14:03:32.37 & +54:21:02.75 & {\it Chandra} & 934 & 98.38 & 2000-03-26\\
31 &&&&&& 2065 & 9.63 & 2000-10-29\\
32 &&&&&& 5337 & 9.94 & 2004-07-05\\
32 &&&&&& 5338 & 28.57 & 2004-07-06\\
32 &&&&&& 5339 & 14.32 & 2004-07-07\\
32 &&&&&& 5340 & 54.42 & 2004-07-08\\
32 &&&&&& 4734 & 35.48 & 2004-07-11\\
33 &&&&&& 6170 & 47.95 & 2004-12-22\\
33 &&&&&& 6175 & 40.66 & 2004-12-24\\
34 &&&&&& 6169 & 29.38 & 2004-12-30\\
35 &&&&&& 4737 & 21.81 & 2005-01-01\\
&&&&&& 4731 & 56.24 & 2004-01-19\\
&&&&&& 5297 & 21.69 & 2004-01-29\\
&&&&&& 5300 & 52.09 & 2004-03-07\\
&&&&&& 5309 & 70.77 & 2004-03-14\\
&&&&&& 4732 & 69.79 & 2004-03-19\\
&&&&&& 6114 & 66.20 & 2004-09-05\\
&&&&&& 6115 & 35.76 & 2004-09-08\\
&&&&&& 4735 & 28.78 & 2004-09-12\\
&&&&&& 4736 & 77.35 & 2004-11-01\\
36 &&&&& {\it XMM-Newton} & 0212480201 & 32.41 & 2005-01-08\\[-2pt]
\hline\\[-9pt]
37 & M\,51 & $8.0 \pm 0.6^g$ & 13:29:43.30 & +47:11:34.7 & {\it Chandra} & 354 & 14.86 & 2000-03-21\\
38 &&&&&& 1622 & 26.81 & 2001-06-23\\
39 &&&&&& 3932 & 47.97 & 2003-08-07\\
40 &&&&&& 13813 & 179.20 & 2012-09-09\\
41 &&&&&& 13812 & 157.46 & 2012-09-12\\
42 &&&&&& 15496 & 40.97 & 2012-09-19\\
43 &&&&&& 13814 & 189.85 & 2012-09-20\\
44 &&&&&& 13815 & 67.18 & 2012-09-23\\
&&&&&& 15816 & 73.10 & 2012-09-26\\
&&&&&& 15553 & 37.57 & 2012-10-10\\[-2pt]
\hline\\[-12pt]
\end{tabular}
\end{center}
\end{scriptsize}
\begin{flushleft}
$^a$ \citet{AntennaeDist...22.3Mpc}; $^b$ \citet{NGC4631dist}; $^c$ \citet{NGC247dist}; $^d$ \citet{NGC300dist}; $^e$ \citet{M81dist}; $^f$ \citet{M101dist}; $^g$ \citet{M51dist} 
\end{flushleft}
\label{obs_table}
\end{table*}

\section{Data Analysis}


We selected a representative sample of {\it Chandra} and {\it XMM-Newton} observations of the seven known ULSs (Table \ref{obs_table}). We considered only observations in which the sources had enough counts for a meaningful colour ($\gtrsim$20 net counts) or at least a rough spectral analysis ($\gtrsim$50 net counts). We do not analyze and discuss in this paper: the observations in which the same sources were not detected or were just at the detection limit; the fraction of time in which each source was detectable over the course of all its observations; and the reason for this on/off behaviour. All that is left to follow-up work.

For the {\it Chandra} observations, we downloaded the data from the public archives and re-processed them with standard tasks in the Chandra Interactive Analysis of Observations ({\small CIAO}) Version 4.7 data analysis system \citep{2006SPIE.6270E..1VF}. We filtered out exposure intervals with high particle background. We defined circular source extraction regions with 2" radii. We extracted the local background from suitably selected nearby regions at least three times as large as the source region. We used the {\small CIAO} task {\it specextract} to extract a spectrum (with its associated background, response and ancillary response files) from each observation. Spectra with $\gtrsim$200 counts were grouped to a minimum of 15 counts per bin so that $\sqchi$ statistics could be used.  Spectra with lower counts were left unbinned and fitted with Cash statistic \citep{1979ApJ...228..939C}. We used {\it dmextract} to build background-subtracted light curves.

We downloaded the {\it XMM-Newton} data from NASA's High Energy Astrophysics Science Archive Research Center (HEASARC) archive. We processed the European Photon Imaging Camera (EPIC) observation data files with the Science Analysis System ({\small SAS}) version 14.0.0 (\verb|xmmsas_20141104_1833|). As with the {\it Chandra} data, we filtered out exposure intervals with high particle background. We defined circular source extraction regions with 20" radii whenever possible, and rectangular regions (of approximately equivalent area) on a couple of occasions when the source was located near a chip gap. We extracted the background from nearby regions at least three times as large, not including any other bright sources or chip gaps, and located at similar distances from readout nodes. We selected single and double events (pattern 0--4 for the pn camera and pattern 0--12 for MOS1 and MOS2 cameras), with the standard flagging criteria \verb|#XMMEA_EP| and \verb|#XMMEA_EM| for pn and MOS, respectively, and \verb|FLAG=0|. After building response and ancillary response files with the {\small SAS} tasks {\it rmfgen} and {\it arfgen}, we used {\it epicspeccombine} to create average EPIC spectra and response files for each source. Finally, we grouped the spectra to a minimum of 20 counts per bin, so that we could use Gaussian statistics. We built background-subtracted light curves using the {\small SAS} tasks {\it evselect} and {\it epiclccorr}.

We used {\small XSPEC} version 12.8.2 for spectral fitting \citep{1996ASPC..101...17A}.  For timing analysis, we used standard {\small FTOOLS} tasks \citep{1995ASPC...77..367B}, such as {\it lcurve}, {\it powspec}, {\it efsearch} and {\it statistics}.

\section{Main Results}\label{results_sect}

\subsection{X-ray spectral properties}
For each source, and each observing epoch with at least $\approx$50 net counts, we fitted the background-subtracted spectrum, starting with one-component models: {\it phabs*phabs*bbody} and {\it phabs*phabs*diskbb}. The first photoelectric absorption component was fixed at the line-of-sight value for the relevant host galaxy; the second component was left free. From this initial level of fitting, two things are apparent: that a single thermal component is already the best fit for most epochs, and, to first order, a satisfactory fit for all epochs; and that for all epochs, blackbody and disk-blackbody models are statistically equivalent. The first statement is another way of saying that ULSs (by definition) are not dominated by a broad-band tail above 1 keV, unlike standard ULXs. The second statement is not surprising, considering that the peak temperature of the thermal component is $\approx$100 eV and the detectors are only sensitive to energies $\ga$300 eV, in the Wien tail; a similar situation occurs when fitting the low-energy soft excess in ULXs. The only difference between the two models is in the definition of blackbody temperature $kT_{\rm bb}$ compared with the colour temperature of the disk $kT_{\rm in}$: for the same thermal spectrum, $kT_{\rm in}$ is always systematically higher than $kT_{\rm bb}$ by $\approx$20\%.

In a few cases, our spectral fitting showed significant residuals in addition to the dominant soft thermal component, particularly around 0.7--2 keV (sometimes with a distinctive residual feature at $\approx$1 keV). It was necessary to add one or more spectral components to account for the harder emission. At this stage we make no specific assumptions on the physical nature of this harder component. The nature of the fit residuals suggests that the harder component could be optically-thin thermal plasma emission, which we modelled with {\it mekal}\footnote{Even at our best signal-to-noise level, there were no statistically significant differences between {\it mekal} and {\it apec} models.} in {\small {XSPEC}}. Spectral residuals in addition to the smooth continuum have been seen in some two-component ULXs (for example, NGC\,5408 X-1 and NGC\,6946 X-1) and were also successfully modelled with thermal plasma emission \citep{2003MNRAS.341L..49R,2006MNRAS.368..397S,2007ApJ...660..580S,2014MNRAS.438L..51M}, with luminosities $\sim$ a few $10^{38}$ erg s$^{-1}$. However, an alternative interpretation was provided by \citet{2014MNRAS.438L..51M}, who showed that similar residuals are also consistent with broadened, blue-shifted absorption in the line of sight of a fast, partially ionized, optically-thin outflow. Our ULS spectra have a substantially lower signal-to-noise ratio than the spectra of NGC\,5408 X-1 and NGC\,6946 X-1 discussed by \citet{2014MNRAS.438L..51M}, making  a test between those two alternative interpretation more difficult. For this paper, we chose to limit our spectral analysis to the thermal-plasma emission model; however, we are aware of the alternative interpretation and leave its exploration to future work.

In summary, when significant residuals were detected above the blackbody/disk-blackbody base model, we tried improving the fit with two methods: by adding one or two thermal plasma components at fixed solar abundance, with characteristic temperatures $\approx$0.6--0.9 keV; and by testing for the presence of an absorption edge at $kT_{\rm edge} \approx 1$ keV.  We found significant (and variable) thermal-plasma contributions in the spectra of the M\,101 ULS, in several epochs: they are discussed in more details in Section 4.2; see also Soria \& Kong (2015, MNRAS, in press, arXiv:1511.04797). Thermal plasma emission is also seen in some of the epochs for the ULSs in NGC\,247, the Antennae Galaxies and M\,51. Adding an absorption edge provides significant improvements to some of the spectra with the high signal-to-noise in four out of seven sources: those in the Antennae, NGC\,247, M\,101 and NGC\,4631. In some sources (most notably those in NGC\,247 and M\,101), the spectrum is best fitted with a blackbody (or disk-blackbody) plus thermal-plasma emission in some epochs, and blackbody (or disk-blackbody) plus an absorption edge in other epochs (Section 4.2).

We stress that the main objective of this work is to determine the properties of the dominant optically-thick thermal component and attempt a physical interpretation; nonetheless, careful treatment of additional first-order components (absorption edges, harder emission above 1 keV, and line residuals consistent with thermal-plasma) is necessary to make sure that the characteristic temperature and radius of the dominant blackbody component is not over- or under-estimated. After finding successful fits for all epochs (with or without additional components), we summarized the physical parameters of the blackbody components in Table \ref{params_tab}; when a disk-blackbody replaces the blackbody component in the same spectra, we obtain the physical parameters listed in Table \ref{params_tab_disk}; the additional thermal-plasma and edge features and their significance are listed in Table \ref{residual_tab}.

The bolometric blackbody luminosity is defined as $L^{\rm bb}_{\rm bol} \equiv 4\pi r_{\rm bb}^2 \sigma T_{\rm bb}^4$, independent of the viewing angle $\theta$. It is obtained directly from one of the {\small XSPEC} {\it bbody} fit parameters, namely the normalization constant, because $L_{\rm bol}^{\rm bb} = \sqrt{N_{\rm bb}} \, d_{10{\rm kpc}} \, 10^{39}$ erg s$^{-1}$. The adopted galaxy distances are quoted in Table \ref{obs_table}.
The bolometric disk luminosity is defined as $L^{\rm diskbb}_{\rm bol}$ $\equiv 4\pi r_{\rm in}^2 \sigma T_{\rm in}^4$, where $r_{\rm in} = \sqrt{N_{\rm diskbb}} \, d_{10{\rm kpc}}/\sqrt{\cos \theta}$ km and $N_{\rm diskbb}$ is the normalization constant in {\small XSPEC}. Physically, the fit parameter $r_{\rm in}$ is related to the true inner-disk radius $R_{\rm in}=1.19r_{\rm in}$ \citep{1998PASJ...50..667K}. We cannot obtain $L^{\rm diskbb}_{\rm bol}$ directly from our {\small XSPEC} {\it diskbb} fit parameters, because of its intrinsic dependence on $\cos \theta$; we can only obtain $L^{\rm diskbb}_{\rm bol} \times \cos \theta$, and $r_{\rm in} \sqrt{\cos \theta}$. As a double check on our bolometric disk-blackbody luminosities, we re-derived them with an alternative method: we defined a dummy response from 0.01 to 10 keV ({\it dummyrsp} in {\small XSPEC}), calculated the model flux $f^{\rm diskbb}_{\rm bol}$ over that (essentially bolometric) energy range, and converted it to a luminosity with the relation $L^{\rm diskbb}_{\rm bol} = 2\pi \left(d^2/\cos \theta\right) f^{\rm diskbb}_{\rm bol}$, suitable for accretion disks. The two methods give the same result, as expected. For plotting purposes, and for consistency with the luminosities given in a comparison sample of ULXs from the literature, we assumed $\cos \theta = 1/2$ throughout this paper. With this conventional choice of viewing angle, disk luminosities are simply obtained as $4\pi d^2$ times the observed fluxes, like for a spherically symmetric emitter\footnote{This is often an implicit assumption in many studies of X-ray binaries and ULXs in the literature, in which the luminosities of all emission components are defined as $L \equiv 4\pi d^2 f$. The alternative choice of $\theta = 0$ is also sometimes adopted in the literature; however, this is not suitable to our sample of sources because they are more likely to be high-inclination objects, as discussed in Section 5.3.}. $L^{\rm diskbb}_{\rm bol}$ for a generic angle $\theta$ is simply obtained from the values listed in Table \ref{params_tab_disk}, divided by $2 \cos \theta$.


\begin{table}
\centering
\caption{Characteristic radii, temperatures and bolometric luminosities of the {\it bbody} component in each ULS observation with enough counts for spectral fitting. The row number identifies the source and the epoch, from the observation list in Table \ref{obs_table}. ObsID 934 of the M\,101 ULS was split into three sub-intervals based on count rate (Section \ref{m101_sect}). Errors are 90\% confidence limits for single parameters.}
\begin{scriptsize}
\begin{tabular}{ccccc}
\hline\hline\\[-9pt]
Row & $n_{\rm H}$ & $r_{{\rm bb}}$ & $kT_{{\rm bb}}$ & $L^{{\rm bb}}_{{\rm bol}}$  \\
& $(10^{22}\,\centi\meter^{-2})$ &$(10^3\,\kilo\meter)$ & $(\ev)$ & $(10^{39}\,\ergs)$\\[-2pt]
\hline 
1 & \U{0.07}{0.20}{0.07} & \U{7.4}{68.6}{3.4} & \U{100}{35}{31} & \U{0.7}{8.0}{0.4}\\[3.5pt]
2 & \U{0.13}{0.17}{0.12} & \U{12.3}{70.0}{5.4} & \U{107}{24}{20} & \U{2.6}{14.0}{0.9}\\[3.5pt]
3 & \U{0.12}{0.32}{0.12} & \U{17.9}{1315}{8.8} & \U{82}{34}{24} & \U{1.9}{72.2}{1.6}\\[3.5pt]
4 & \U{0.36}{0.31}{0.17} & \U{15.4}{75.0}{7.1} & \U{119}{33}{22} & \U{6.1}{22.3}{4.3}\\[3.5pt]
5 & \U{0.06}{0.17}{0.06} & \U{7.5}{53.9}{3.0} & \U{118}{26}{27} & \U{1.4}{6.3}{0.7}\\[3.5pt]
6 & \U{0.21}{0.43}{0.18} & \U{23.0}{127.1}{11.1} & \U{98}{30}{31} & \U{6.3}{75.9}{5.1}\\[3.5pt]
7 & \U{0.22}{0.29}{0.17} & \U{38.2}{162.2}{18.5} & \U{87}{25}{20} & \U{10.7}{63.2}{9.0}\\[3.5pt]
8 & \U{0.18}{0.20}{0.13} & \U{6.6}{59.8}{3.0} & \U{127}{50}{32} & \U{1.5}{7.3}{0.9}\\[3.5pt]
9 & \U{0.22}{0.26}{0.17} & \U{10.0}{198.3}{4.7} & \U{119}{47}{33} & \U{2.6}{26.4}{1.9}\\[3.5pt]
10 & \U{0.22}{0.26}{0.17} & \U{6.8}{45.6}{3.1} & \U{132}{53}{41} & \U{1.3}{8.2}{0.4}\\[3.5pt]
11 & \U{0.05}{0.06}{0.05} & \U{11.2}{17.8}{5.6} & \U{68}{9}{8} & \U{0.3}{0.6}{0.2}\\[3.5pt]
12 & \U{0.07}{0.05}{0.04} & \U{14.8}{17.4}{7.4} & \U{59}{5}{5} & \U{0.3}{0.5}{0.2}\\[3.5pt]
13 & \U{0.11}{0.16}{0.11} & \U{87.5}{389.9}{42.8} & \U{43}{12}{10} & \U{3.4}{14.0}{3.2}\\[3.5pt]
14 & \U{0.03}{0.12}{0.03} & \U{2.1}{3.2}{0.7} & \U{168}{30}{34} & \U{0.5}{0.7}{0.1}\\[3.5pt]
15 & \U{0.05}{0.02}{0.02} & \U{22.3}{5.3}{3.6} & \U{78}{3}{3} & \U{2.3}{0.7}{0.5}\\[3.5pt]
16 & \U{0.16}{0.13}{0.09} & \U{52.2}{133.5}{26.3} & \U{70}{7}{7} & \U{8.3}{25.4}{5.5}\\[3.5pt]
17 & \U{0.01}{0.09}{0.01} & \U{15.4}{33.1}{3.3} & \U{79}{6}{10} & \U{1.2}{2.6}{0.3}\\[3.5pt]
18 & $<0.15$ & \U{2.4}{11.8}{0.6} & \U{109}{13}{21} & \U{0.1}{0.4}{0.03}\\[3.5pt]
19 & $<0.04$ & \U{8.3}{1.9}{1.4} & \U{97}{6}{7} & \U{0.8}{0.5}{0.1}\\[3.5pt]
20 & \U{0.08}{0.10}{0.08} & \U{28.6}{77.6}{14.4} & \U{73}{12}{9} & \U{3.0}{8.1}{2.0}\\[3.5pt]
21 & \U{0.03}{0.07}{0.03} & \U{13.1}{15.7}{3.4} & \U{91}{8}{9} & \U{1.5}{1.9}{0.5}\\[3.5pt]
22 & \U{0.34}{0.28}{0.21} & \U{38.9}{660.0}{19.6} & \U{84}{22}{16} & \U{9.5}{130.3}{8.2}\\[3.5pt]
23 & \U{0.11}{0.09}{0.07} & \U{41.6}{72.8}{20.9} & \U{69}{8}{7} & \U{5.1}{10.2}{3.2}\\[3.5pt]
24 & \U{0.05}{0.11}{0.05} & \U{23.0}{95.9}{11.5} & \U{73}{16}{12} & \U{1.9}{6.7}{1.2}\\[3.5pt]
25 & $<0.14$ & \U{29.6}{183.6}{14.9} & \U{49}{13}{12} & \U{0.7}{2.4}{0.4}\\[3.5pt]
26 & $<0.06$ & \U{12.7}{24.2}{4.1} & \U{66}{9}{9} & \U{0.4}{0.7}{0.2}\\[3.5pt]
27 & \U{0.02}{0.26}{0.02} & \U{8.7}{48.1}{4.4} & \U{71}{16}{19} & \U{0.2}{0.7}{0.1}\\[3.5pt]
29 & $<0.07$  & \U{14.3}{51.2}{7.2} & \U{69}{14}{13} & \U{0.6}{1.6}{0.3}\\[3.5pt]
29 & \U{0.13}{0.31}{0.13} & \U{19.9}{115.2}{10.0} & \U{71}{23}{18} & \U{1.3}{8.6}{1.1}\\[3.5pt]
30-high & \U{0.04}{0.01}{0.01} & \U{10.2}{0.4}{0.3} & \U{135}{3}{4} & \U{4.6}{0.2}{0.3}\\[3.5pt]
30-med & \U{0.04}{0.01}{0.01} &\U{10.5}{0.3}{0.3} & \U{119}{2}{2} & \U{2.9}{0.2}{0.2}\\[3.5pt]
30-low & \U{0.04}{0.01}{0.01} &\U{18.4}{0.9}{1.0} & \U{90}{2}{2} & \U{2.9}{0.2}{0.2}\\[3.5pt]
31 & \U{0.08}{0.07}{0.05} & \U{29.0}{32.7}{14.4} & \U{77}{11}{10} & \U{3.6}{6.8}{2.0}\\[3.5pt]
32 & \U{0.10}{0.04}{0.04} & \U{47.0}{32.6}{17.0} & \U{69}{7}{7} & \U{6.6}{5.9}{2.9}\\[3.5pt]
33 & \U{0.20}{0.13}{0.12} & $>54$ & $<49$ & $>2.1$\\[3.5pt]
34 & \U{0.12}{0.06}{0.05} & \U{43.5}{29.3}{17.2} & \U{75}{6}{6} & \U{7.7}{8.6}{3.8}\\[3.5pt]
35 & \U{0.13}{0.03}{0.03} & \U{22.5}{10.3}{4.7} & \U{100}{13}{10} & \U{6.6}{1.2}{1.0}\\[3.5pt]
36 & \U{0.13}{0.05}{0.04} & \U{101.5}{82.7}{43.1} & \U{56}{5}{5} & \U{12.8}{17.9}{6.9}\\[3.5pt]
37 & \U{0.01}{0.04}{0.01} & \U{6.0}{9.0}{1.3} & \U{114}{11}{16} & \U{0.8}{1.0}{0.2}\\[3.5pt]
38 & $<0.07$ & \U{3.5}{4.8}{0.8} & \U{135}{17}{20} & \U{0.5}{0.5}{0.1}\\[3.5pt]
39 & \U{0.003}{0.084}{0.003} & \U{5.6}{10.0}{1.0} & \U{122}{12}{23} & \U{0.9}{1.2}{0.1}\\[3.5pt]
40 & $<0.04$ & \U{7.8}{13.1}{1.7} & \U{98}{10}{10} & \U{0.7}{1.1}{0.1}\\[3.5pt]
41 & \U{0.03}{0.15}{0.03} & \U{10.4}{30.1}{3.1} & \U{92}{12}{8} & \U{1.1}{3.1}{0.4}\\[3.5pt]
42 & $<0.20$ & \U{6.2}{56.3}{1.2} & \U{115}{8}{25} & \U{0.9}{5.3}{0.2}\\[3.5pt]
43 & \U{0.12}{0.09}{0.02} & \U{25.1}{41.3}{9.5} & \U{84}{8}{7} & \U{4.5}{9.3}{2.5}\\[3.5pt]
44 & $<0.21$ & \U{7.5}{16.3}{1.9} & \U{105}{16}{18} & \U{0.9}{1.6}{0.1}\\
\hline
\end{tabular}
\end{scriptsize}
\label{params_tab}
\end{table}

\begin{table}
\centering
\caption{Characteristic inner-disk radii, peak colour temperatures and bolometric luminosities of the {\it diskbb} component in each ULS observation with enough counts for spectral fitting. Row numbers are defined as in Table \ref{params_tab}. An angle $\theta = 60^{\circ}$ was adopted in the expressions for $r_{\rm in}$ and $L^{{\rm diskbb}}_{{\rm bol}}$. Errors are 90\% confidence limits for single parameters.}
\begin{scriptsize}
\begin{tabular}{ccccc}
\hline\hline\\[-9pt]
Row & $n_{\rm H}$ & $r_{{\rm in}}$ & $kT_{{\rm in}}$ & $L^{{\rm diskbb}}_{{\rm bol}}$  \\
&$(10^{22}\,\centi\meter^{-2})$ & $(10^3\,\kilo\meter)$ & $(\ev)$ & $(10^{39}\,\ergs)$\\[-2pt]
\hline 
1 & \U{0.09}{0.19}{0.09} & \U{7.9}{20.4}{3.9} & \U{120}{62}{41} & \U{1.7}{18.9}{1.3}\\[3.5pt]
2 & \U{0.14}{0.16}{0.11} & \U{16.3}{169.5}{7.4} & \U{123}{31}{26} & \U{8.9}{52.6}{6.5}\\[3.5pt]
3 & \U{0.09}{0.29}{0.09} & \U{22.9}{374.3}{11.5} & \U{95}{49}{30} & \U{5.5}{28.5}{4.8}\\[3.5pt]
4 & \U{0.23}{0.36}{0.15} & \U{13.4}{101.6}{6.3} & \U{153}{53}{36} & \U{12.8}{56.9}{9.1}\\[3.5pt]
5 & \U{0.10}{0.16}{0.10} & \U{8.6}{84.4}{4.0} & \U{140}{49}{36} & \U{3.6}{19.6}{2.6}\\[3.5pt]
6 & \U{0.29}{0.43}{0.21} & \U{43.5}{848.7}{21.3} & \U{106}{42}{35} & \U{30.9}{2475.8}{27.4}\\[3.5pt]
7 & \U{0.26}{0.29}{0.18} & \U{58.7}{*}{29.1} & \U{97}{33}{24} & \U{38.6}{836.4}{33.3}\\[3.5pt]
8 & \U{0.22}{0.19}{0.14} & \U{6.8}{81.7}{3.3} & \U{157}{116}{45} & \U{3.2}{19.2}{2.5}\\[3.5pt]
9 & \U{0.29}{0.24}{0.17} & \U{14.5}{335.8}{6.9} & \U{134}{61}{38} & \U{8.7}{99.6}{6.8}\\[3.5pt]
10 & \U{0.29}{0.24}{0.17} & \U{9.3}{248.9}{4.4} & \U{151}{42}{49} & \U{5.8}{67.0}{4.0}\\[3.5pt]
11 & \U{0.06}{0.06}{0.06} & \U{14.5}{28.4}{7.3} & \U{77}{11}{9} & \U{1.0}{1.7}{0.6}\\[3.5pt]
12 & \U{0.08}{0.05}{0.04} & \U{21.2}{29.7}{10.7} & \U{66}{7}{6} & \U{1.1}{1.6}{0.6}\\[3.5pt]
13 & \U{0.12}{0.16}{0.12} & \U{153}{999}{76} & \U{47}{16}{11} & \U{14.1}{624}{13.4}\\[3.5pt]
14 & \U{0.10}{0.13}{0.09} & \U{2.9}{9.9}{1.0} & \U{203}{70}{50} & \U{1.1}{2.5}{0.6}\\[3.5pt]
15 & \U{0.08}{0.02}{0.2} & \U{29.0}{7.4}{5.1} & \U{89}{3}{3} & \U{6.8}{2.0}{1.5}\\[3.5pt]
16 & \U{0.20}{0.13}{0.10} & \U{89.0}{278.3}{35.3} & \U{76}{8}{8} & \U{35.6}{122.1}{24.3}\\[3.5pt]
17 & \U{0.03}{0.09}{0.03} & \U{19.9}{50.2}{7.0} & \U{91}{12}{13} & \U{3.5}{8.2}{1.7}\\[3.5pt]
18 & \U{0.03}{0.17}{0.03} & \U{2.7}{28.3}{1.0} & \U{129}{16}{28} & \U{0.3}{1.5}{0.1}\\[3.5pt]
19 & $<0.05$ & \U{6.7}{2.0}{1.3} & \U{124}{10}{9} & \U{1.4}{0.2}{0.2}\\[3.5pt]
20 & \U{0.09}{0.10}{0.08} & \U{35.9}{108.7}{15.3} & \U{84}{16}{12} & \U{8.3}{23.0}{5.7}\\[3.5pt]
21 & \U{0.07}{0.07}{0.06} & \U{17.0}{24.6}{6.0} & \U{104}{13}{11} & \U{4.4}{6.3}{2.3}\\[3.5pt]
22 & \U{0.40}{0.28}{0.21} & \U{66.2}{*}{32.0} & \U{92}{27}{19} & \U{40.5}{669}{36.1}\\[3.5pt]
23 & \U{0.13}{0.08}{0.07} & \U{57.8}{111.2}{27.4} & \U{78}{10}{9} & \U{16.2}{32.9}{10.3}\\[3.5pt]
24 & \U{0.07}{0.10}{0.07} & \U{25.6}{135.1}{11.4} & \U{86}{26}{17} & \U{4.6}{17.4}{3.3}\\[3.5pt]
25 & $<0.01$ & \U{36.7}{309.2}{16.8} & \U{56}{18}{14} & \U{1.7}{7.8}{1.2}\\[3.5pt]
26 & $<0.06$ & \U{13.8}{34.4}{4.9} & \U{77}{13}{12} & \U{0.9}{1.7}{0.4}\\[3.5pt]
27 & \U{0.05}{0.26}{0.05} & \U{6.8}{10.7}{2.6} & \U{88}{20}{15} & \U{0.4}{0.3}{0.2}\\[3.5pt]
28 & $<0.08$ &\U{13.4}{70.5}{5.2} & \U{85}{19}{19} & \U{1.2}{3.5}{0.5}\\[3.5pt]
29 & \U{0.17}{0.17}{0.17} & \U{30.1}{*}{17.1} & \U{80}{35}{21} & \U{4.6}{194}{4.2}\\[3.5pt]
30-high & \U{0.06}{0.01}{0.01} & \U{8.3}{0.4}{0.4} & \U{174}{6}{7} & \U{8.2}{0.4}{0.4}\\[3.5pt]
30-med & \U{0.06}{0.01}{0.01} & \U{6.1}{0.3}{0.3} & \U{178}{2}{2} & \U{4.8}{0.4}{0.4}\\[3.5pt]
30-low & \U{0.06}{0.01}{0.01} & \U{18.4}{1.6}{1.1} & \U{109}{2}{2} & \U{6.2}{0.8}{0.8}\\[3.5pt]
31 & \U{0.10}{0.06}{0.06} & \U{33.9}{48.5}{17.4} & \U{90}{14}{13} & \U{9.8}{21.2}{5.6}\\[3.5pt]
32 & \U{0.11}{0.04}{0.04} & \U{39.7}{30.4}{16.5} & \U{81}{9}{6} & \U{17.0}{14.2}{8.0}\\[3.5pt]
33 & \U{0.22}{0.13}{0.12} & $>88$ & $<54$ & $>8.0$\\[3.5pt]
34 & \U{0.15}{0.06}{0.06} &\U{66.0}{48.5}{17.4} & \U{84}{8}{6} & \U{27.8}{32.4}{15.0}\\[3.5pt]
35 & \U{0.11}{0.03}{0.03} &\U{12.2}{21.1}{4.9} & \U{147}{6}{6} & \U{8.8}{15.8}{2.2}\\[3.5pt]
36 & \U{0.14}{0.05}{0.04} & \U{146.7}{146.4}{63.1} & \U{62}{5}{5} & \U{41.0}{70.8}{-21}\\[3.5pt]
37 & \U{0.05}{0.08}{0.05} & \U{6.8}{12.9}{2.6} & \U{135}{25}{21} & \U{2.0}{2.8}{1.0}\\[3.5pt]
38 & \U{0.004}{0.093}{0.004} & \U{2.1}{7.9}{0.6} & \U{194}{32}{49} & \U{0.8}{0.1}{0.1}\\[3.5pt]
39 & \U{0.04}{0.03}{0.04} & \U{6.1}{12.4}{2.2} & \U{145}{27}{27} & \U{2.1}{2.3}{0.9}\\[3.5pt]
40 & \U{0.001}{0.084}{0.001} & \U{5.8}{23.7}{1.7} & \U{128}{24}{29} & \U{1.3}{2.5}{0.4}\\[3.5pt]
41 & \U{0.04}{0.01}{0.04} & \U{11.2}{40.2}{4.1} & \U{111}{18}{17} & \U{2.4}{4.7}{1.1}\\[3.5pt]
42 & \U{0.06}{0.26}{0.06} & \U{8.9}{214.4}{3.3} & \U{129}{23}{35} & \U{2.9}{36.7}{1.6}\\[3.5pt]
43 & \U{0.13}{0.17}{0.08} & \U{23.9}{54.4}{8.0} & \U{104}{11}{9} & \U{8.7}{8.5}{1.5}\\[3.5pt]
44 & $<0.09$ & \U{6.5}{18.0}{2.0} & \U{131}{25}{22} & \U{1.6}{1.4}{0.3}\\
\hline
\end{tabular}
\end{scriptsize}
\label{params_tab_disk}
\end{table}

\begin{figure}
\includegraphics[width=84mm]{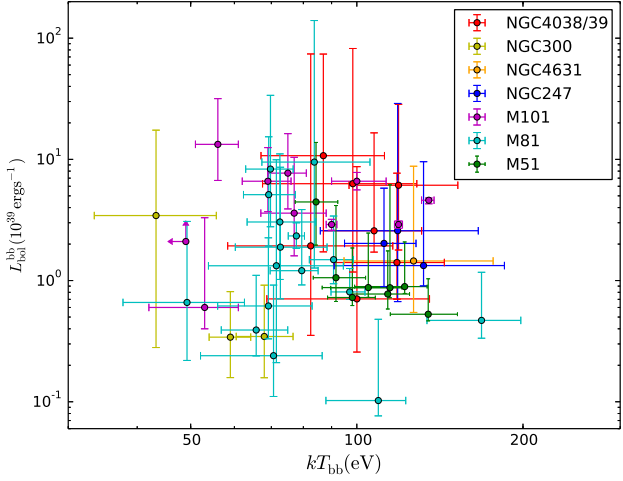}
 \caption{Bolometric luminosity inferred from single-temperature blackbody fits, for each observation, plotted against the corresponding best-fitting blackbody temperature. No significant correlation exists between the two quantities, either for individual sources or for the population as a whole.}
  \label{ULS_LvsT_fig}
\end{figure}

\begin{figure}
\includegraphics[width=84mm]{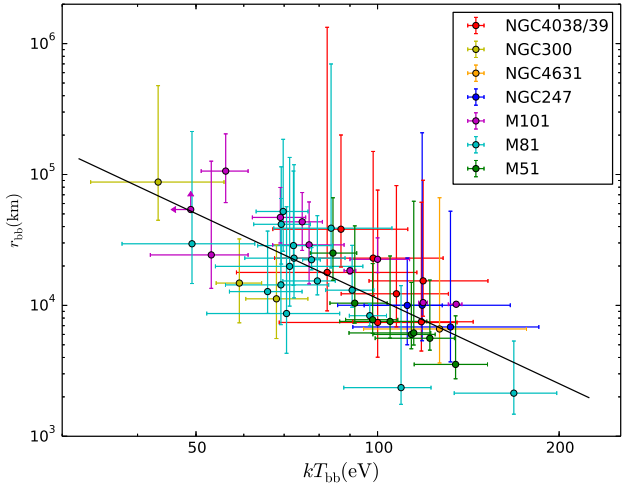}
 \caption{Best-fitting blackbody radius, plotted against the best-fitting blackbody temperature, for each observation with spectral information. The best-fitting inverse correlation $r_{\rm bb}\propto T_{\rm bb}^{-2.16\pm0.46}$ is overplotted.}
  \label{SSS_fig}
\end{figure}

\begin{figure}
\includegraphics[width=84mm]{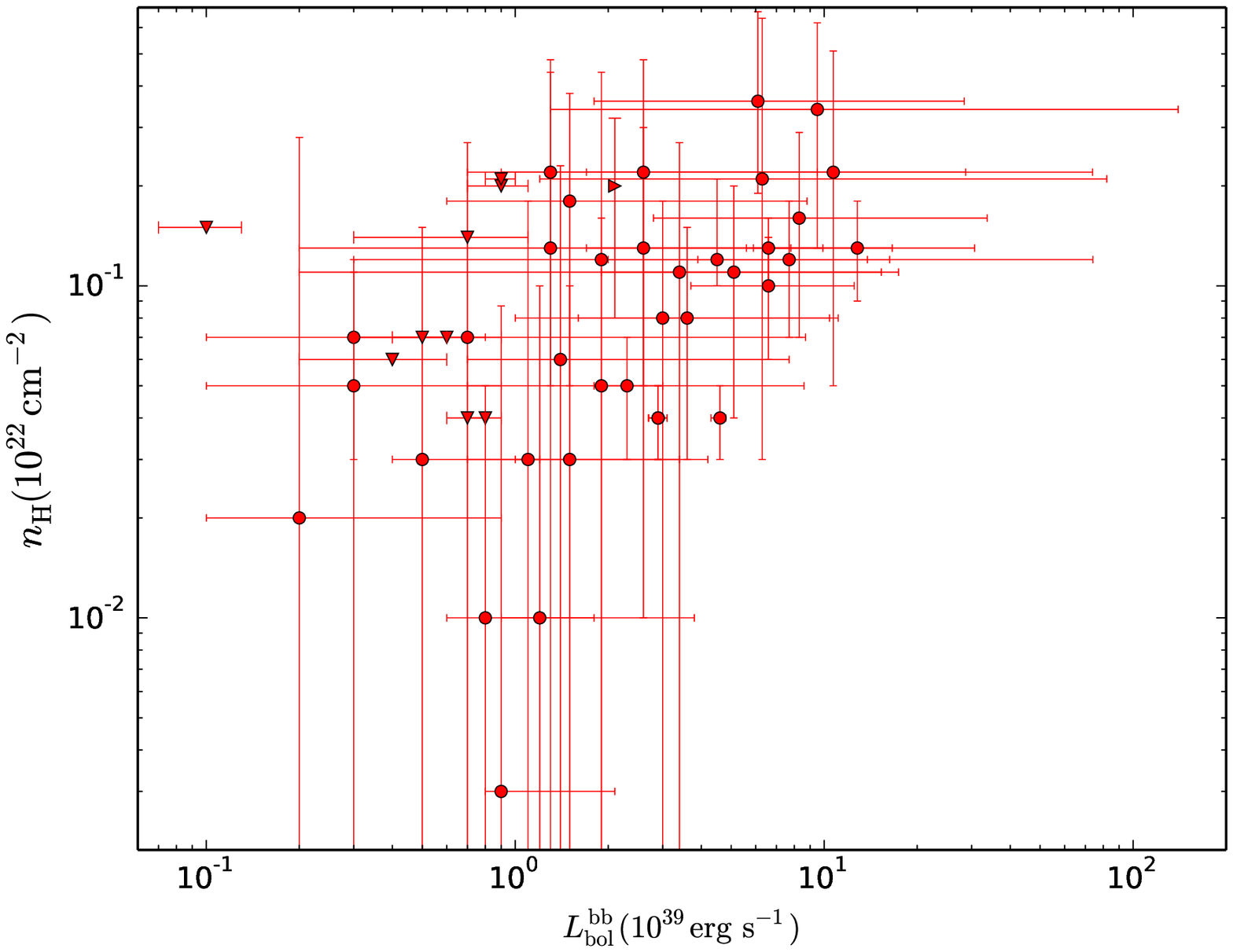}\\
\includegraphics[width=84mm]{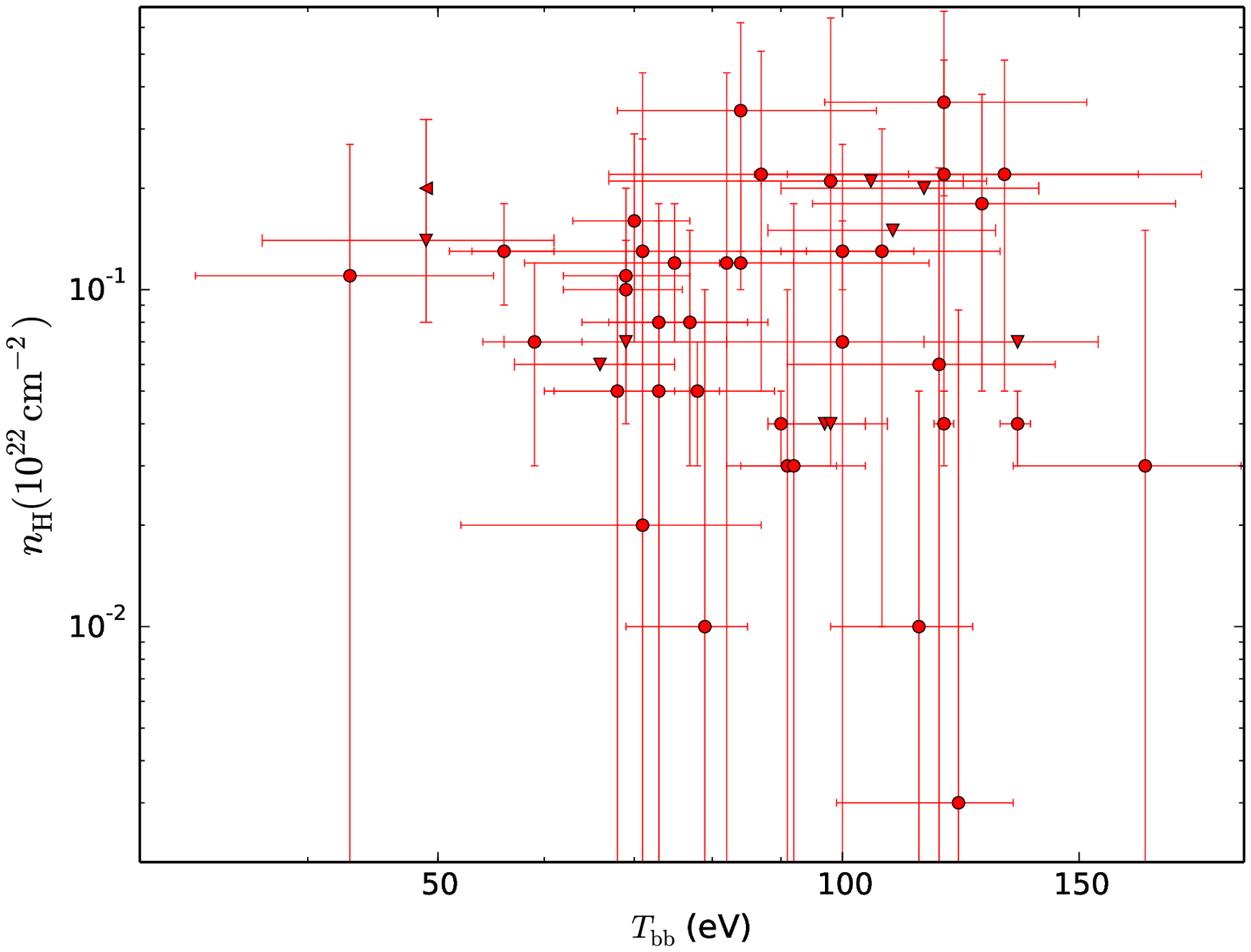}\\
\includegraphics[width=84mm]{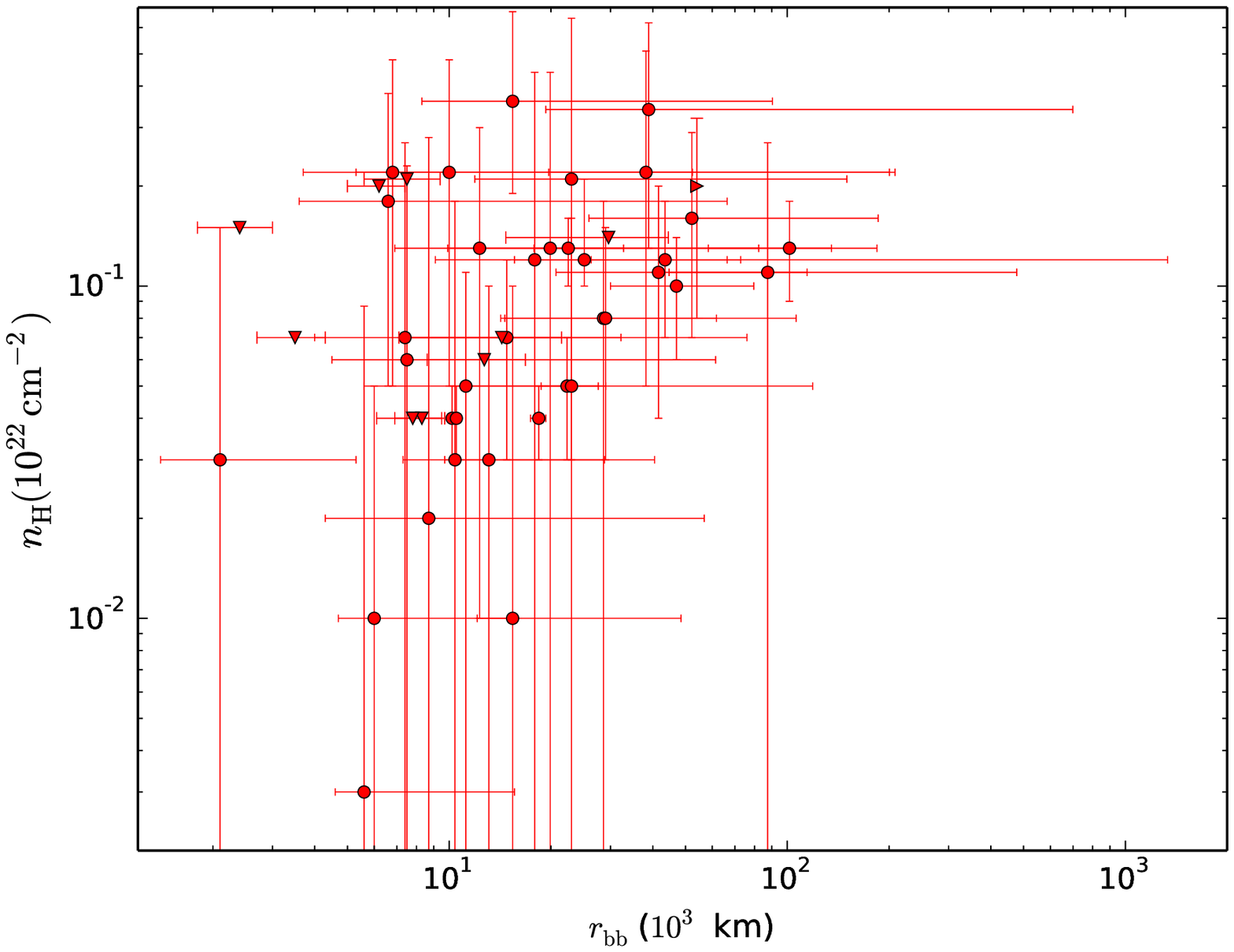}
 \caption{Top panel: best-fitting column density $n_{\rm H}$ plotted against the extrapolated blackbody bolometric luminosity. Middle panel: $n_{\rm H}$ plotted against the best-fitting blackbody temperature. Bottom panel: $n_{\rm H}$ plotted against the best-fitting blackbody radii. In all panels, triangles indicate values that are only upper or lower limits in one of the parameters.}
  \label{NH_fig}
\end{figure}



The first, most obvious result of our spectral modelling is that there is no $L\propto T^4$ trend in any of the sources (when observed multiple times) or in the sample as a whole (Figure \ref{ULS_LvsT_fig}, and Tables \ref{params_tab} and \ref{params_tab_disk}). This is equivalent to saying that the radius of the emitting region is not constant between sources or for the same source at various epochs. A constant inner-disk radius $R_{\rm in} \approx 1.19 r_{\rm in} \approx R_{\rm ISCO}$ (radius of the innermost stable circular orbit) is usually taken as the defining property of the high/soft state in luminous BHs at accretion rates $\ga$ a few percent of the Eddington limit \citep{2002MNRAS.337L..11K, 2006ARA&A..44...49R}. The IMBH scenario was based on the interpretation of the thermal component as disk emission in the high/soft state; however, we see that the thermal component in ULSs does not behave like a standard disk in the canonical high/soft state. Instead, temperature variations at approximately constant bolometric luminosity are often found in systems with expanding photospheres, such as some novae and classical SSSs, which have been observed to oscillate between an X-ray bright (interpreted as a hotter, smaller photosphere) and a UV-bright (cooler, larger photosphere) state \citep{1992A&A...262...97V, 1993A&A...278L..39P}. 

Another important result of our fits is that there is an anticorrelation between fitted radii and temperatures (Figure \ref{SSS_fig}, and Tables \ref{params_tab} and \ref{params_tab_disk}). For example, using blackbody models, we find that for the ULS in M\,101 $r_{\rm bb}\propto T_{\rm bb}^{-1.88\pm0.87}$ \citep[Spearman rank $r = -0.82$ ignoring errors and $r=-0.87\pm0.10$ when considering errors using a Monte Carlo perturbation error analysis outlined in][]{2014arXiv1411.3816C}. For the whole sample of ULSs $r_{\rm bb}\propto T_{\rm bb}^{-2.16\pm0.46}$ (Spearman rank $r = -0.73$ ignoring errors and $r=-0.46\pm0.09$ with errors). This is again consistent with the behaviour of a moving photosphere.

For completeness, we also plot the best-fitting values of the absorbing column density $n_{\rm H}$ versus the bolometric luminosity, the temperature and the radius (Figure \ref{NH_fig}). There is always, inevitably, a degree of degeneracy between column density and the other fit parameters for supersoft thermal components. We do not find any systematic trend between $n_{\rm H}$ and blackbody temperature, which supports our claim that the fitted spread in temperatures is a real effect. Some of the observations with low $n_{\rm H}$ also have low fitted radii, but this needs not be a fitting degeneracy: it is physically plausible that $n_{\rm H}$ increases when our line of sight goes through a thicker, larger outflow. In particular, for ObsID 934 of the M\,101 ULS, it was shown by Soria \& Kong (2015, MNRAS, in press, arXiv:1511.04797) that the strong intra-observation evolution (in the sense of smaller radii for hotter temperatures) is recovered both when the column density is locked between all three sub-intervals, and when it is left free; therefore at least in that case it is not due to a fitting degeneracy.

\begin{figure}
\includegraphics[width=84mm]{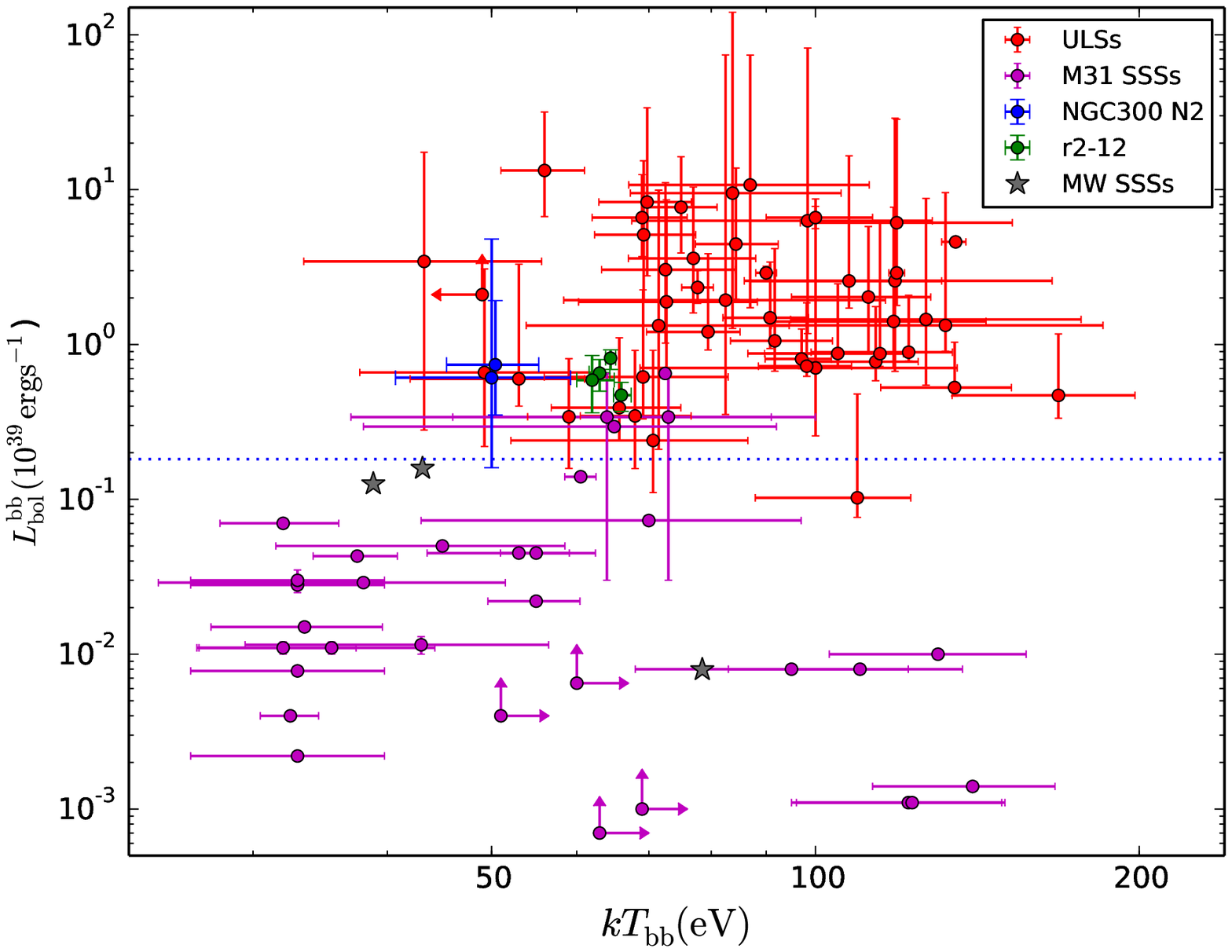}
 \caption{Red datapoints: bolometric luminosity versus blackbody temperature distribution of the seven ULSs in our sample at various epochs (values from Table \ref{params_tab}). Magenta datapoints: same, for a sample of classical SSSs in M\,31, from \citet{2010ApJ...717..739O}; stars mark representative locations of the three best-known SSSs in the Local Group (Cal83, Cal87 and RXJ0513.9$-$6951), from \citet{2013A&A...559A..50N}. Green and blue datapoints are the luminosity and temperature of the (variable) supersoft sources r2-12 in M\,31 \citep[data from][]{2008ApJ...676.1218T} and NGC\,300 SSS2 (from our own spectral analysis) respectively; both sources are brighter than classical SSSs but narrowly missed out on our threshold $10^{39}$ erg s$^{-1}$ luminosity for the definition of ULSs.
The dashed line shows the Eddington luminosity of a 1.4-$M_{\odot}$ accretor, which should provide an approximate upper limit to the luminosity of classical SSSs if powered by nuclear-burning WDs. }
  \label{SSS_LvT_galac_fig}
\end{figure}

\begin{figure}
\includegraphics[width=84mm]{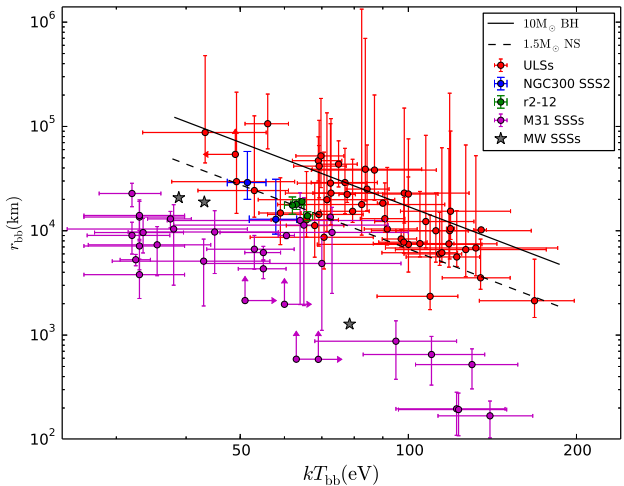}
 \caption{Red datapoints: fitted inner-disk radius versus blackbody temperature for our sample of ULSs. Magenta datapoints: same, for the classical SSSs in M\,31 \citep{2010ApJ...717..739O}. Green and blue datapoints are the luminosity and radius of r2-12 in M\,31 and NGC\,300 SSS2 respectively. Stars are representative values for Cal83, Cal87 and RXJ0513.9$-$6951. Solid line: radii and temperatures predicted by the super-Eddington outflow model described in Soria \& Kong (2015, MNRAS, in press, arXiv:1511.04797), for a 10-$M_{\odot}$ BH. Dashed line: same, for a 1.5-$M_{\odot}$ NS.}
  \label{SSS_RvT_galac_fig}
\end{figure}

Two classes of physical sources show soft thermal emission at temperatures $\sim$0.1 keV, comparable with those fitted to ULSs. The first class is that of classical SSSs. For our comparison, we used a sample of SSSs in M31 with well-constrained distance and spectral parameters \citep{2010ApJ...717..739O}, and we added the three best-studied SSSs in the Local Group (Cal87, Cal83 and RXJ0513.9$-$6951), with fit parameters from \citet{2013A&A...559A..50N}. We find (Figure \ref{SSS_LvT_galac_fig} and Figure \ref{SSS_RvT_galac_fig}) that ULSs and classical SSSs represent two separate populations. Although their temperature range largely overlaps, ULSs have larger radii, and bolometric blackbody luminosities at least an order of magnitude higher. We are aware of the uncertainty in the extrapolation of a blackbody spectrum from soft X-rays to (unobserved) UV; however, the difference in luminosity between the two populations is a robust result, and applies also to the directly observed luminosity above 0.3 keV. Only ULSs reach luminosities one order of magnitude above the Eddington limit of a nuclear-burning WD. As we mentioned in Section 2.2, a couple of sources  
(r2-12 in M\,31 and SSS2 in NGC\,300)
seem to have intermediate properties, and it is not clear whether they should be classified as ULSs or as extremely bright classical SSSs. In summary, apart from those few ambiguous cases, the luminosity distribution of ULSs and SSSs is consistent with the scenario that the former group contains accreting black holes and the latter group  nuclear-burning white dwarfs.

\begin{figure}
\includegraphics[width=84mm]{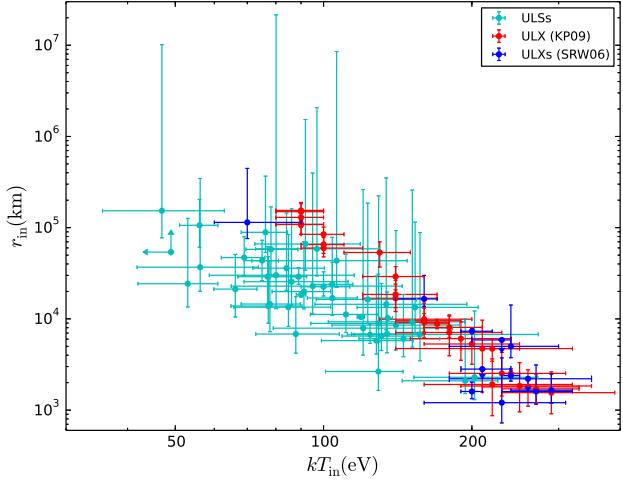}
 \caption{Light blue datapoints: best-fitting disk-blackbody radius versus best-fitting peak colour temperature, for each ULS observation with spectral information. Red datapoints: disk-blackbody radius versus peak temperature for the soft thermal component in a sample of ULX spectra fitted by \citet{2009MNRAS.398.1450K}. Blue datapoints: disk-blackbody radius versus peak temperature for the soft thermal component in a sample of ULXs studied by \citet{2006MNRAS.368..397S}. This diagram suggests a physical connection between the soft thermal component of ULSs and ULXs---although, in the latter case, a dominant harder component is also present.}
  \label{SSS_ULX_fig}
\end{figure}
\begin{figure}
\centering
\includegraphics[width=84mm]{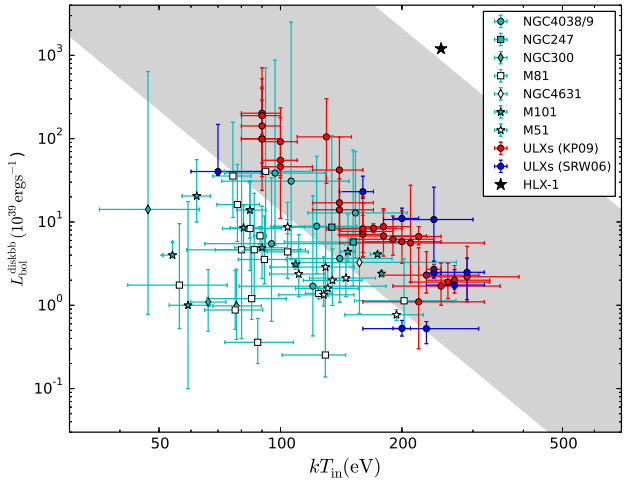}
\caption{Bolometric disk-blackbody luminosity versus peak colour temperatures for ULSs and ULXs, showing no sign of the $L \propto T^4$ standard disk relation. (See \citet{2007Ap&SS.311..213S} and \citet{2009MNRAS.398.1450K}, for a discussion of the lack of such correlation in ULXs.) The grey shaded area represents the range of temperature and luminosities in which the spectrum of an accreting BH is expected to be in the canonical high/soft state, dominated by a thermal disk component ({\it i.e.}, for $0.02 \la \dot{m} \la 1$): almost all ULSs fall outside this region, effectively ruling out an IMBH interpretation. The star near the top of the diagram represents the characteristic luminosity and peak temperature of the IMBH candidate HLX-1 \citep{2009Natur.460...73F} at the peak of its recurrent outbursts.}
\label{LvsT_ULX_fig}
\end{figure}

The second class of accreting sources with a cool thermal component is that of ULXs with a soft excess \citep[][for a review]{2011NewAR..55..166F}. The physical origin of the soft excesses in ULXs is still disputed. Currently, perhaps the most popular interpretation is that the soft excess is emitted from a super-critical disk wind, launched from near the spherization radius \citep{2003MNRAS.345..657K, 2007MNRAS.377.1187P, 2013MNRAS.435.1758S, 2015MNRAS.447.3243M}. An alternative scenario \citep{2013ApJ...776L..36M} is that it comes from the disk---either from a full standard disk around an IMBH or from the outer part of the disk around a super-Eddington stellar-mass BH. In addition to the thermal component, ULXs have a (dominant) harder component, with a downturn at energies $\ga$5 keV. The origin of the hard component (absent or very faint in ULSs) is also disputed, but is likely to be associated with the hot inner region of the inflow---either a non-standard disk or a warm corona \citep{2007Ap&SS.311..203R, 2014ApJ...785L...7M, 2015MNRAS.447.3243M}. Regardless of the true physical origin of the two components, ULX spectra have traditionally been fitted with a disk-blackbody model plus a Comptonized component. For our comparison of ULX and ULS properties, we used two samples of well-studied ULXs in nearby galaxies, from \citet{2006MNRAS.368..397S} and \citet{2009MNRAS.398.1450K}. This is by no means a statistically complete sample; however, it is representative of the general appearance of ULXs with a broad-band component and a soft thermal component. We compared the disk-blackbody temperatures and radii of our sample of ULSs (Table \ref{params_tab_disk}) with the disk-blackbody temperatures and radii of those ULXs from the published literature\footnote{Similar results are expected if we compare single-temperature blackbody rather than disk-blackbody fit parameters in ULS and ULX spectra; we leave this exercise to further work.}. We find (Figure \ref{SSS_ULX_fig} and Figure \ref{LvsT_ULX_fig}) that thermal components in ULXs lie at the high-temperature end of the ULS population. There is a degree of overlapping in temperature and luminosity for sources with a thermal component at $kT_{\rm in} \approx 0.10$--$0.15$ keV. Above those temperatures, we find almost exclusively ULXs with the additional, dominant harder component; below those temperatures, most sources are ULSs without hard tails. The lack of ULXs with a dominant hard tail and a thermal component cooler than $\approx$100 eV may be partly due to the fact that it is more difficult to identify a statistically significant soft excess at such low temperatures if the spectrum is dominated by the broad-band component. However, the lack of a population of soft sources with no hard tail and a thermal component hotter than $\approx$150 eV must be due instead to a real threshold in the physical structure of the inflow/outflow.

\begin{table*}
\centering
\caption{Best-fitting parameters and F-test significance of additional spectral features (absorption edges or line emission) at all epochs in which they significantly improve the fit. Errors are 90\% confidence limits for single parameters.}
\begin{tabular}{lrrrrrr}
\hline\hline
\multicolumn{1}{c}{Galaxy \& Obs ID}&\multicolumn{3}{c}{Thermal Plasma}&\multicolumn{3}{c}{Edge}\\
\cline{2-7}\\[-6pt]
& $kT$ (keV) & Normalization & Significance & $E$ (keV) & $\tau_{\rm max}$ & Significance\\
\hline\\
Antennae-3042 &&&& \U{1.04}{0.03}{0.04} & \U{3.68}{3.06}{2.41} & $>99.9\%$\\[5pt]
NGC\,4631-0110900201 &&&& \U{0.94}{0.03}{0.02} & \U{4.18}{4.30}{1.82} &  $>99.999\%$\\[5pt]
NGC\,247-12437 &&&& \U{0.95}{0.03}{0.02} & \U{4.45}{2.98}{1.93} & $>99.999\%$\\[5pt]
NGC\,247-17547 & \U{1.50}{*}{0.50} & \U{9.8}{7.8}{6.9}$\E{-5}$ & $>95\%$ &&&\\[5pt]
M\,51-1622 &&&& \U{0.98}{0.12}{0.08} & \U{1.6}{*}{1.0} & $>95\%$\\[5pt]
M\,51-3932 & \U{0.61}{0.11}{0.12} & \U{6.6}{5.3}{3.1}$\times10^{-6}$ & $>99\%$ &&&\\[5pt]
M\,51-13813 & \U{0.60}{0.11}{0.14} & \U{4.9}{1.8}{1.4}$\times10^{-6}$ & $>99.99\%$ &&&\\[5pt]
M\,51-13812 & \U{0.56}{0.06}{0.06} & \U{5.2}{2.3}{1.6}$\times10^{-6}$ & $>99.99\%$ &&&\\[5pt]
& \U{5.1}{*}{3.2} & \U{1.5}{1.1}{1.0}$\times10^{-6}$ & $>95\%$ &&&\\[5pt]
M\,51- 13814 & \U{0.64}{0.08}{0.04} & \U{1.6}{0.4}{0.3}$\times10^{-5}$ & $>99.99\%$ & \U{1.02}{0.05}{0.05} & \U{0.52}{0.30}{0.24} & $>99.9\%$\\[5pt]
M\,51- 13815 & \U{0.67}{0.10}{0.08} & \U{6.9}{3.4}{2.9}$\times10^{-6}$ & $>95\%$ & \U{1.30}{0.06}{0.04} & \U{5.3}{*}{3.9} & $>99\%$\\[5pt]
M\,101-934-high & \U{0.61}{0.06}{0.06} & \U{1.7}{1.0}{1.1}$\E{-5}$ & $>99.9\%$ &&&\\[5pt]
& \U{0.98}{0.16}{0.17} & \U{3.9}{1.4}{1.4}$\E{-5}$ & $>95\%$ &&&\\[5pt]
& \U{2.5}{*}{1.2} & \U{1.9}{1.9}{1.6}$\E{-5}$ & $>95\%^1$ &&&\\[5pt]
M\,101-934-med & \U{0.61}{0.06}{0.06} & \U{2.6}{0.5}{0.4}$\E{-5}$ & $>99\%$ & \U{1.07}{0.03}{0.03} & \U{2.1}{1.3}{0.8} & $>99.9\%$\\[5pt]
M\,101-row 32 & \U{0.59}{0.21}{0.26} & \U{2.9}{4.6}{1.5}$\E{-6}$ & $>99\%$ & \U{0.93}{0.05}{0.04} & \U{2.1}{1.6}{0.9} & $>99\%$\\[5pt]
M\,101-4737 & \U{0.70}{0.17}{0.13} & \U{3.1}{1.5}{1.5}$\E{-5}$ & $>95\%$ &&&\\[5pt]
 & \U{1.30}{0.20}{0.20} & \U{4.3}{1.7}{1.6}$\E{-5}$ & $>99\%$ &&&\\
\hline 
\end{tabular}
\begin{flushleft}
$^1$ This thermal plasma component is statistically equivalent to a {\it bremsstrahlung} or {\it comptt} component.
\end{flushleft}
\label{residual_tab}
\end{table*}

\subsection{Transition ULSs in M\,101 and NGC\,247}\label{m101_sect}
In some epochs, the two ULSs in M\,101 and NGC\,247 overlap with the spectral parameters typical of the ULX population. In both sources, a harder component appears in epochs when their blackbody temperature is highest (Soria \& Kong 2015, MNRAS, in press, arXiv:1511.04797), approaching 150 eV (corresponding to the brightest X-ray state for both sources, although not necessarily the most luminous in bolometric terms). Both ULSs show an interesting and somewhat similar spectral change at those high count rates. 

\begin{figure*}
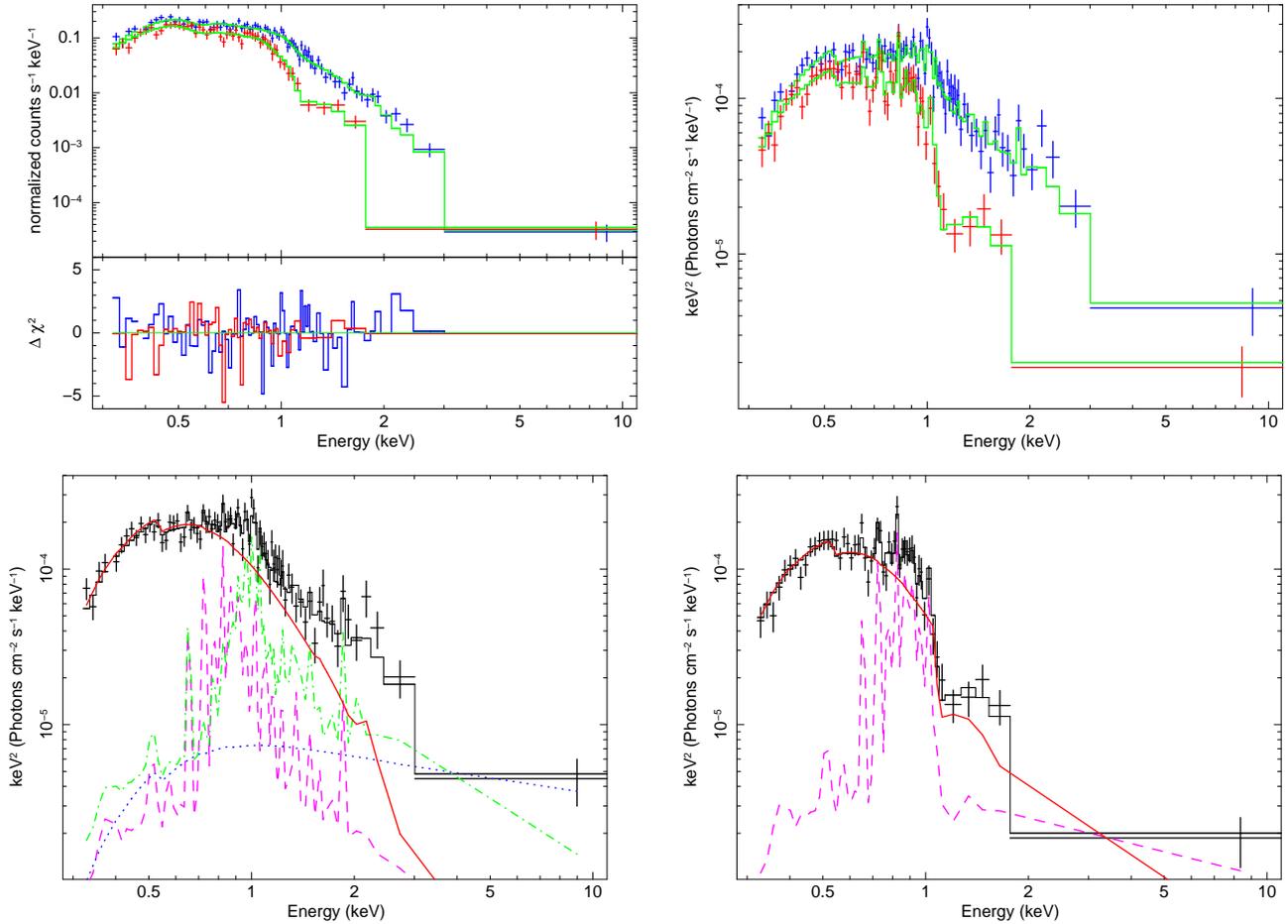

  \centering
  \includegraphics[width=60mm,angle=270]{m101_1.ps}
  \includegraphics[width=60mm,angle=270]{m101_unfold_plot.ps}\\[8pt]
  \hspace{0.05cm} 
  \includegraphics[width=60mm,angle=270]{test_m101_high_unfolded.ps}
\hspace{0.7cm}
  \includegraphics[width=60mm,angle=270]{test_m101_med_unfolded.ps}\\
\caption{Top left panel: {\it Chandra}/ACIS-S spectral data, model fits and $\chi^2$ residuals for the M\,101 ULS during the high count-rate (blue datapoints) and medium count-rate (red datapoints) time intervals of ObsID 934 (2000 March 26). The intervals were originally defined by \citet{2003ApJ...582..184M} and were also used by \citet{2005ApJ...632L.107K} and by Soria \& Kong (2015, MNRAS, in press, arXiv:1511.04797).
Both spectra were grouped to $> 15$ counts per bin before $\chi^2$ fitting. 
Top right panel: unfolded spectra of the M\,101 ULS during the high and medium count-rate intervals of ObsID 934 (blue and red datapoints, respectively), corresponding to the spectral fits shown in the top left panel.
Bottom left panel: detailed view of the model components required to fit the high count-rate interval: a blackbody component at $T_{\rm bb} \approx  135$ eV (solid red curve), and three mekal components with $T_1 \approx 0.6$ keV (dashed magenta curve), $T_2 \approx 1.0$ keV (dot-dashed green curve) and $T_3 \approx 2.5$ keV (dotted blue curve).
Bottom right panel: model components required to fit the medium count-rate interval: a blackbody component at $T_{\rm bb} \approx 119$ eV (solid red curve), a single mekal component with $T \approx 0.6$ keV (dashed magenta curve), and an absorption edge at $E \approx 1.05$ keV with optical depth $\tau \approx 2$.}
\label{lightcurves_fig}
\end{figure*}

For M\,101, we fitted two sub-intervals (at ``high" and ``medium" count rate) of the long {\it Chandra} observation from 2000 March 26 (ObsID 934); those time intervals were empirically defined by \citet{2003ApJ...582..184M}, and were later analyzed and discussed also by \citet{2005ApJ...632L.107K} and Soria \& Kong (2015, MNRAS, in press, arXiv:1511.04797).
The fitted blackbody temperature in the medium count-rate interval is $kT_{\rm bb} \approx 120$ eV, and there is a highly significant (Table \ref{residual_tab}) thermal-plasma components at $kT_1 \approx 0.6$ keV (red datapoints and residuals in Figure \ref{lightcurves_fig}). The thermal-plasma emission contributes $\approx 10\%$ of the unabsorbed luminosity in the 0.3--10 keV band, that is $\approx$ $3\E{38}\,\ergs$, compared with a total (blackbody plus mekal components) 0.3--10 keV emitted luminosity of $\approx$$2.5\times 10^{39}$ erg s$^{-1}$ in that time interval. Above 1 keV, the spectrum is significantly modified by an absorption edge at $E \approx 1.05 \pm 0.05$ keV, with optical depth $\tau \approx 2$ (Table \ref{residual_tab}).

In the high count-rate interval, the temperature of the optically thick thermal component increases to $kT_{\rm bb} \approx 135$ eV; the thermal-plasma component at $kT_1 \approx 0.6$ keV is still detected and another one is required at $kT_1 \approx 1.0$ keV, suggesting a spread of temperatures in the thermal plasma. No edge is present in the spectrum this time; instead, an additional, harder emission component (Table \ref{residual_tab}) provides significant flux in the 1--5 keV band (blue datapoints and residuals in Figure \ref{lightcurves_fig}). This hard tail can be equally well fitted with an additional {\it mekal} or bremsstrahlung component at $kT \ga 2$ keV, or with inverse-Compton emission ({\it e.g., comptt}). In fact, if we had no other knowledge of this source from other epochs, the inverse-Compton model would have been the default choice, and this accreting source would have been classified as a standard broad-band ULX with a strong blackbody soft excess at $kT_{\rm bb} \approx 0.14$ keV (or a disk temperature $kT_{\rm in} \approx 0.18$ keV). The 0.3--10 keV luminosity emitted in the three mekal components (or in the two mekal plus comptonized component) is $\approx$$7\times 10^{38}$ erg s$^{-1}$ during the high count-rate interval. The total emitted luminosity (blackbody plus all other components) in the same band is $\approx$$4\times 10^{39}$ erg s$^{-1}$ during the same interval.

Another epoch when the M\,101 ULS spectrum displays a significant harder component is 2005 January 1 (ObsID 4737), as shown in Soria \& Kong (2015, MNRAS, in press, arXiv:1511.04797). The hard excess above the blackbody component is again well fitted with mekal components, providing once again an emitted luminosity $\approx$$7\times 10^{38}$ erg s$^{-1}$ in the 0.3--10 keV band out of a total in the same band of $\approx$$4\times 10^{39}$ erg s$^{-1}$.

\begin{figure*}
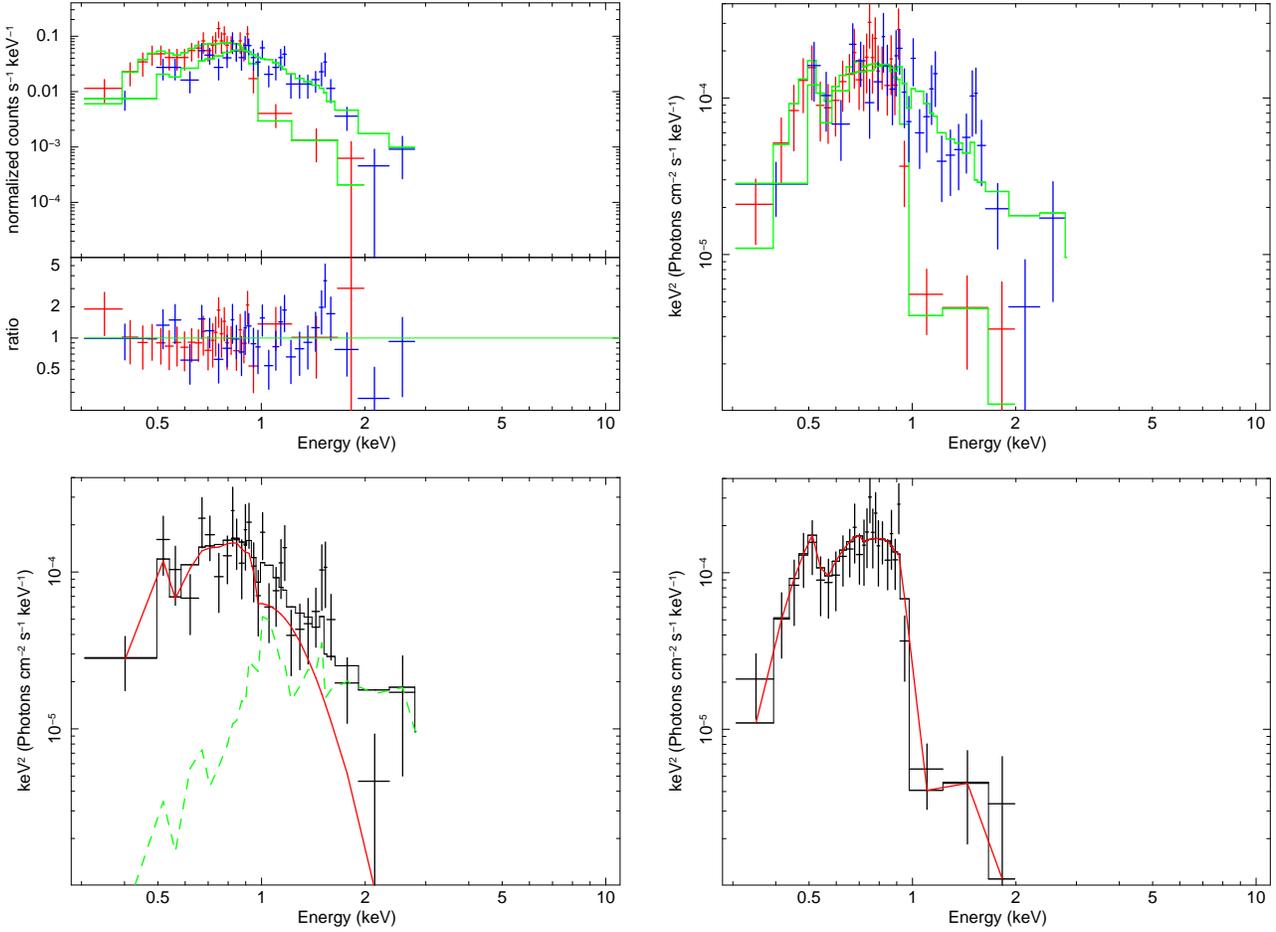

  \centering
  \includegraphics[width=60mm,angle=270]{ngc247_finalplot1.ps}
  \includegraphics[width=60mm,angle=270]{ngc247_finalplot2.ps}\\[8pt]
  \hspace{0.05cm} 
  \includegraphics[width=60mm,angle=270]{ngc247_finalplot4.ps}
  \includegraphics[width=60mm,angle=270]{ngc247_finalplot3.ps}\\   
\caption{Top left panel: {\it Chandra}/ACIS-S spectral data, model fits and data/model ratios for the NGC\,247 ULS during two different observations: blue datapoints are for ObsID 17547 (2014 November 12), red datapoints for ObsID 12437 (2011 February 1). As for the M\,101 ULS (see Figure \ref{lightcurves_fig}), both spectra are dominated by blackbody emission (see Table \ref{params_tab} for the fit parameters), but during ObsID 17547 there is additional harder emission, while during ObsID 12437 there is an absorption edge. Model ratios rather than residuals were plotted because both spectra were fitted (unbinned) with the Cash statistics and were later rebinned for plotting purposes only. Top right panel: unfolded spectra of the NGC\,247 ULS during ObsID 17547 (blue datapoints) and ObsID 12437 (red datapoints), corresponding to the fits shown in the top left panel.
Bottom left panel: detailed view of the model components required to fit ObsID 17547: a blackbody component at $T_{\rm bb} \approx  130$ eV, and a mekal component with $T_1 \approx 1.3$ keV.
Bottom right panel: model components required to fit ObsID 12437: a blackbody component at $T_{\rm bb} \approx  120$ eV, and an absorption edge at $E \approx 0.95$ keV with optical depth $\tau \approx 4$.}
\label{lightcurves_ngc247_fig}
\end{figure*}

For the ULS in NGC\,247, we compared the {\it Chandra} spectra from 2011 February 1 (ObsID 12437) and from 2014 November 12 (ObsID 17547). In the first epoch (red datapoints and ratios in Figure \ref{lightcurves_ngc247_fig}), the fitted blackbody temperature is $kT_{\rm bb} \approx 120$ eV. Adding thermal-plasma components here does not improve the fit, although this may be due to the low signal-to-noise ratio and few counts available (we had to use the Cash statistics for spectral fitting). An absorption edge is detected at $E = 0.95 \pm 0.02$ keV, with optical depth $\tau \approx 4$. In the 2014 spectrum (blue datapoints and ratios in Figure \ref{lightcurves_ngc247_fig}), the fitted blackbody temperature has increased to $kT_{\rm bb} \approx 140$ eV. The absorption feature has disappeared, and a hard tail is now significantly detected above the blackbody spectrum, up to $\approx$2 keV. The hard component can be equally well modelled as thermal-plasma emission at $kT \approx 1.5$ keV, or as an inverse-Compton component with a high-energy cut-off at $\approx$2 keV. Based on this epoch alone, this source could also be classified as a standard ULX with soft excess. The unabsorbed 0.3--10 keV luminosity of the harder component is $\approx$2 $\times 10^{38}$ erg s$^{-1}$, out of a total emitted luminosity in the same band of $\approx$$1.5\times 10^{39}$ erg s$^{-1}$.

\begin{figure*}
\centering
\includegraphics[width=86mm]{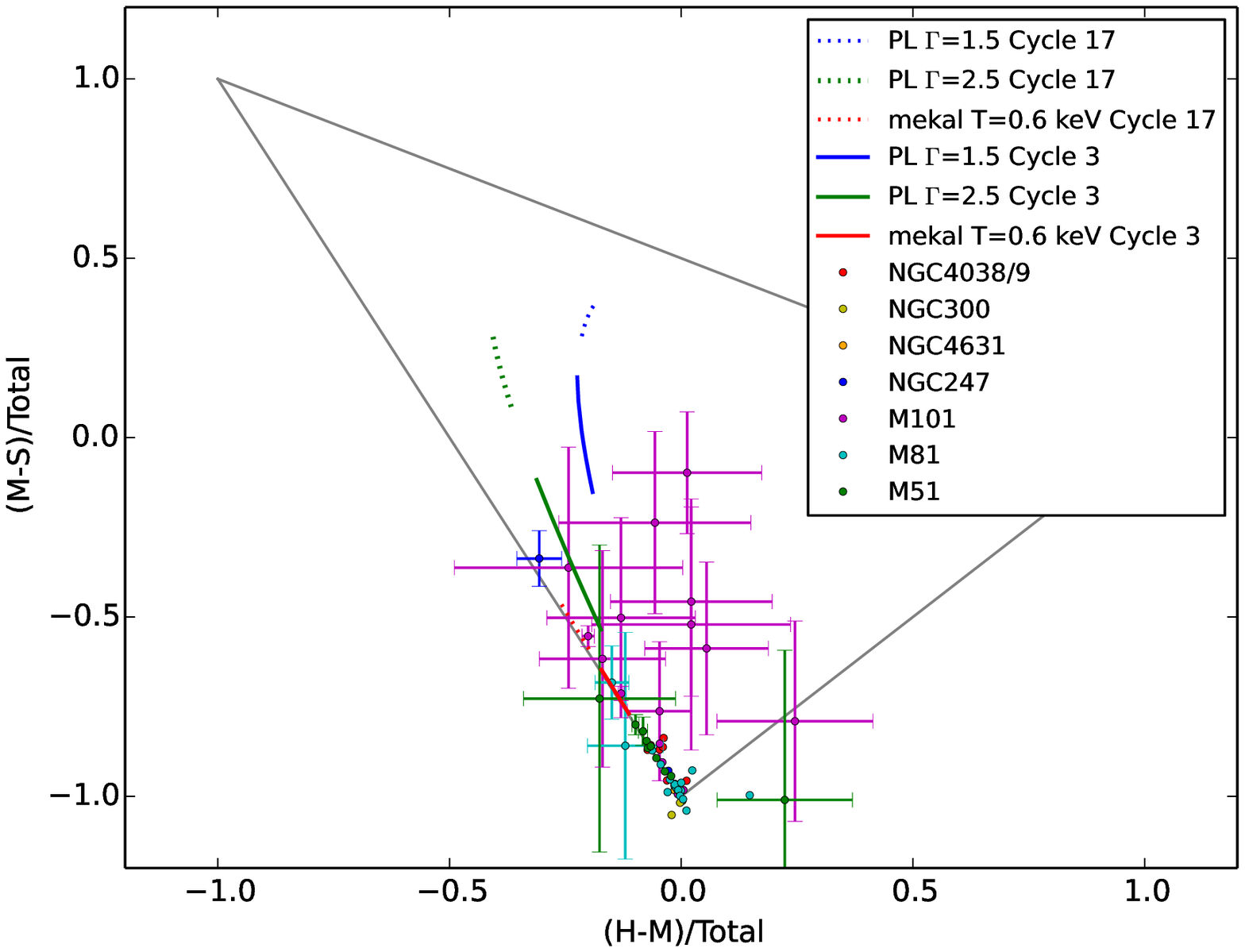}
\includegraphics[width=86mm]{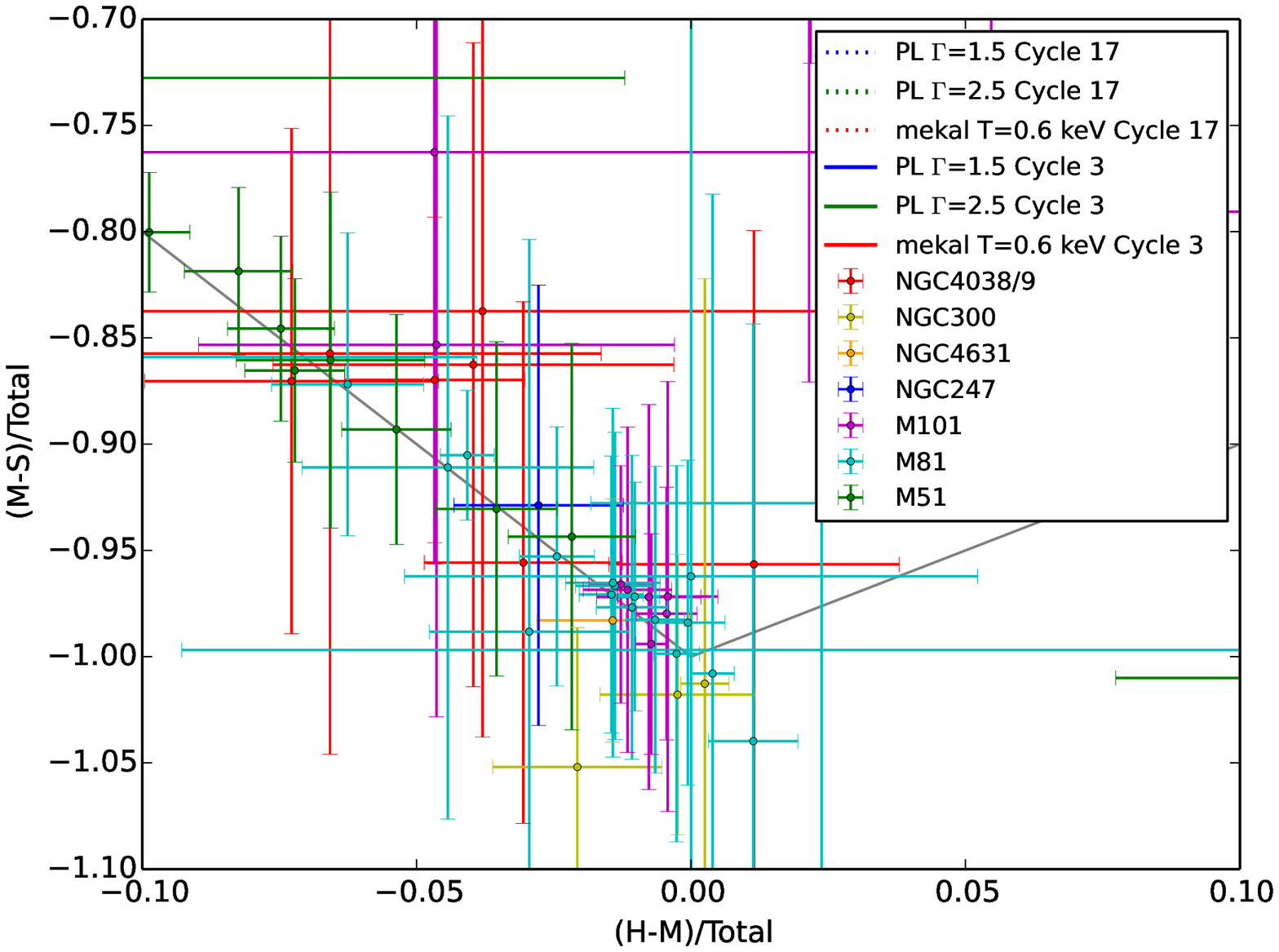}
\caption{Left panel: X-ray colour-colour plot of all ULS observations considered in our study (Table \ref{obs_table}). Error bars are plotted only for datapoints not clustered around $(0.0, -1.0)$; error bars for the clustered datapoints are shown in the zoomed-in right panel. The colours are defined based on the background-subtracted count rates in the following bands: S = 0.3--1.1 keV; M = 1.1--2.5 keV; H = 2.5--7.0 keV. Purely supersoft thermal spectra would cluster exactly at $(0.0, -1.0)$. As an indicative comparison, we plot the expected colours of sources with different spectral models. Red solid line: colours of a thermal-plasma spectrum with $kT = 0.5$ keV, and absorbing column density ranging from $2 \times 10^{20}$ cm$^{-2}$ to $2 \times 10^{21}$ cm$^{-2}$, if observed with {\it Chandra}/ACIS-S during Cycle 3. Red dotted line: same as above but for {\it Chandra}/ACIS-S observations during Cycle 17. Green solid line: colours of a power-law spectrum with photon index $\Gamma = 2.5$, and absorbing column density ranging from $2 \times 10^{20}$ cm$^{-2}$ to $2 \times 10^{21}$ cm$^{-2}$, if observed with {\it Chandra}/ACIS-S during Cycle 3. Green dotted line: same as above but for {\it Chandra}/ACIS-S observations during Cycle 17. Blue solid line: as above, for a photon index $\Gamma = 1.5$ during {\it Chandra} Cycle 3. Blue dotted line: as above, for {\it Chandra} Cycle 17. 
Right panel: X-ray colour-colour plot with error bars, zoomed in around $(0.0, -1.0)$ for clarity. 
These plots show that in most of the epochs, the observed colours are consistent with a pure supersoft thermal spectrum (essentially consistent with the predicted regions plotted in Figure \ref{colour_plot} for temperatures of 70--150 eV), but in a few cases there is a hint of additional harder components.}
\label{hardness_fig}
\end{figure*}

Among the observations listed in Table \ref{obs_table} for all seven ULSs, only a few have enough counts to allow multi-component fitting. However, the presence of harder spectral components in several observations (in addition to those already discussed for M\,101 and NGC\,247) may be indirectly inferred from model-independent X-ray colour-colour plots. Using the energy bands defined in Section 2.1, we expect that supersoft sources with a pure blackbody spectrum at $kT_{\rm bb} \la 100$ eV should cluster around (H$-$M)/T$ \approx 0$, (M$-$S)/T $\approx -1$. Instead, we find (compare Figure \ref{hardness_fig} with Figure \ref{colour_plot}) that the distribution of all the observations suggests a possible small scatter towards slightly harder colours \citep[a type of colours characteristic of what were called ``quasi-soft sources'',][]{2003ApJ...592..884D,2004ApJ...609..710D}. Such colours are consistent with the presence of either a thermal-plasma emission or a soft inverse-Compton or power-law tail in many epochs.

\subsection{Preliminary study of short-term variability}

Although the focus of this paper is on the spectral properties of the ULS population, we also briefly examined the intra-observation variability of the individual sources. The main objectives are to search for possible eclipses and to test whether there is a difference between the variability of the harder and softer photons. Finding eclipses in a ULS would confirm that it is viewed at high inclination, and would provide other useful constraints on the size of the system and the type of donor star, hence it would also help us understand the ULS class as a whole. Finding different variability behaviour at lower and higher energies would confirm the presence of two separate emission components, strengthening the link with the behaviour of standard ULXs \citep{2011MNRAS.411..644M, 2013MNRAS.435.1758S}.

Possible detection of an eclipse was claimed for M\,81 N1 in the {\it Chandra} ObsID 390 \citep{2002ApJ...574..382S}. At the beginning of that observation, the observed count rate and flux were at the highest recorded value for this source, $\approx$0.2 counts s$^{-1}$, but dropped to a value consistent with zero in $\approx$ 200 s. That drop happened simultaneously in a softer (0.3--0.7 keV) and harder (0.7--1.5 keV) sub-band (Figure \ref{M81_eclipse}, top panel). Possibilities such as a sudden change in accretion rate, or in the size of the photosphere, were examined in \citet{2002ApJ...574..382S}, but the most likely explanation was considered to be an eclipse by the companion star.  Therefore, we searched for evidence of similar sudden flux changes in other epochs. We examined all 20 {\it Chandra} observations, and found that sharp drops of the count rate to effectively zero for a few hundred or a few 1000 s occur several times, with different durations, no regular pattern, and sometimes interspersed with small flarings. We also found no relation between the hardness of the source (ratio between the count rates at 0.7--1.5 keV and 0.3--0.7 keV) during an observation and the likelihood of flux dips. In short, those dips appear as random fluctuations, as far as we can ascertain. We show (Figure \ref{M81_eclipse}, top panel) the original \citet{2002ApJ...574..382S}'s claim of an eclipse, together with a similar episode during ObsID 5940 (Figure \ref{M81_eclipse}, middle panel), and, conversely, a sudden increase of the count rate in ObsID 5944 (Figure \ref{M81_eclipse}, bottom panel) after an initial part of the observation consistent with no observed flux. Dips in the count rate from an average baseline value to zero often occur over timescales as short as $\approx$100 s. The morphology of these dips (and a few other dips of shorter duration during some of the other epochs) suggests that we are not seeing an eclipse from the donor star. Instead, we argue that we are looking at occultations from cold, optically thick ``clouds" (or equivalent structures) located near or above the outer disk. In the super-Eddington outflow interpretation of ULSs, we do indeed expect the emitting region to be surrounded by denser, cooler clouds, possibly falling onto the outer disk in a ``failed wind" scenario. Another possibility is that the obscuring material is associated with the accretion stream from the donor star, or where such stream impacts the outer disk.
In the case of M\,81 N1, assuming that it is a stellar-mass X-ray binary, with typical outer radius of the accretion disk $\sim$10$^{11}$--10$^{12}$ cm, we can easily verify that obscuring clouds or clumps in virial motion or Keplerian rotation, located at the outer edge of the disk, would be able to occult the emitting region (size $r_{\rm bb} \approx$ a few $10^9$ cm) in a time $t \sim 2r_{\rm bb}/v_{\rm K} \la 100$ s.

We found analogous energy-independent dips in the {\it Chandra} lightcurves of the M\,51 ULS, at some epochs. The two most striking examples are shown in Figure \ref{M51lightcurves_fig}. In the top panel, we show the full-band lightcurve from ObsID 13814. We examined other point-like sources and background regions in the same field to verify that the dip that occurred at observation time $\approx$1.5 $\times 10^5$ s is real, and not a glitch or detector artifact. The full duration of the dip is $\approx$8000 s. We also show (Figure \ref{M51lightcurves_fig}, middle panel) a zoomed-in view of the same dip in the 0.3--0.7 keV and 0.7--1.5 keV bands. Only 3 days later (ObsID 13815), in the subsequent {\it Chandra} observation of M\,51, the ULS appeared at a very low flux level at the start of the observation (Figure \ref{M51lightcurves_fig}, bottom panel) and then jumped up to an average flux an order of magnitude higher in less than an hour; the initial low-flux interval lasted for at least 18 ks. Clearly, we cannot explain both dips as due to an eclipse by the donor star, because of their different duration. On the other hand, it is also unlikely (especially for the dip in ObsID 13814) that such sharp, energy-independent flux suppressions and recoveries are due to state transition in the emitting region of the flow. Thus, occulting material in the outer disk or outer part of the BH Roche lobe passing in front of the emitting region seems to be the most likely explanation, especially if ULSs are seen at high inclination angles.

Somewhat similar dips were previously seen in a ULX in NGC\,55 \citep{2004MNRAS.351.1063S} and were also interpreted as orbiting clumps of obscuring material passing in front of our line of sight. In that case, the relative depth of the flux dips increased at higher energies \citep{2004MNRAS.351.1063S}; instead, in the M\,81 and M\,51 ULSs, the dips appear energy independent. However, the energy bands over which we detected the dips in those two ULSs are only 0.3--0.7 keV and 0.7--1.5 keV, because of a lack of higher-energy photons. Another ULX that showed non-periodic dipping behaviour, found from {\it Swift} monitoring, is NGC\,5408 X-1 \citep{2013MNRAS.433.1023G}; incidentally, both ULXs in NGC\,55 and NGC\,5408 are in the soft-ultraluminous regime \citep{2013MNRAS.435.1758S}, which is relevant for what we shall discuss in Section 5.3. A possible  comparison can also be made with the transitions between Compton-thin and Compton-thick spectra in some AGN, most notably NGC\,1365, which have been interpreted \citep{2005ApJ...623L..93R,2007ApJ...659L.111R} as occultations by rotating clouds or similar circumnuclear absorbers.

A discussion of the complex high intra-observation variability of the ULS in M\,101 is presented in Soria \& Kong (2015, MNRAS, in press, arXiv:1511.04797); here we shall only mention a couple of issues more directly relevant to the questions addressed in this paper. We did not find conclusive evidence of sharp energy-independent dips at any epoch, although there are possible hints of similar behaviour, for example the interval between 4000 and 6000 s into {\it Chandra} ObsID 4737 (Figure \ref{M101lightcurves_fig}, top panel). Instead, M\,101 ULS is notable for the short-term variability and irregular flaring of the emission, especially in the harder energy band (0.7--1.5 keV). The different behaviour of the soft-band and hard-band lightcurves (Figure \ref{M101lightcurves_fig}, both panels) is consistent with the presence of two emission components, which is what we have already shown from spectral analysis (Section 4.2). For a more quantitative estimate, we  calculated the root-mean-square (rms) fractional variability \citep{2002ApJ...568..610E, 2003ApJ...598..935M, 2003MNRAS.345.1271V, 2005MNRAS.363.1349G, 2011MNRAS.411..644M} in the two individual observations with the highest signal-to-noise ratio: ObsID 934 and ObsID 4737. The lightcurves from both observations were binned to 10-s intervals. The results are summarized in Table \ref{rms_table} (see also Soria \& Kong 2015, MNRAS, in press, arXiv:1511.04797). The huge fractional variability is another strong argument against IMBH models and more generally against the standard disk emission. Accretion disks in a high/soft state have low ($<$ a few $\%$) variability \citep{2010LNP...794...53B}.  The increase in rms variability at higher energies is similar to what is found in ULXs in the soft-ultraluminous state \citep{2013MNRAS.435.1758S, 2011MNRAS.411..644M, 2015MNRAS.447.3243M}.

\begin{table}
\caption{rms fractional variability in different energy bands, for the two observations of the M\,101 ULS with higher signal-to-noise ratio.}
\begin{center}
\begin{tabular}{ccc}
\hline\hline\\[-7pt]
Parameter & ObsID 934 & ObsID 4737 \\[-1pt]
\hline\\[-9pt]
Band (Hz) & $10^{-5}$--0.05  & $5 \times 10^{-5}$--0.05\\[3pt]
rms (0.3--0.7 keV) & $< 36\%$ & $(58\pm12)\%$ \\[3pt]
rms (0.7--1.5 keV) & $(70\pm4)\%$ & $(89\pm6)\%$ \\[3pt]
rms (1.5--7.0 keV) & $(100\pm24)\%$ & $(162\pm17)\%$ \\
\hline\\[-12pt]
\end{tabular}
\end{center}
\label{rms_table}
\end{table}

Although the lightcurves in the 0.3--0.7 and 0.7--1.5 keV bands for the M\,101 ULS appear different and probably dominated by two distinct emission components, such components cannot be independent of each other. As we discussed in Section 4.2 (see also Soria \& Kong 2015, MNRAS, in press, arXiv:1511.04797), the harder component appears only when the soft thermal component is warmer (blackbody temperature $\ga$100 eV) and its emitting area smaller (blackbody radius $\la$20,000 km). In the framework of the optically thick wind model, the harder component becomes visible when the outflow photosphere shrinks. Another caveat is that the variability of the observed count rate in the softer band is not necessarily a good proxy for the variability of the thermal component. We know that its radius and temperature vary in anticorrelation within a single observation (as is the case in ObsID 934) and between epochs, with smaller changes in the emitted flux (the larger emitting area somewhat compensating for the lower temperature). The observed changes in the 0.3--0.7 keV count rate mostly track the shift of the blackbody peak between soft X-rays and far-UV (in and out of the detector's sensitivity) rather than changes in the intrinsic bolometric luminosity of the thermal component; conversely, the dramatic flaring of the lightcurve above 0.7 keV probably tracks real flux changes in the harder component.

Finally, we examined the remaining four ULSs of our sample, looking for further examples of interesting short-term variability in the 0.3--0.7 and 0.7--1.5 keV bands, at least at epochs with better signal-to-noise ratio (Figure \ref{extralightcurves_fig}). We already showed that NGC\,247 ULS has two emission components during {\it Chandra} ObsID 17547: its lightcurve shows the harder component decline and disappear over the 5000-s duration of that observation. Independent short-term variability is seen in both the 0.3--0.7 and 0.7--1.5 keV bands for the Antennae ULS, but the low signal-to-noise ratio prevents more quantitative analysis. Only emission in the 0.3--0.7 keV band was detected in the NGC\,300 ULS, the softest of all the sources in our sample. The {\it XMM-Newton} observation of the NGC\,4631 ULS provides another possibly different type of variability, with a characteristic timescale of $\approx$15-ks independently seen both in the softer and harder band. This quasi-periodic variability was discovered by \citet{2007A&A...471L..55C} and interpreted as the binary period of the system; alternatively, \citet{2009ApJ...696..287S} suggested it might also be explained as a pulsation period of a B-type donor star. Considering that the periodicity is based only on two dips during an $\approx$40-ks good-time-interval of a single observation, we cannot rule out a non-periodic flaring behaviour, similar to flaring behaviour of other ULSs.  

\section{Physical interpretation}

\subsection{Alternative models for ULSs}



Our results show that ULSs share common properties and deserve to be identified as a distinct sub-class of accreting sources. Their characteristic blackbody temperatures are $\approx$50--140 eV, their fitted radii span a range $\approx$5,000--100,000 km, and the extrapolated bolometric luminosities are $\approx$ a few times $10^{39}$ erg s$^{-1}$. We have presented a sample of observations large enough to show a significant anticorrelation between temperature and luminosity, but no trend between bolometric luminosities and temperatures. Based on the population properties presented in Section \ref{results_sect}, we can now address some of the unsolved problems.

The most exotic and intriguing scenario proposed for ULSs is that we are seeing the optically thick, geometrically thin disk of an accreting IMBH in the high/soft state, by analogy with the thermal state of stellar-mass BHs. This implies that the fitted disk-blackbody radius corresponds to the inner radius of the accretion disk and should be a good approximation for the innermost stable circular orbit. Assuming non-spinning Schwarzschild BHs, the characteristic range of radii observed in our sample translates to a black hole mass range  $\approx$500--10,000\,$\Msol$. However, there are several problems with this interpretation. Firstly, we see the fitted radius change from epoch to epoch for the same source, indicating that it cannot be fixed at $r_{\rm ISCO}$. This point is reinforced by the fact that none of the sources follows the $L\propto T^4$ trend that is expected of accretion disks in the high/soft state \citep[\textit{e.g.}, ][]{2006ARA&A..44...49R}, nor the flatter $L\propto T^2$ trend observed when the accretion rate approaches the Eddington limit \citep{2004ApJ...601..428K}. Instead, we observe no significant trend in luminosity associated with changes in radius and temperature (Figures \ref{ULS_LvsT_fig} and \ref{SSS_fig}).

Secondly, the high/soft state scenario is not self-consistent, in the sense that the inferred luminosities and temperatures are too low for the fitted radii. This can be better illustrated in the following way. The characteristic relation between peak colour temperature and dimensionless accretion rate (in Eddington units) for a standard disk is 
\begin{equation}
kT_{\rm in} \approx 230\,(\dot{m}/M_4)^{1/4}\  {\mathrm{eV}} 
\end{equation}
\citep{1973A&A....24..337S, 1998PASJ...50..667K, 2007Ap&SS.311..213S, 2012MNRAS.420.1848D}, where $M_4$ is the BH mass in units of $10^4 M_{\odot}$. For a BH to be in the canonical high/soft state, the accretion rate must be (on average) $\dot{m} \ga 0.02$ \citep{2003A&A...409..697M}; we can also take $\dot{m} \la 1$ as a safe upper limit above which stellar-mass BHs are in the very high state or ultraluminous state, with an additional power-law component or broad-band inverse-Compton emission, and will appear non-thermal \citep{2006ARA&A..44...49R,2009MNRAS.397.1836G}. Taken together, those relations imply that for each fitted colour temperature, we can determine the range of BH masses for which the accretor is in the thermal-dominant state: 
\begin{equation}
M_{\rm min,4} \equiv 0.02 \left(\frac{230{\mathrm{~eV}}}{kT_{\rm in}}\right)^4 \la M_4 \la \left(\frac{230{\mathrm{~eV}}}{kT_{\rm in}}\right)^4 \equiv M_{\rm max,4}
\end{equation}
From this mass range, we can then determine the corresponding luminosity range of the thermal-dominant state at a given temperature, because $L \approx \dot{m} \, L_{\rm Edd}$ in that state. Thus: 
\begin{equation}
L_{\rm min} \equiv 0.02 L_{\rm Edd}(M_{\rm min}) \la L \la  L_{\rm Edd}(M_{\rm max}) \equiv L_{\rm max}, 
\end{equation}
that is 
\begin{equation}
5.2 \times 10^{38} \, \left(\frac{230{\mathrm{~eV}}}{kT_{\rm in}}\right)^4 
\la \frac{L}{{\mathrm{~erg~s}}^{-1}} 
\la 1.3 \times 10^{42} \left(\frac{230{\mathrm{~eV}}}{kT_{\rm in}}\right)^4.\label{lum_eq}
\end{equation}
The luminosity range of Equation \eqref{lum_eq} is plotted as a grey shaded area in Figure \ref{LvsT_ULX_fig}. For example, for $kT_{\rm in} = 100$ eV, the high/soft state condition $\dot{m} \ga 0.02$ requires a BH mass $M \ga 5600 M_{\odot}$ and a bolometric disk luminosity $L \ga 1.5 \times 10^{40}$ erg s$^{-1}$. The extrapolated disk luminosity of all the ULSs in our sample at almost all epochs falls below the shaded area; therefore, if they were IMBHs, their accretion rates would be too low to be in the high/soft state. 
The M\,101 ULS is the only source in our sample for which a direct mass estimate has been attempted, based on the optical spectra and time variability properties. \citet{2013Natur.503..500L} argued that they have determined strong constraints to its dynamical mass, suggesting that it is a stellar-mass BH. Some of the results of that study are still partly disputed (Soria \& Kong 2015, MNRAS, in press, arXiv:1511.04797), but the evidence in favour of a stellar-mass accretor appears to be stacking up, as it is also for standard ULXs \citep{2014Natur.514..198M}.

The second possible physical interpretation for ULSs is that they are the high-luminosity tail of the classical SSS population, which is thought to be powered by surface-nuclear-burning WDs \citep{1992A&A...262...97V, 1996LNP...472.....G, 1997ARA&A..35...69K, 2002A&A...387..944G, 2009ApJ...692.1532S}. ULSs and SSSs largely overlap in their temperature range; however, ULSs are on average an order of magnitude more luminous than the upper luminosity threshold of SSSs (Figure \ref{SSS_LvT_galac_fig}), with very few detections in the gap between the two populations. In itself, this does not prove that classical SSSs and ULSs are distinct physical systems, because ULSs have been specially selected over a large volume of space based on their extreme luminosity. (A study of the luminosity distribution of SSSs and ULSs in an identical, complete sample of galaxies is beyond the scope of this work.) Models of soft, thermal X-ray emission from the expanding atmosphere of a WD have been successfully used to reproduce nova outburst behaviour \citep[\textit{e.g.}, ][]{2010AN....331..179N, 2012ApJ...756...43V, 2013A&A...559A..50N}. The anticorrelation between temperatures and radii observed in ULSs (Figure \ref{SSS_RvT_galac_fig}) is at least qualitatively in agreement with this scenario, corresponding to the transition between optically bright and X-ray bright phases of SSSs \citep{1996ApJ...470.1065S, 2005MNRAS.364..462M, 2013MNRAS.432.2886R}. The fast variability (typical of clumpy ejecta), the occasional dips possibly caused by cold absorbers transiting in front of the emitting source, and the detection of additional thermal-plasma components may also be consistent with an optically thick outflow, which cools as it expands, launched from the WD surface. What is more problematic is that ULSs are persistently seen at luminosities an order of magnitude higher than the Eddington limit of the most massive WDs. It is hard to explain how quasi-steady nuclear burning can be sustained at such luminosities for decades, instead of triggering runaway processes leading to nova-like outbursts on timescales of a few tens of days \citep[\textit{e.g.}, as observed in MAXIJ0158$-$744: ][]{2012ApJ...761...99L, 2013ApJ...779..118M}.

The third possible interpretation for ULSs is that they are powered by either BHs or NSs accreting at highly super-critical rates. In this scenario, massive radiatively-driven outflows are launched from the disk; if the accretion is high enough and our viewing angle sufficiently high, such outflows may be optically thick and create a large photosphere that shrouds the source of X-ray photons near the compact object \citep{2003MNRAS.345..657K, 2007MNRAS.377.1187P, 2010MNRAS.402.1516K, 2015MNRAS.447L..60S}. The expansion and contraction of the wind photosphere accounts for the anticorrelated changes in the observed radius and luminosity, similarly to the expanding and contracting WD atmosphere in the thermonuclear burning scenario. Density inhomogeneities in the outflow can explain the highly variable nature of ULSs and the coexistence of optically-thin and optically-thick thermal emission components at some epochs, and absorption edges at other epochs. Clumpy, optically thick outflows are indeed predicted by MHD simulations of super-critically accreting BHs \citep{2014ApJ...796..106J,2013PASJ...65...88T, 2014PASJ...66...48T}. It was shown (Soria \& Kong 2015, MNRAS, in press, arXiv:1511.04797) that with plausible assumptions on the super-Eddington luminosity and launching radius of the outflow (near the spherization radius), the observed temperatures and blackbody radii of ULSs suggest a highly super-Eddington $\dot{m} \sim$ a few 100 (Figure \ref{SSS_RvT_galac_fig}), which for a 10-$M_{\odot}$ BH corresponds to $\dot{M} \approx 10^{-4} M_{\odot}$ yr$^{-1}$. This is certainly extreme 
but similar to the accretion rate inferred for example in SS\,433 \citep{2004ASPRv..12....1F}. Mass transfer rates $> 10^{-4} M_{\odot}$ yr$^{-1}$ from a donor star onto a BH or NS, on a thermal timescale, were shown to be viable \citep{2015arXiv150308745W} at various stages of stellar evolution, most notably when intermediate-mass stars pass through the Hertzsprung gap.

\begin{figure}
    \centering
    \includegraphics[width=84mm]{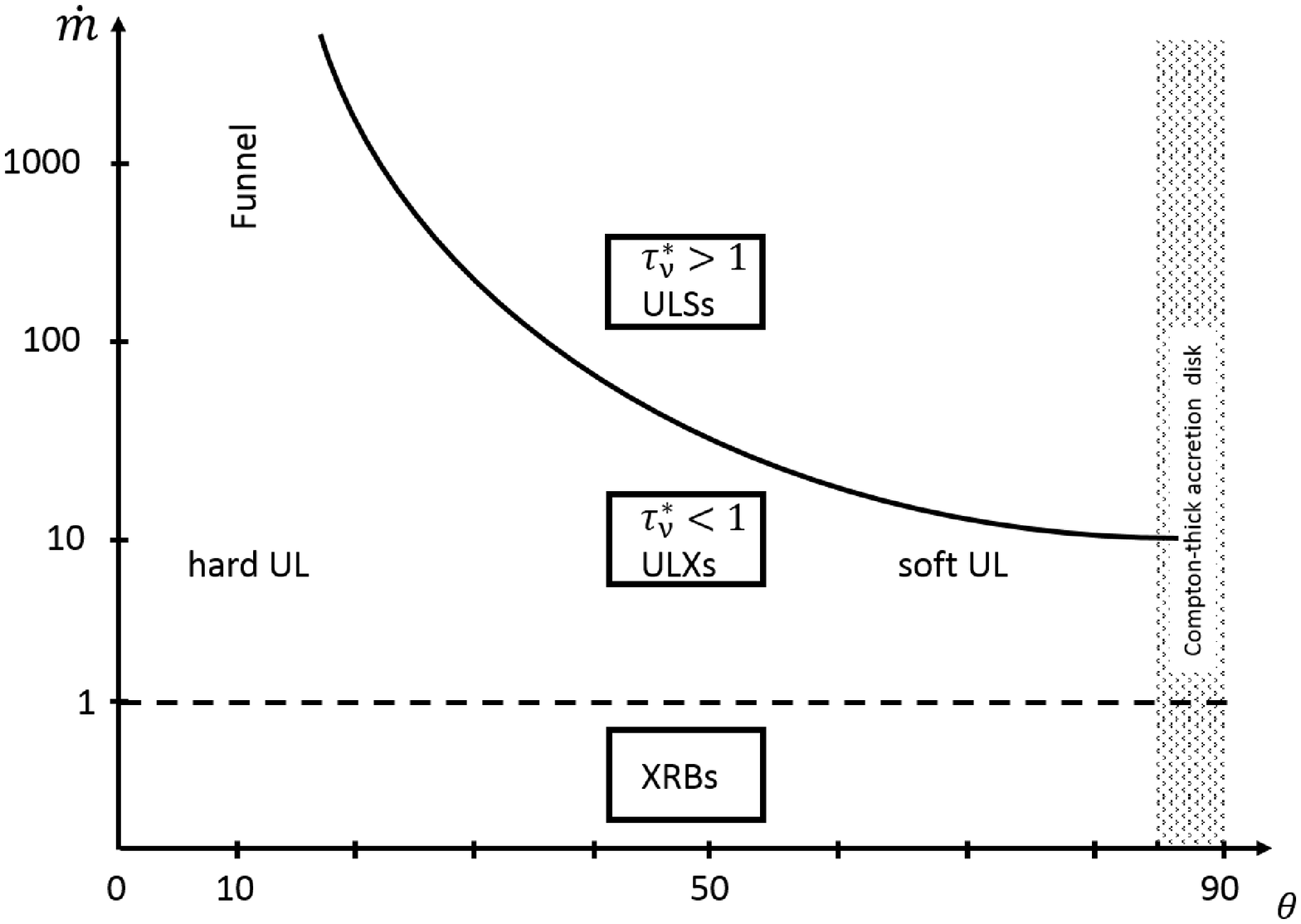}\\[8pt]
    \includegraphics[width=84mm]{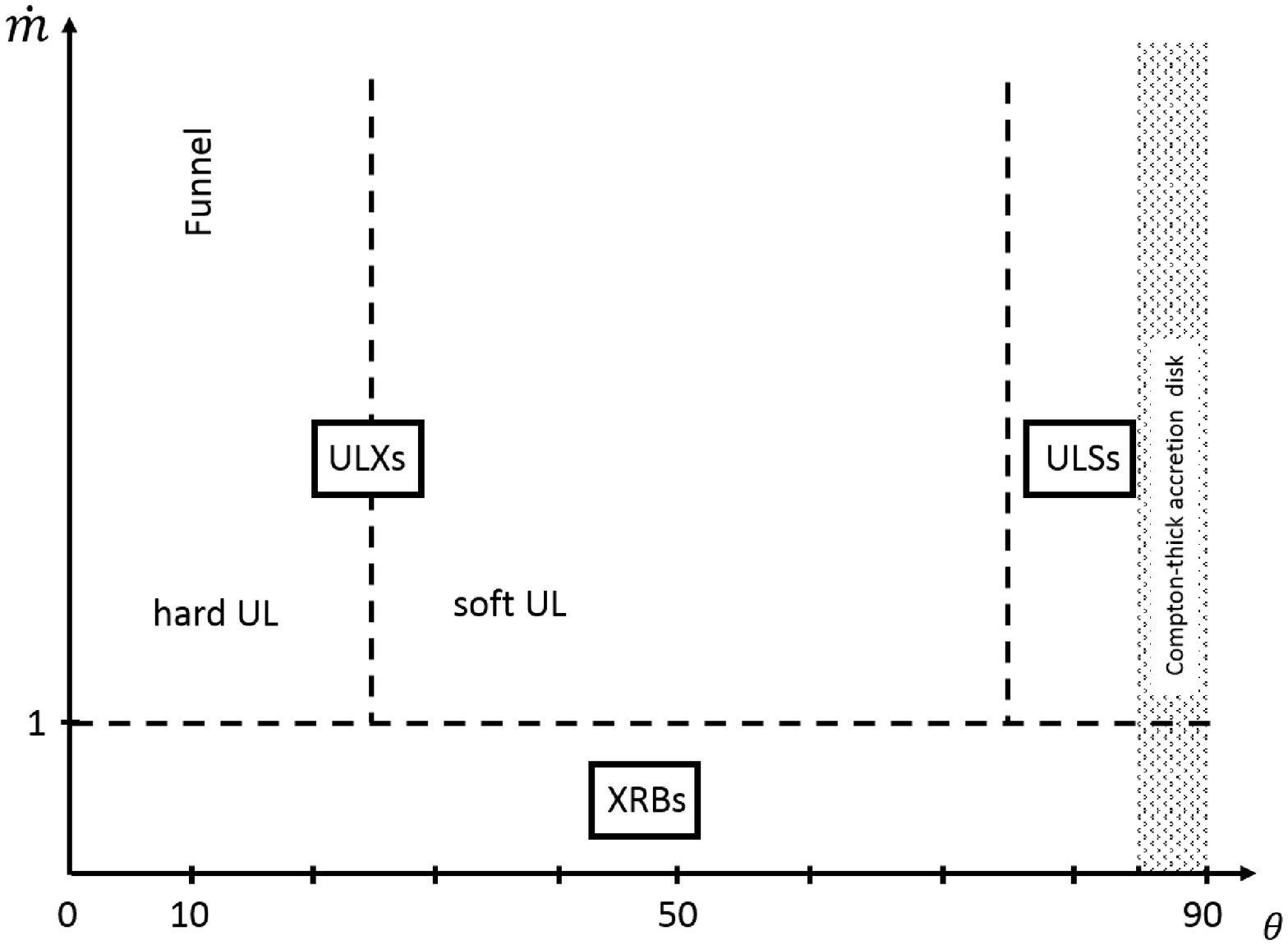}
\caption{Top panel: cartoon sketch of our proposed classification of ULXs and ULSs, function both of the accretion rate $\dot{m}$ and of the viewing angle $\theta$.  For super-critical accretors, at most lines-of-sight (except for the polar funnel), there will be an accretion rate threshold at which the outflow becomes dense enough to be effectively optically thick to the X-ray emission from the inner accretion disk. However, this threshold will be reached at lower values of $\dot{m}$ for sources seen at higher inclination angles, because the wind is thicker at higher $\theta$. Bottom panel: an alternative classification in which the difference between ULXs and ULSs depends only on $\theta$. This is not our preferred scenario, because it ignores the fact that the optical thickness of the outflow increases with $\dot{m}$ \citep{2007MNRAS.377.1187P}, and the opening angle of the funnel decreases with $\dot{m}$ \citep[e.g.,][]{2009MNRAS.393L..41K}.}
\label{cartoon_model}
\end{figure}

We suggest that the accretion rate and viewing angle determine the average temperature, radius and luminosity over a viscous timescale, while the clumpiness of the outflow causes the observed short-term variability and possible occultation episodes, on timescales of a few 100 to a few 1000 s.
The simple model proposed by \citet{2015MNRAS.447L..60S} and Soria \& Kong (2015, MNRAS, in press, arXiv:1511.04797) is based on a spherically symmetric approximation of the outflow. In reality \citep[as already noted in][]{2007MNRAS.377.1187P}, the viewing angle is also important, because the outflow is expected to be denser closer to the disk plane, and may have an open funnel along the polar axis. Therefore, a super-Eddington accreting BH may appear as a ULS (in which the source of hard photons is completely masked by the wind photosphere) when seen at high inclination angles {\it {and}} extremely high accretion rates (effectively optically thick outflow); or as a ULX in the soft-ultraluminous regime (in which there is a soft inverse-Compton tail) when seen at slightly lower inclination angles and/or accretion rates (outflow only optically thick to scattering); or as a ULX in the hard-ultraluminous regime (in which there is a harder, more extended inverse-Compton tail) when seen at low inclination angles (looking into the polar funnel). See Figure \ref{cartoon_model} for a cartoon description of this spherically-symmetric scenario.

\begin{figure}
    \hspace{-0.65cm}
    \centering
    \includegraphics[width=55mm,angle=270]{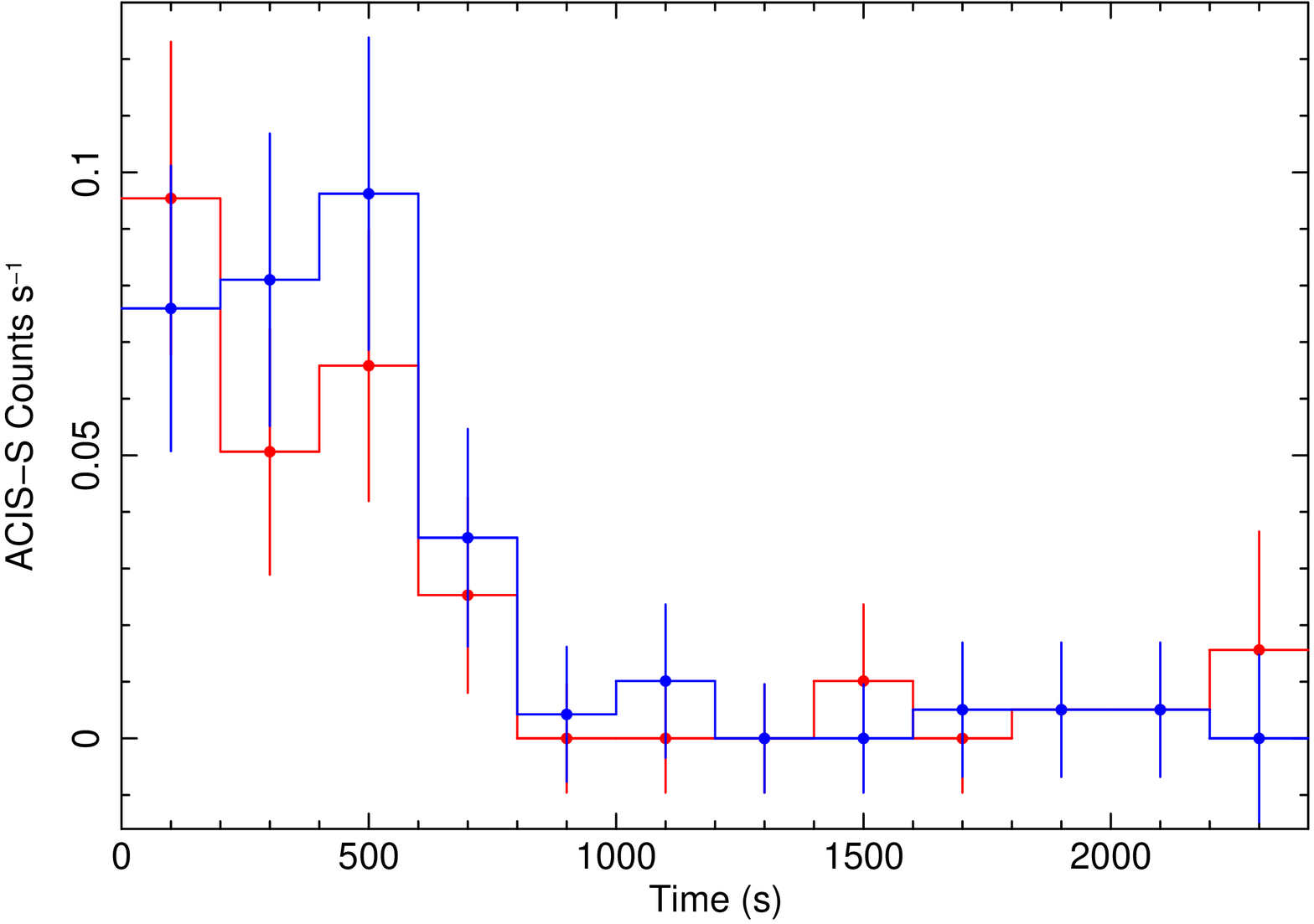}\\[8pt]
    \hspace{-0.65cm} 
    \includegraphics[width=55mm,angle=270]{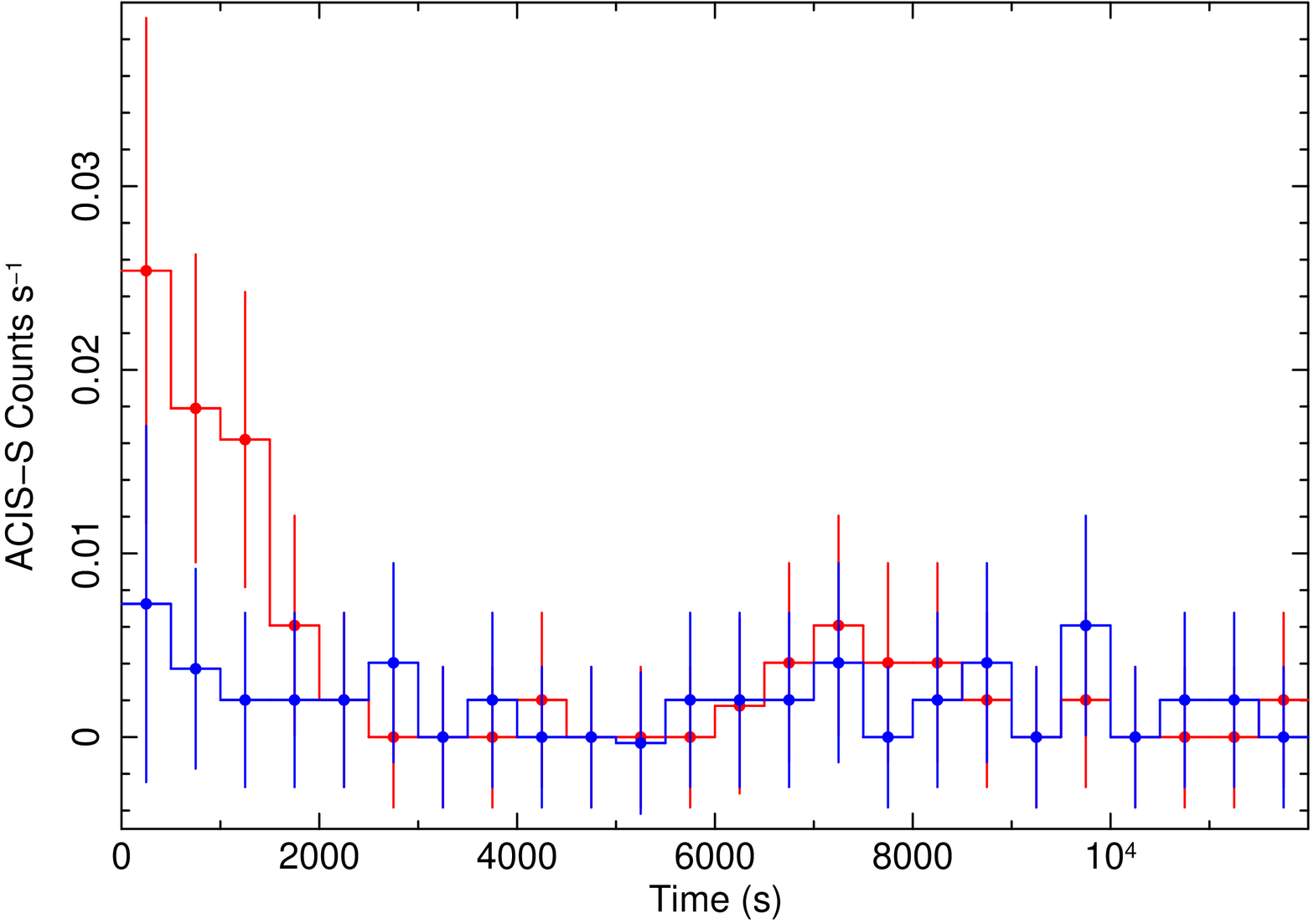}\\[8pt]
    \hspace{-0.65cm} 
    \includegraphics[width=55mm,angle=270]{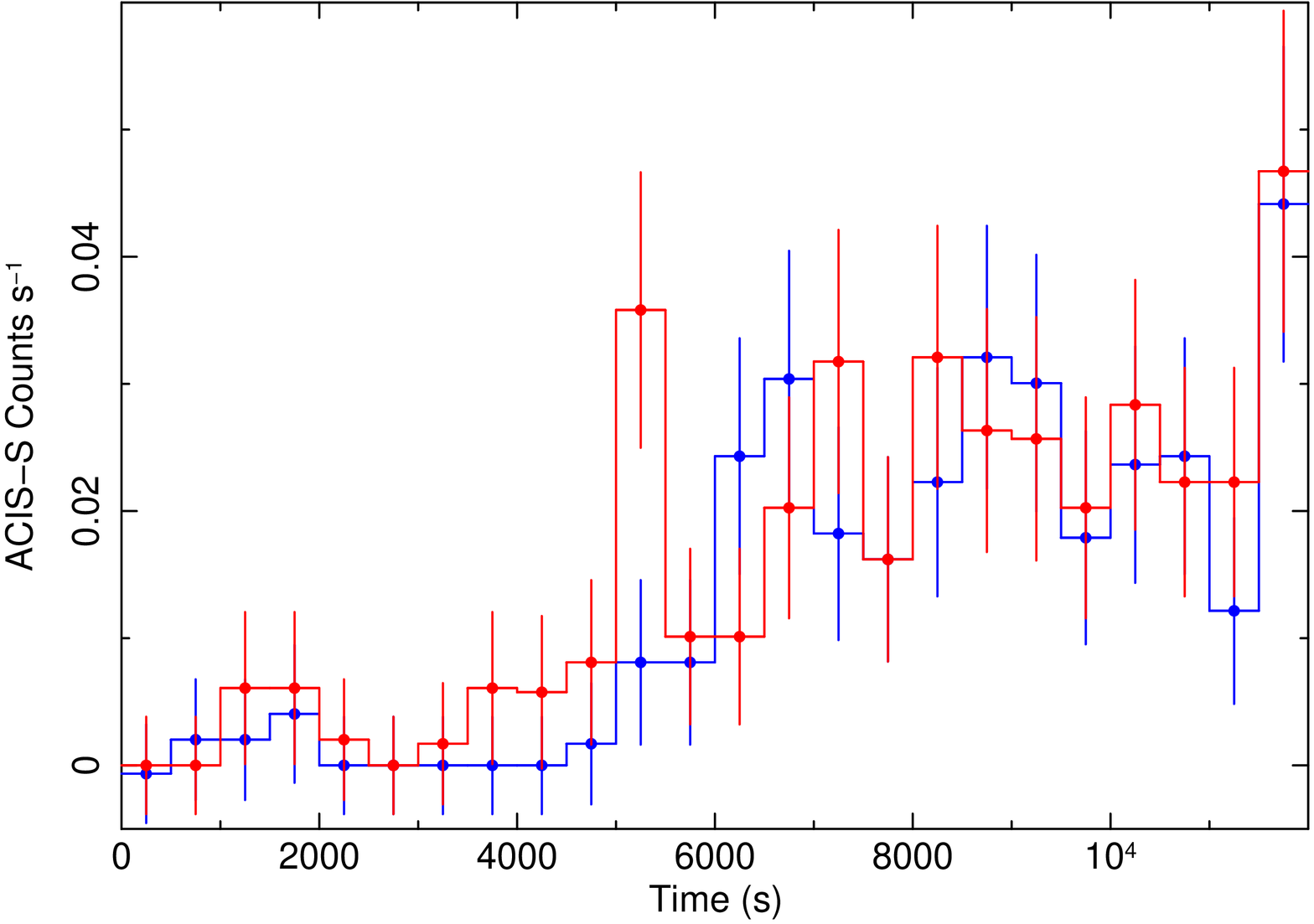}
\caption{Top panel: {\it Chandra}/ACIS-S background-subtracted lightcurve of M\,81 N1 during ObsID 390. Red datapoints = 0.3--0.7 keV band; blue datapoints = 0.7--1.5 keV band. The data have been binned to 200-s intervals.  Middle panel: as above, for ObsID 5940 and 500-s bins. Bottom panel: as above, for ObsID 5944 and 500-s bins. Error bars in all plots have been calculated using Gehrels statistics \citep{1986ApJ...303..336G}. (See Table \ref{obs_table} for details of the observing epochs.)}
\label{M81_eclipse}
\end{figure}



\subsection{ULS accretors: neutron stars or black holes?}

For a large enough accretion rate, NS accretors should be just as viable as BH accretors in powering ULSs. If the magnetic field of the NS is too strong, it will impede the accretion flow and truncate the accretion disk at the magnetospheric radius. However, at near- or super-Eddington accretion rates, the pressure of the inflow can push the magnetospheric radius down to $r_{\rm ISCO}$ or just outside the NS surface. The magnetospheric radius is defined as
\begin{gather}
r_{\rm M} = 5.1\E{8}\dot{M}_{16}^{-2/7}m_1^{-1/7}\mu_{30}^{4/7}\,\centi\meter,\label{mag_field_eq}
\end{gather}
\citep{2002apa..book.....F} where $\dot{M}_{16}$ is the accretion rate in $10^{16}\,\gram\,\second^{-1}$, $m_1$ is the NS mass in $\Msol$, $\mu_{30}$ is the magnetic moment $\mu=Br^3$ in units of $10^{30}\,\guass\,\centi\meter^3$. 
For a 1.4-$M_{\odot}$ NS, $r_{\rm ISCO} \approx 1.2 \times 10^6$ cm $\approx r_{\rm NS}$. Hence, the magnetospheric radius is pushed down to $r_{\rm ISCO}$ for $\dot{M} \ga 4.4 \times 10^{19} B_9^2$ g s$^{-1}$ $\approx 7 \times 10^{-7} B_9^2 M_{\odot}$ yr$^{-1}$ (with $B_9$ the magnetic field in units of $10^9$ G). Such values of the accretion rates are plausible not only for Hertzsprung-gap stars but even for massive donors evolving on their nuclear timescale. For comparison, the Eddington accretion rate for a NS is $\approx 10^{18}$ g s$^{-1}$. Moreover, if a super-Eddington disk outflow develops and forms a large, optically thick photosphere at $\dot{M} \sim$ a few $10^{-5} M_{\odot}$ yr$^{-1}$, the details of what happens near the inner edge of the accretion disk or at the surface of the NS become irrelevant, and a NS would look very similar to a BH, scaled to their respective Eddington luminosities. Since extremely super-critical accretion is expected to happen both in NS and BH systems \citep{2015arXiv150308745W}, we suggest that some ULSs may contain NS, as is also the case for some ULXs \citep{2014Natur.514..202B}. 
When the fitted blackbody radii and temperatures of the 7+2 supersoft sources in our sample are compared with the predictions of the optically thick outflow model of Soria \& Kong (2015, MNRAS, in press, arXiv:1511.04797), we see (Figure \ref{SSS_RvT_galac_fig}) that there is an observational spread between the expected location of 10-$M_{\odot}$ BHs and 1.4-$M_{\odot}$ NSs. Systems such as r2-12 and NGC\,300 SSS2 are indeed consistent with super-Eddington accreting NSs.

\subsection{Link between ULXs and ULSs}


The X-ray spectra of most ULXs can be described as a slightly curved broad-band component with an additional, soft thermal component (sometimes known as soft excess) at $kT \approx 0.15$--0.30 keV \citep[][for a review]{2011NewAR..55..166F}. The origin of both components is still unclear. In a super-Eddington accretion scenario \citep{2003MNRAS.345..657K, 2007MNRAS.377.1187P, 2015MNRAS.447.3243M}, the soft emission comes from the wind, near or just outside the spherization radius. Smeared line-like residuals are sometimes seen at energies $\sim$0.7--2 keV particularly for soft-ultraluminous ULXs; those features are consistent either with thermal plasma emission from a collisionally ionized wind, or with absorption of the smooth broad-band continuum in a partially ionized region of the outflow \citep{2014MNRAS.438L..51M}. The broad-band component becomes steeper, and truncated at lower energies (typically, $E \approx 5$ keV) for sources seen at higher inclination angles, which are probably seen through a thicker disk wind \citep[soft ultraluminous regime:][]{2013MNRAS.435.1758S}. This interpretation is consistent with theoretical models of super-Eddington accretion flows \citep[{\it e.g.},][]{2012ApJ...752...18K}. Conversely, ULXs seen at low inclination angles ({\it i.e.}, looking down the polar funnel) have a harder spectrum, caused by higher-energy photons emitted in the innermost part of the inflow.

By contrast, ULSs are dominated by the thermal component, at even lower temperatures than ULXs; in most observations, that is in fact the only component significantly detected. At some epochs, a non-dominant harder emission in the 1--5 keV band is also detected (Section 4.2). This harder component is consistent with multi-temperature thermal-plasma emission; however, given the relatively low signal-to-noise ratio, we are unable to rule out the alternative interpretation of a soft Comptonized component with superposed absorption lines from partially ionized plasma. In the former case, the harder component may be due to emission from shock-heated gas just outside the outflow photosphere; in the latter case, we may be seeing occasional glimpses of the hard emission from the innermost part of the inflow, down-scattered and partly absorbed through the outflow. The presence of absorption edges at $\approx$ 1 keV in some sources at some epochs (most notably, the M101\,ULS in the {\it Chandra} ObsID 934, the NGC\,247 ULS in the {\it Chandra} ObsID 12437, and the NGC\,4631 ULS in the {\it XMM-Newton} observation) is another indication of absorption through clumpy, variable wind.

\begin{figure}
    \hspace{-0.65cm}
    \centering
    \includegraphics[width=55mm,angle=270]{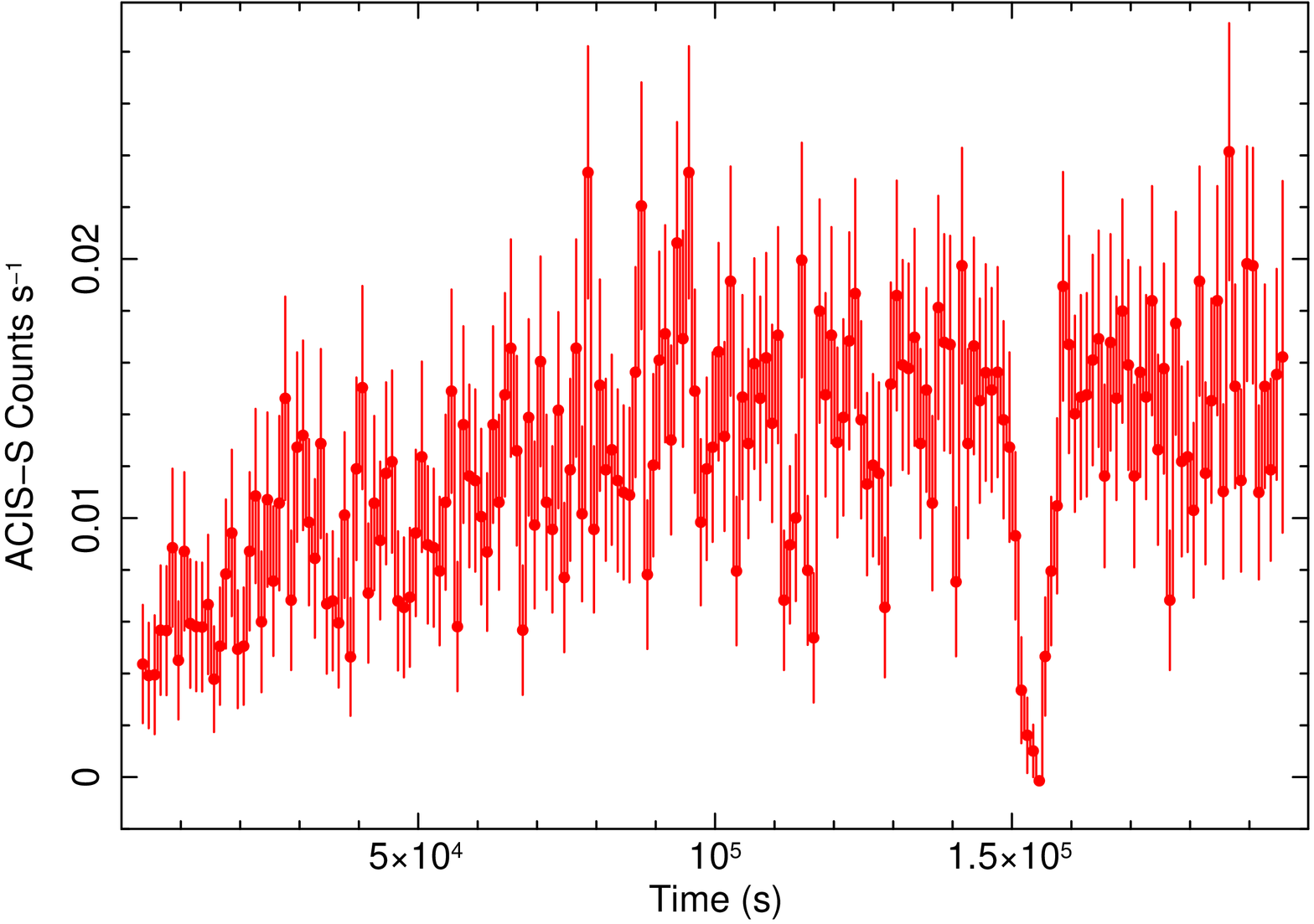}\\[8pt]
    \hspace{-0.65cm}     
    \includegraphics[width=55mm,angle=270]{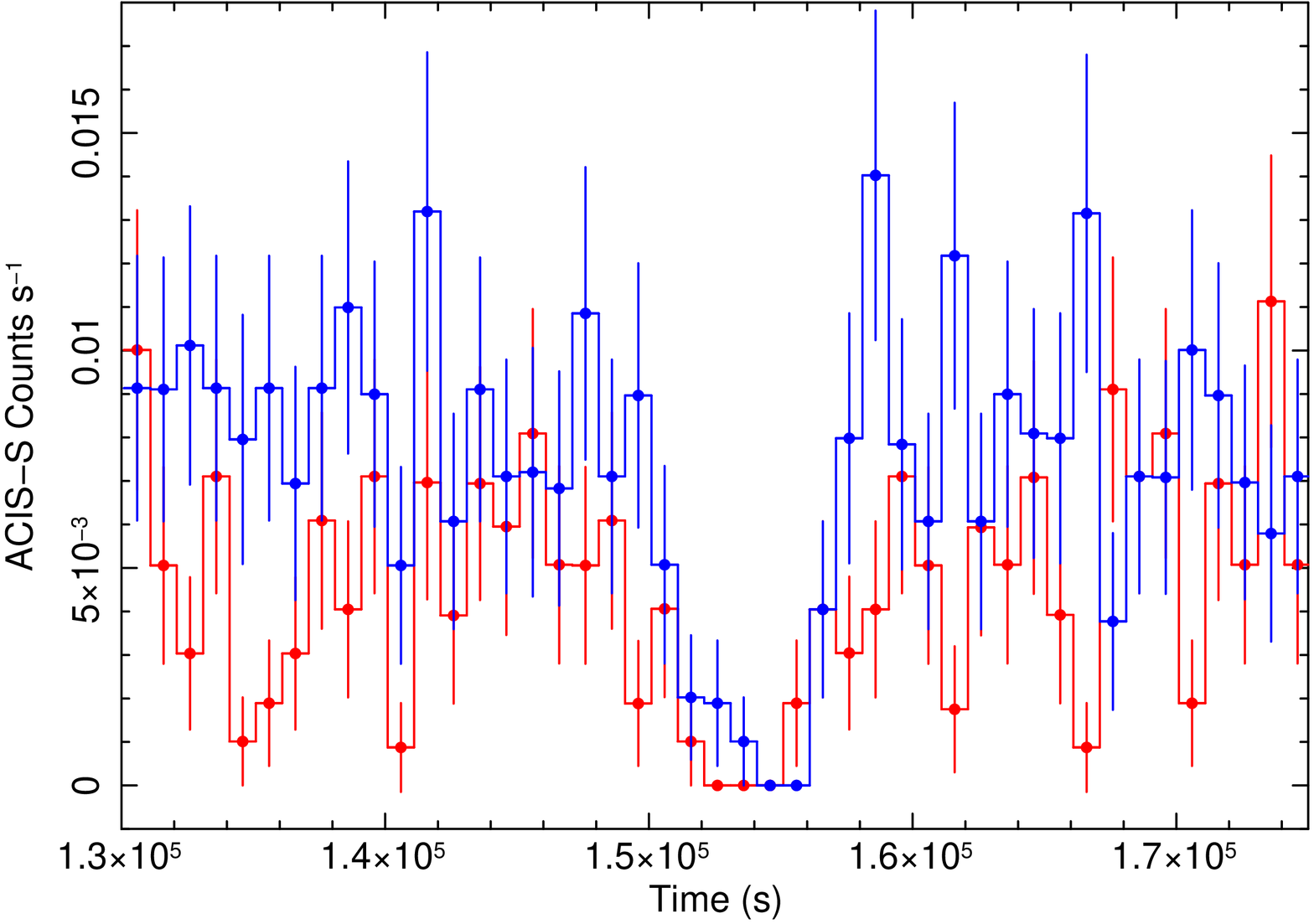}\\[8pt]
    \hspace{-0.65cm} 
    \includegraphics[width=55mm,angle=270]{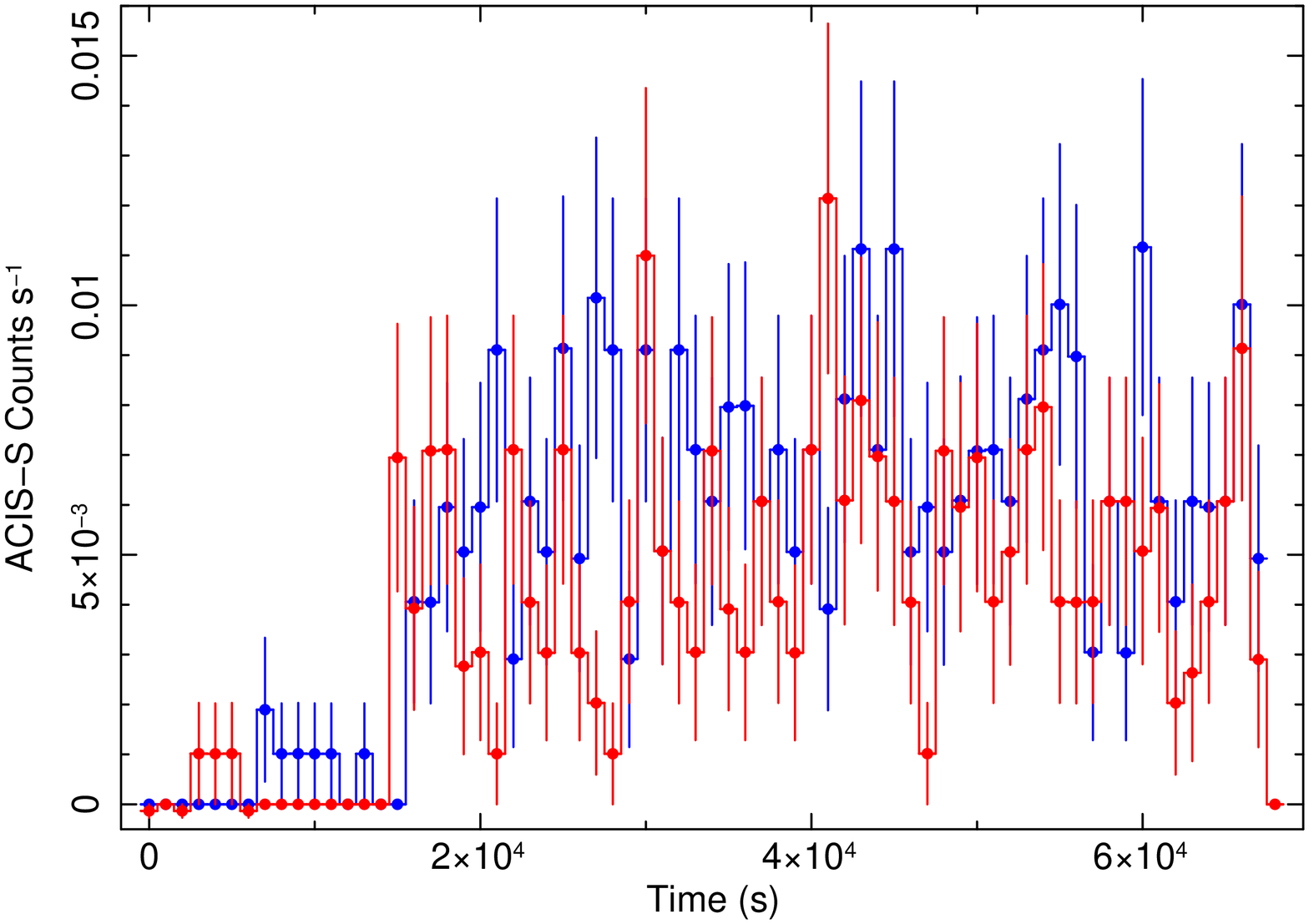}
\caption{Top panel: {\it Chandra}/ACIS-S background-subtracted lightcurve of the M\,51 ULS during ObsID 13814, in the 0.3--1.5 keV band (1000-s bins). Middle panel: zoomed-in view of the time around the dip, during ObsID 13814 (1000-s bins). Red datapoints = 0.3--0.7 keV band; blue datapoints = 0.7--1.5 keV band. Bottom panel: {\it Chandra}/ACIS-S 0.3--1.5 keV lightcurve of the M\,51 ULS during ObsID 13815 (1000-s bins). Red datapoints = 0.3--0.7 keV band; blue datapoints = 0.7--1.5 keV band.}
\label{M51lightcurves_fig}
\end{figure}

\begin{figure}
    \hspace{-0.65cm}
    \centering
    \includegraphics[width=55mm,angle=270]{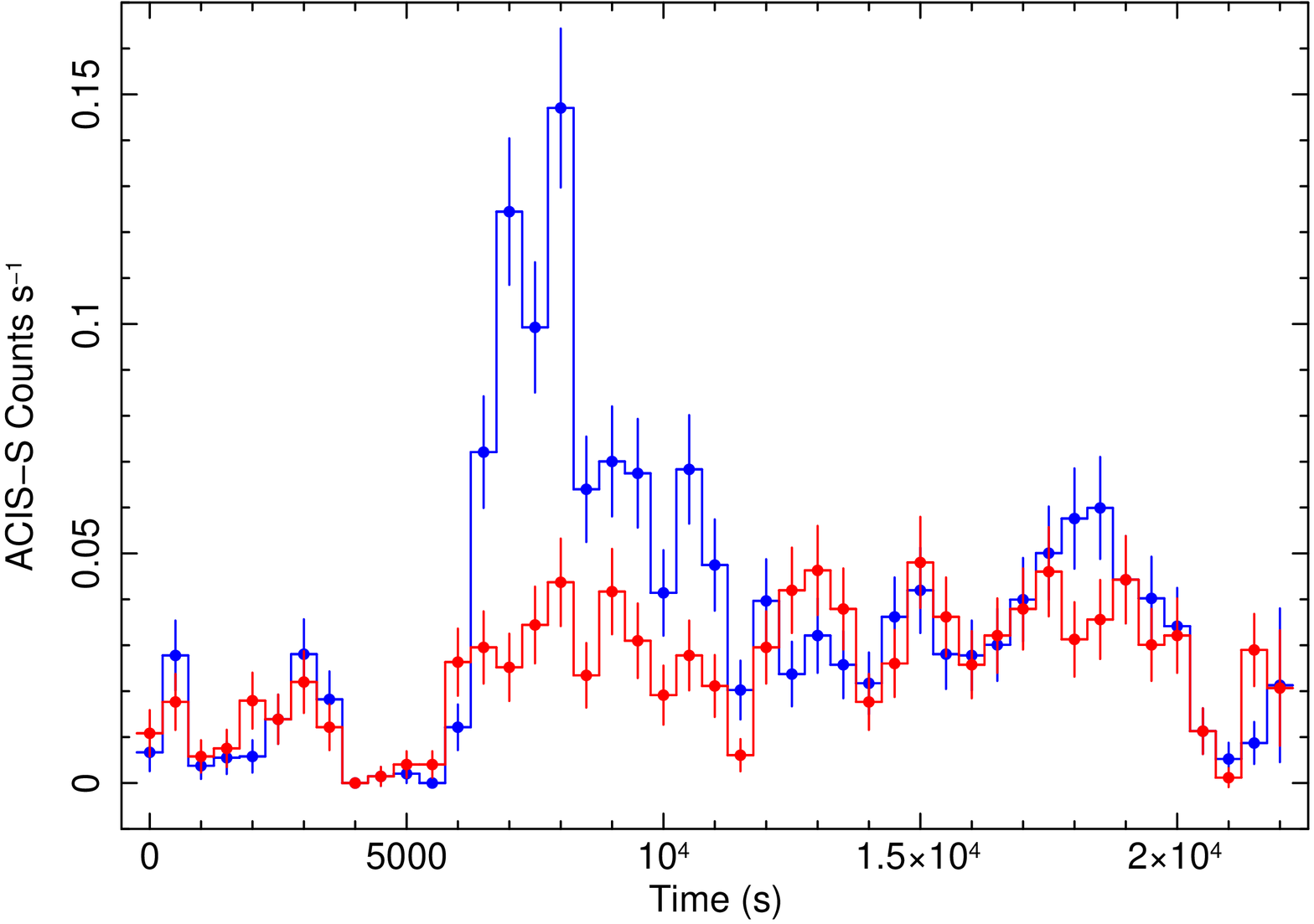}\\[8pt]
    \hspace{-0.65cm} 
    \includegraphics[width=55mm,angle=270]{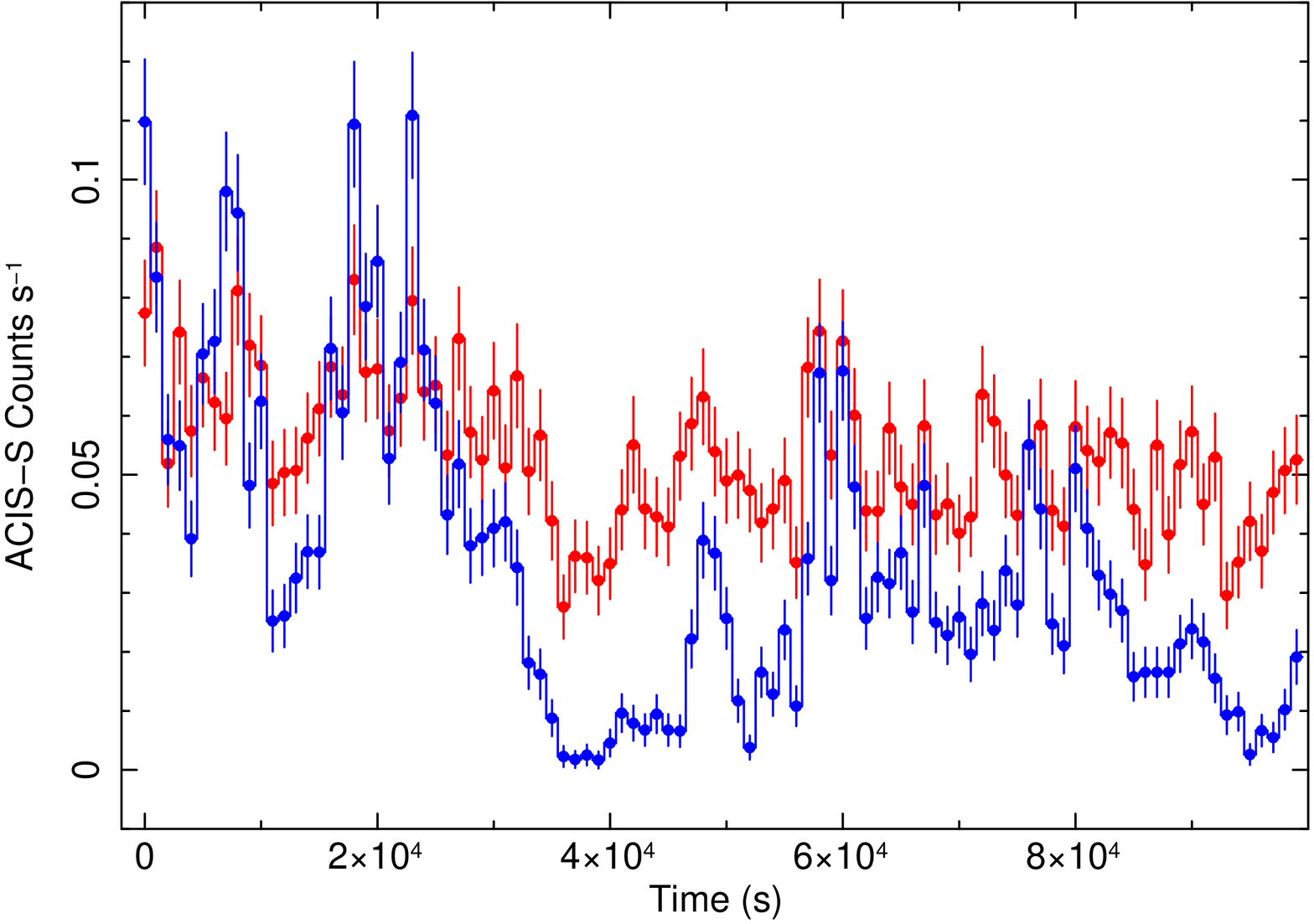}
\caption{Top panel: {\it Chandra}/ACIS-S background-subtracted lightcurve of the M\,101 ULS during ObsID 4737 (500-s bins). Red datapoints = 0.3--0.7 keV band; blue datapoints = 0.7--1.5 keV band. Bottom panel: {\it Chandra}/ACIS-S background-subtracted lightcurve of the M\,101 ULS during ObsID 934 (1000-s bins). Red datapoints = 0.3--0.7 keV band; blue datapoints = 0.7--1.5 keV band.}
\label{M101lightcurves_fig}
\end{figure}

The intense short-term variability seen in many ULSs (Table \ref{rms_table}, and Figures \ref{M81_eclipse}--\ref{extralightcurves_fig}) provides a further observational test of the relation between ULSs and ULXs. Low variability is seen from (sub-Eddington) standard disks in the high/soft state \citep{2010LNP...794...53B}. Instead, high variability is seen in ULXs in the soft-ultraluminous regime \citep{2013MNRAS.435.1758S}, in which the fractional variability appears to increase at higher energy bands \citep{2011MNRAS.411..644M,2013MNRAS.435.1758S}.
For ULXs, this finding was interpreted as evidence that most of the variability is associated with the high-energy tail rather than the soft thermal emission; \citet{2013MNRAS.435.1758S} argued that the variability is due to clumpy material at the edge of the super-Eddington outflow intermittently occulting our direct view of the hot central regions of the inflow. For the brightest ULS in our sample, the one in M\,101, we also find that when a harder component is present, its emission is more variable and distinct from than that of the soft component (Figure \ref{M81_eclipse}), which is consistent with the ULX interpretation. However, we stress again that with the data at hand, we cannot yet determine whether the harder ULS component is optically-thin thermal plasma emission in the outflow, or a glimpse of direct hard emission from the innermost part of the inflow, down-scattered and absorbed through the outflow. 

The soft thermal component is itself variable from observation to observation and sometimes within individual observations. We showed (Section 4.1) that there is a correlation between lower blackbody temperatures and higher blackbody radii. For example, the fitted temperature of the thermal component in the M\,101 ULS went from $< 50$ eV (90\% confidence limit) on 2004 December 22--24, up to $75\pm6$ eV on 2004 December 30, then $100 \pm 10$ eV on 2005 January 1, and down to $56\pm5$ eV on 2005 January 8 (Table \ref{params_tab}). At the same epochs, the fitted radius changed from $> 54,000$ km (90\% confidence limit) to $43,500^{+29,300}_{-17,200}$ km, to $22,500^{+10,300}_{-4,700}$ km, and back up to $100,000^{+80,000}_{-40,000}$ km. This behaviour suggests that the soft emission comes from the moving photosphere of a variable outflow rather than a fixed-sized structure like an accretion disk, consistent with the interpretation for ULXs. We speculate that we are seeing day-to-day stochastic variability in the density and kinetic energy of the outflow, analogous to the stochastic variability in X-ray luminosity seen in bright X-ray binaries.

Based on the empirical properties summarized above, we propose that ULSs are the most extreme form of soft-ultraluminous regime, in which the hard photons from the innermost regions are completely (or almost completely) masked and reprocessed by the optically-thick outflow.
Like soft-ultraluminous ULXs, ULSs are likely to be seen at higher inclination angles. We propose that the key qualitative and quantitative difference between soft-ultraluminous ULXs and ULSs is that in the former, the outflow is optically thick to scattering but still effectively optically thin; in ULSs, it is effectively optically thick \citep[see][for a definition of effective optical thickness, $\tau_{\nu}^*$]{1979AstQ....3..199R}. In quantitative terms, we can express this condition as follows. Let us assume that the wind (with density $\rho$) is launched from radius $R$ and observed from infinity; let us define an absorption opacity $\kappa^a_\nu$ and a scattering opacity $\kappa_{\rm s}$. For ULSs we have:
\begin{displaymath}
\tau^\ast_\nu(R) = \int_{R}^{\infty} \rho 
                \sqrt{\kappa^a_\nu \left(\kappa^a_\nu + \kappa_{\rm s}\right)} \, {\mathrm{d}}r 
                \approx \int_{R}^{\infty} \rho  \sqrt{\kappa^a_\nu \kappa_{\rm s}} \, {\mathrm{d}}r
                 > 1 
\end{displaymath}\\[-20pt]
\begin{equation}
\tau_{\rm s}(R) =  \int_{R}^{\infty} \rho \, \kappa_{\rm s} \, {\mathrm{d}}r > 1,
\end{equation}
while for ULXs it is:
\begin{displaymath}
\tau^\ast_\nu(R) = \int_{R}^{\infty} \rho 
                \sqrt{\kappa^a_\nu \left(\kappa^a_\nu + \kappa_{\rm s}\right)} \, {\mathrm{d}}r 
                \approx \int_{R}^{\infty} \rho  \sqrt{\kappa^a_\nu \kappa_{\rm s}} \, {\mathrm{d}}r
                 < 1 
\end{displaymath}\\[-20pt]
\begin{equation}
\tau_{\rm s}(R) =  \int_{R}^{\infty} \rho \, \kappa_{\rm s} \, {\mathrm{d}}r > 1.
\end{equation}

We suggest that this is the main reason why ULXs still have a dominant broad-band tail (rolling over at $E \ga 5$ keV) even in the soft ultraluminous state, consisting of hard photons down-scattered multiple times but not absorbed by the outflow. Instead, ULSs lose their tail and are dominated by soft thermal photons emitted near the photosphere. In order for the outflow to be effectively optically thick, at a given viewing angle, the mass outflow rate (likely to be proportional to the mass accretion rate at the outer edge of the disk) must exceed a characteristic threshold, which in a spherical approximation can be estimated as $\dot{m} \ga$ a few 100 \citep[][Soria \& Kong 2015, MNRAS, in press, arXiv:1511.04797]{2015MNRAS.447L..60S}. Soft-ultraluminous ULXs are more likely to be in a regime with $10 \la \dot{m} \la 100$.  These numbers are purely indicative, and depend on our viewing angle: we do not expect the wind to be optically thick along near-face-on line-of-sight regardless of accretion rate, while more moderate accretion rates may be enough to produce an optically thick outflow if the source is viewed almost edge on. For the same reason, the fitted photospheric radius will appear larger when a system is seen more edge-on. Then, epoch-to-epoch variability in ULSs may occur because of changes in the accretion rate (and consequently in the outflow density), or because of changes in our viewing angle, due to disk precession.

If our scenario is correct and ULSs have a highly super-critical mass accretion rate, we might wonder why their extrapolated bolometric luminosities (Table \ref{params_tab}) are only $\sim$ a few $10^{39}$ erg s$^{-1}$, barely reaching $10^{40}$ erg s$^{-1}$ in the most luminous epochs. This is a few times lower than the most luminous ULXs found in the same volume of space. This can be justified as follows. 
Firstly, given the logarithmic dependence of emitted luminosity on $\dot{m}$, with (in the outflow-dominated case) $L \sim L_{\rm Edd} \,(1+0.6\ln \dot{m})$ \citep{1973A&A....24..337S, 2007MNRAS.377.1187P}, the supposedly higher accretion rate of ULSs does not make a large difference: for example, for $\dot{m} = 30$, $L \approx 3L_{\rm Edd}$, while for $\dot{m} = 300$, $L \approx 4.4L_{\rm Edd}$. Secondly, in the model of Soria \& Kong (2015, MNRAS, in press, arXiv:1511.04797), the main parameter that determines whether or not the outflow is effectively optically thick along a given line of sight (and therefore whether the source will appear as a ULS or a ULX along that line of sight) is the dimensionless mass accretion rate $\dot{m}$ (normalized by BH mass), not the absolute accretion rate $\dot{M}$. The absolute value of $\dot{M}$ available for accretion is limited by the evolutionary stage of the donor star, but for a fixed $\dot{M}$, smaller BHs (and NSs) will have a higher value of $\dot{m}$ and therefore a higher chance to be seen as ULSs than more massive BHs. For example, a 50-$M_{\odot}$ BH with $\dot{M} \approx 2 \times 10^{21}$ g s$^{-1}$ will have $\dot{m} \approx 30$ (probably not high enough to make the outflow effectively optically thick) and an intrinsic luminosity $L \approx 2 \times 10^{40}$ erg s$^{-1}$ (typical of the brightest ULXs). Instead, a 10-$M_{\odot}$ BH with $\dot{M} \approx 4 \times 10^{21}$ g s$^{-1}$ will have $\dot{m} \approx 300$ and an intrinsic luminosity $L \approx 6 \times 10^{39}$ erg s$^{-1}$ (typical of ULSs). Thirdly, if the outflow is denser and effectively optically thick as we suggest it is in ULSs, a larger fraction of radiative power is absorbed and converted to thermal and kinetic energy of the outflow: the photospheric luminosity may be only $\sim 20$--$50$ per cent of the intrinsic radiative power \citep{1999AstL...25..508L, 2007MNRAS.377.1187P}. Finally, the higher apparent luminosity of hard-ultraluminous ULXs is boosted by geometric collimation along the polar funnel, while the apparent luminosity of ULSs is more likely to be reduced, if they are seen at high inclination angles.

Another corollary of our proposed scenario is that ULSs may become similar to soft-ultraluminous ULXs when their photospheric radius seen along our line of sight decreases (corresponding to a decrease in $\dot{m}$ and in the total mass in the outflow or to a change in viewing angle), thus revealing the underlying inverse-Compton scattering region and/or inner disk region, sources of the harder emission component. This is consistent with our observed temperature distribution of the thermal component in ULSs (Figure \ref{SSS_ULX_fig}): sources with $kT_{\rm bb} \la 150$ eV are mostly ULSs, while sources with $kT_{\rm bb} \ga 150$ eV generally contain a harder component and are classified as standard ULXs. For temperatures $\approx 100$--150 eV we expect a degree of overlapping between the two classes, as for the same accretion rate and outflow structure, a system can appear as a ULX if seen more face-on, or as a ULS if seen more edge-on. The observed appearance of harder emission components in some ULSs (particularly those in M\,101 and NGC\,247) only when their blackbody temperature reached $\approx$130 eV and their radius shrank below $\approx$20,000 km may be additional evidence in favour of our proposed connection.

\begin{figure*}
    \hspace{-0.8cm}
     \centering
     \includegraphics[width=55mm,angle=270]{lc_Ant-3042_gehrels.ps}
     \includegraphics[width=55mm,angle=270]{lc_NGC247-17547_gehrels.ps}\\[8pt]
     \hspace{-0.62cm} 
     \includegraphics[width=55mm,angle=270]{lc_NGC300-9883_gehrels.ps}
     \includegraphics[width=55mm,angle=270]{NGC4631_0110900201_lc.ps}
\caption{Top left panel: {\it Chandra}/ACIS-S background-subtracted lightcurve of the Antennae ULS during ObsID 3042 (2000-s bins). Top right panel: {\it Chandra}/ACIS-S background-subtracted lightcurve of the NGC\,247 ULS during ObsID 17547 (250-s bins); see Figure \ref{lightcurves_ngc247_fig} for a spectrum of that same observation.  Bottom left panel: {\it Chandra}/ACIS-S background-subtracted lightcurve of the NGC\,300 ULS during ObsID 9883 (500-s bins); only soft (0.3--0.7 keV) emission was detected. Bottom right panel: {\it XMM-Newton}/EPIC-pn background-subtracted lightcurve of the NGC\,4631 ULS during ObsID 0110900201 (500-s bins). In all panels, red datapoints = 0.3--0.7 keV band; blue datapoints = 0.7--1.5 keV band. Error bars were calculated using Gehrels statistics.}
\label{extralightcurves_fig}
\end{figure*}

\section{Conclusions}

Individually identified ULSs in nearby galaxies have often been considered odd sources, difficult to place in the X-ray binary and ULX classification schemes. Here we have re-examined {\it Chandra} and {\it XMM-Newton} data for a sample of seven previously identified ULSs, showing that they share common properties: a thermal spectrum with a characteristic range of temperatures, luminosities, sizes, and an anticorrelation between radius and temperature, consistent with $r_{\rm bb} \sim T_{\rm bb}^{-2}$. Thus, they may represent a new sub-type or state of accreting systems. 

We discussed the main alternative interpretations of ULSs presented in the literature, and showed that the IMBH scenario is strongly disfavoured for several reasons: rapid changes in the fitted radius of the thermal component; inconsistency of temperatures and luminosities with the standard disk parameters; strong short-term flux variability. Steady nuclear burning on the surface of a WD (with a possibly inflated or outflowing atmosphere) is usually invoked for classical SSSs at luminosities $\la 10^{38}$ erg s$^{-1}$, but there are no convincing models able to explain steady sustained luminosities $\sim 10$--$50 L_{\rm Edd}$ (for a WD) over decades without the triggering of nova-like outbursts. Instead, we argued that optically-thick outflows from super-critically accreting BHs (and possibly also NSs) are the most likely explanation for this class. In particular, a clumpy optically-thick wind with an expanding or contracting photosphere (function of the mass density in the outflow and therefore also related to the accretion rate) seems a natural explanation for the  anti-correlation between observed temperatures and radii, and for the strong short-term variability.

We showed that in some ULSs, a harder emission component is detected alongside the dominant thermal component at some epochs, resulting in a spectrum that sometimes resembles those of typical ULXs; more specifically, those in the soft-ultraluminous regime \citep{2013MNRAS.435.1758S}, which are probably seen through a dense wind at high inclination. The appearance of a harder component in ULSs happens preferentially at epochs when the fitted blackbody radius of the thermal component is smallest ($r_{\rm bb} \la 20,000$ km) and the fitted blackbody temperature is highest ($T_{\rm bb} \ga 100$ eV), consistent with a reduced size of the photosphere and therefore a more direct view of the inner regions. Based on this finding, and on the analogies between the thermal component in ULXs and ULSs, we proposed that ULSs are simply the regime of the ULX population seen through the densest outflow: a state in which any direct, harder emission from the inner disk and the inverse-Compton region has been completely masked and thermalized by the optically thick wind. In our scenario, the difference between ULSs and ULXs in the soft-ultraluminous regime is that the outflow is effectively optically thick in ULSs (leading to complete absorption and thermal reprocessing of the harder photons from the inner region) and only optically thick to scattering in ULXs, sources in which a harder tail is still carrying most of the flux. This may be due to a higher viewing angle and/or a higher accretion and outflow rate in ULSs. As previously discovered in soft-ultraluminous ULXs \citep{2013MNRAS.435.1758S, 2015MNRAS.447.3243M}, we found at least for one ULS in our sample (M\,101 ULS, the only one with sufficiently high signal-to-noise ratio for this kind of analysis) that the harder component has a higher fractional variability than the softer component. 

The nature of the harder component remains unclear, given the relatively low signal-to-noise ratio of even the brightest sources, and the fact that only CCD-resolution spectra are available for ULSs. Such a low spectral resolution, compounded by the degradation in the soft response of {\it Chandra}/ACIS-S, makes it almost impossible to identify individual emission or absorption features in the soft and medium energy bands. However, even at this low resolution, we confirmed that several ULSs have strong absorption edges at $E \approx 1$ keV \citep[as noted by][]{2004ApJ...617L..49K}, which appear and disappear at different epochs. Their physical interpretation remains unclear, and is beyond the scope of this work, but it is another piece of evidence in favour of massive, optically thick outflows in those sources. As for the excess emission above the dominant thermal component, particularly in the 0.7--1.5 keV band, we showed that it is consistent with thermal plasma emission, but we cannot rule out more complex interpretations, such as an inverse-Compton component with superposed absorption features caused by absorption in a partially ionized outflow, as suggested by \citet{2014MNRAS.438L..51M} for analogous spectral features in some ULXs.

Using a spherically symmetric approximation and standard assumptions about the launching radius of the wind, we estimated that  $\dot{m} \sim$ a few 100 is required to make the outflow effectively optically thick (although, this value is likely to be an overestimate for more edge-on systems). Thus, if our interpretation is correct, ULSs are some of the best case studies of super-critical accretion and super-Eddington outflows in the local Universe. Their extrapolated bolometric luminosities are between a few $\times 10^{39}$ erg s$^{-1}$ and $\approx$1 $\times 10^{40}$ erg s$^{-1}$, a few times lower than for the most luminous ULXs. We explained (Section 5.3) why this is not in contradiction with our suggestion that ULSs have extremely high $\dot{m}$. This is partly because a substantial fraction of the initial radiative power is converted to mechanical and thermal energy of the optically-thick outflow; partly because ULSs may have higher $\dot{m}$ but lower BH mass and lower intrinsic luminosity than the most luminous ULXs; and partly because at least some of the (apparently) most luminous ULXs may benefit from geometric collimation of the emission along the polar direction. 

Another consequence of our proposed scenario is that when the photosphere expands beyond $\approx$100,000 km, and the blackbody temperature decreases below $\approx$50 eV (either because of real physical changes in the two quantities, or because of a change in our line of sight), a ULS should become undetectable in the X-ray band, its thermal emission having shifted to the far UV. ULSs that were detected as bright sources in some observations are sometimes not detected in others. M\,101 ULS provides a typical example, with a detected {\it Chandra}/ACIS-S 0.3--2 keV count rate $\approx$0.1 ct s$^{-1}$ in the brightest observation, but $< 5 \times 10^{-4}$ ct s$^{-1}$ in 14 of the 25 {\it Chandra} observations examined in Soria \& King (2015, submitted). The NGC\,300 ULS is another good example: detected in a bright state in one {\it Chandra} and two {\it Xmm-Newton} observations (listed in Table \ref{obs_table} and analyzed in this paper), it was not detected in another three {\it Chandra} and three {\it XMM-Newton} observations. By analogy with the behaviour of transient X-ray binaries, we could interpret those non detections (or extremely faint detections) as the off (or low) state of ULSs. However, this would be hard to reconcile with our classification, because it is very contrived to imagine that an accreting BH can only switch between extremely super-Eddington ($\dot{m} > 100$) and low sub-Eddington ($\dot{m} < 0.01$) accretion rates, without ever being seen in other canonical accretion states ({\it e.g.}, hard intermediate state, high/soft state, broadened-disk ULX state). Our scenario implies that a ULS non-detection at a certain epoch is not due to a real drop in luminosity, and is caused instead by one of the following three reasons: (a) its photosphere has physically expanded to the point that the source is now in an ultraluminous UV regime (perhaps similar to the ``ultraluminous UV source" in NGC\,6946 discussed by \citealt{2010ApJ...714L.167K}); or (b) our view of the source is completely occulted by cold absorbing material (as is briefly the case during the dips observed at some epochs for the M\,81 and the M\,51 sources); or (c) our line-of-sight has changed (become more edge-on) due to systemic precession. 

In future work we will explore other aspects of the relation between ULSs and ULXs that remain unclear. For example, we do not know yet whether ULSs are preferentially located in younger or older stellar environments: in the former case, the mass donor may be an OB star, while in the latter case they may be fed by an intermediate-mass star going through the Hertzsprung Gap (Wiktorowicz et al.~2015). Many ULXs are surrounded by large collisionally-ionized bubble nebulae, possibly powered by jets (Feng \& Soria 2011); no systematic search for ULS bubbles has been done yet. For example, a lack of ULS bubbles could indicate that fast collimated jets are suppressed at the highest super-Eddington rates, or that ULSs are located in older environments with lower-density interstellar medium.

\section*{Acknowledgments}

We thank Albert Kong, Jan-Uwe Ness, Doug Swartz, Manfred Pakull, Chris Done, Yan-Fei Jiang, Rosanne Di Stefano, Peter Curran, James Miller-Jones, Thomas Russell, Laura Shishkovsky, Vlad Tudor for useful suggestions and discussions. We also thank Jan-Uwe Ness for providing the fit parameters in ascii form, for his sample of classical supersoft sources.  
RS acknowledges support from a Curtin University Senior Research Fellowship. 
He is also grateful for support and hospitality at Strasbourg Observatory during part of this work. 
This paper benefitted from discussions at the 2015 International Space Science Institute 
workshop ``The extreme physics of Eddington and super-Eddington accretion onto Black Holes''
in Bern, Switzerland (team PIs: Diego Altamirano \& Omer Blaes).

\label{lastpage}

\end{document}